\documentstyle[12pt]{report}
\begin{document}
% ***************    NEW COMMANDS   *******************
\newcommand{\cuq}{{\cal Q}}
\newcommand{\bcc}{\begin{center}}
\newcommand{\ecc}{\end{center}}
\def \inbar{\vrule height1.5ex width.4pt depth0pt}
\def \C{\relax\hbox{\kern.25em$\inbar\kern-.3em{\rm C}$}}
\def \R{\relax{\rm I\kern-.18em R}}
\newcommand{\Z}{\ Z \hspace{-.08in}Z}
\newcommand{\be}{\begin{equation}}
\newcommand{\ee}{\end{equation}}
\newcommand{\bea}{\begin{eqnarray}}
\newcommand{\eea}{\end{eqnarray}}
\newcommand{\p}{\psi}
\renewcommand{\l}{\epsilon}
\newcommand{\f}{\phi}
\newcommand{\g}{\gamma}
\newcommand{\G}{\Gamma}
\newcommand{\e}{\eta}
\newcommand{\z}{\pi}
\newcommand{\m}{\mu}
\newcommand{\n}{\nu}
\newcommand{\s}{\sigma}
\renewcommand{\t}{\tau}
\renewcommand{\a}{\alpha}
\renewcommand{\b}{\beta}
\newcommand{\k}{\kappa}
\renewcommand{\d}{\delta}
\newcommand{\r}{\rho}
\newcommand{\th}{\theta}
\renewcommand{\pt}{\frac{\partial}{\partial t}}
\newcommand{\ppt}{\frac{\partial^{2}}{\partial t^{2}}}
\newcommand{\nn}{\nonumber}
\renewcommand{\ll}{\left[ }
\newcommand{\rr}{\right] }
\newcommand{\kt}{\rangle}
\newcommand{\br}{\langle}
\newcommand{\fs}{\small}
\newcommand{\so}{S_{0}}
\newcommand{\I}{\mbox{1}_{m\times m}}
\newcommand{\In}{\mbox{1}_{n\times n}}
\newcommand{\xo}{x_{0}}
\newcommand{\po}{\psi_{0}}
\newcommand{\eo}{\e_{0}}
\newcommand{\ts}{\tilde{S}}
\newcommand{\pss}{\frac{\partial}{\partial s}}
\newcommand{\tcuf}{\tilde{\cal F}}
\newcommand{\cuf}{{\cal F}}
\newcommand{\oot}{\mbox{\fs$\frac{1}{2}$}}
\newcommand{\iot}{\mbox{\fs$\frac{i}{2}$}}
\newcommand{\iv}{\imath\! v}
\newcommand{\ib}{\int_{0}^{\b}}
\newcommand{\lll}{\left(}
\newcommand{\rrr}{\right)}
\newcommand{\llc}{\left\{ }
\newcommand{\rrc}{\right\}}
\newcommand{\lpt}{\left.}
\newcommand{\rpt}{\right.}
\newcommand{\rar}{\longrightarrow}
\newcommand{\lar}{\longleftarrow}
\newcommand{\cinf}{C^{\infty}\!}
\newcommand{\vol}{d\mbox{\bf\fs$\Omega$}}
\newcommand{\sx}{\mbox{\fs$(x)$}}
\newcommand{\hD}{\hat{D}}
\newcommand{\V}{(V)}
\newcommand{\La}{\Lambda}
\newcommand{\ph}{{\cal P}({\cal H})}
\newcommand{\cp}{\C\! P}
%\documentstyle[12pt]{report}
%\begin{document}
%\onehalfspacing
%\onehalfspacequote
%\longtocentry
\title{Supersymmetry, Path Integration, and \vspace{2.4mm} \\
       the Atiyah-Singer Index Theorem}
%\supervisor{Bryce S.~DeWitt}
%\committeesize{5}
\author{Ali Mostafazadeh\thanks{I would like to thank my Ph.~D.
supervisor Prof.~Bryce S.~DeWitt for his invaluable guidance
and support. I am also indebted to Profs. Cecile DeWitt-Morette,
Luis J.~Boya, and Gary Hamrick for
many constructive comments and suggestions. This dissertation is
dedicated to my parents: Afsar Mosannen-Amini and Ebrahim Mostafazadeh.}
\\ The University of Texas at Austin\\
Department of Physics\\ Austin, Texas 78712}
%\previousdegrees{B.S.}
%\degree{Doctor of Philosophy}
%\address{UT-Physics, RLM 5.214 \\ Austin, Texas 78712 \\ U.S.A.}
%\graduationmonth{May}
%\graduationyear{1994}
%\begin{document}
%\signaturepage
%\copyrightpage
%\begin{dedication}
%This dissertation is dedicated to \\
%my parents Afsar Mosannen-Amini and Ebrahim Mostafazadeh.
%\end{dedication}
\date{May 1994}
\maketitle

%\end{document}
%\acknowledgements

%\hspace{1.27cm}I would like to thank my advisor Prof. Bryce S.~DeWitt
%for his guidance,
%invaluable discussions on different aspects of the subject of this
%dissertation, and his
%support and kindness. I consider it a great honor and privilege to have him as
%my advisor.

%I would also like to acknowledge Profs. Cecile DeWitt-Morette and
%Luis J.~Boya for many constructive comments, and my friends Ertu\u{g}rul
%Demircan, Teoman Turgut, Siamak Sadat-Gousheh, and Karman Saririan
%for their support and encouragement. Finally, I am greatly indebted to my
%parents Afsar Mosannen-Amini and Ebraim Mostafazadeh who made many
%sacrifices in their lives to support me,
%particularly since I left home some ten years ago.

%\title{Supersymmetry, Path Integration, and \vspace{0.8mm} \\
%     the Atiyah-Singer Index Theorem}

%\begin{center}
%This dissertation is dedicated to my parents:\\
%Afsar Mosannen-Amini and Ebrahim Mostafazadeh.
%\end{center}

%\newpage
\begin{abstract}

A new supersymmetric proof of the Atiyah-Singer index theorem is presented.
The Peierls bracket quantization scheme is used to quantize the supersymmetric
classical system corresponding to the index problem for
twisted Dirac operator. The problem of factor ordering is addressed and
the unique quantum system that is relevant to the index theorem is
analyzed in detail. The Hamiltonian operator for the supersymmetric
system is shown to include a scalar curvature factor,
$\hbar^2 R/8$. The path integral formulation
of quantum mechanics is then used to obtain a formula for the index.
For the first time, the path integral ``measure'' and the Feynman
propagator of the system are exactly computed. The derivation of
the index formula relies solely on the definition of a Gaussian
superdeterminant. The two-loop analysis of the path integral is also
carried out. The results of the loop and the heat kernel expansions of
the path integral are in complete agreement. This confirms the
existence of the scalar curvature factor in the Schr\"odinger equation
and validates the supersymmetric proof of the index theorem. Many
other related issues are addressed. Finally, reviews of the index theorem
and the supersymmetric quantum mechanics are presented.
\end{abstract}

\tableofcontents
\chapter{Introduction}
\label{I-1}
\vspace{1cm}
\bcc
{\bf Abstract}
\ecc
\bcc
\parbox[b]{4.5in}{\small
The Atiyah-Singer index theorem is introduced and its relation to
supersymmetric quantum mechanics is discussed.}
\ecc
\vspace{.5in}

In 1916, in his quest to unravel the mystery of {\em Gravitation},
Einstein proposed one of the most unsual and ingenious theories
of his time. He named this new theory of gravity the {\em General
Theory of Relativity.}\footnote{In this regard, W.~Pauli said: ``The
general theory of relativity then completed and -- in contrast to the
special theory -- worked out by Einstein alone without simultaneous
contributions by other researchers, will forever remain the classic
example of a theory of perfect beauty in its mathematical structure.''
Also the remarks of M.~Born is worthy of quoting: ``(The general theory of
relativity) seemed and still seems to me at present to be the greatest
accomplishment of human thought about nature; it is a most remarkable
combination of philosophical depth, physical intuition and mathematical
ingenuity. I admire it as a work of art.'' \cite[p.~77]{straumann}.}
Although, general relativity
turned out not to be the ultimate theory of gravitation, it
opened many new horizons in theoretical physics. Probably, the most important
of these was the introduction of differential geometric methods
in theoretical physics.

In fact, before Einstein differential
geometric concepts were used by Poincar\'e in the study
of the foundations of classical mechanics. Poincar\'e, among his numerous
discoveries, had introduced certain qualitative techniques to study
the differential
equations of classical  mechanics. These were associated with the
spaces of solutions of these equations and  gave important
information about the systems under investigation. The ``qualitative
techniques''
of Poincar\'e has since been grown into a vast area of modern mathematics
called {\em topology}.

An important problem in topology is to study the topological properties
of manifolds and fibre bundles. These structures
are widely used in other branches
of mathematics and the natural sciences. Particularly, the use
of manifolds and fibre bundles in theoretical physics are clearly
indispensable. The topological properties of manifolds and fibre
bundles play substantial roles in the construction and the
understanding of physical theories. Among the examples of the applications
of topology in theoretical physics are {\em monopoles} \cite{monopole},
{\em instantons} \cite{instanton}, {\em anomalies} \cite{anomaly}
in field theory, and {\em geometric phase} \cite{berry} and {\em quantum
Hall effect} \cite{Hall} in quantum mechanics and many others \cite{egh,cd1}.

The topological or global properties of a  manifold
are the properties that are not sensitive to continuous deformations
of the manifold. Thus they do not directly depend on the geometry of
the manifold. The same is true for fibre bundles, namely the topology
of a fibre bundle is independent of the details of a geometry (connection)
on the bundle. It is precisely this point that makes it so difficult to
obtain quantitative measures of the topological properties of these
spaces. The quatities that give topological information about a
manifold or a fibre bundle are called {\em topological invariants}.
The Atiyah-Singer index theorem is one of the few mathematical results
that provides closed formulas for a large class of topological
invariants.

A simple  example of a topological invariant is the Euler-Poincar\'e
characteristic of a Riemann surface \cite{egh,cd1}.
The Euler-Poincar\'e characteristic is related to
the genus of the Riemann surface. On the other hand, a result known as
the Gauss-Bonnet theorem  yields the Euler-Poincar\'e characteristic
as an integral of the  Gaussian curvature over the whole surface.
The Gauss-Bonnet theorem
and its generalization to arbitrary closed Riemannian manifolds
are the best known examples of index theorems.\footnote{See
Appendix~\ref{z2-A9} for the details.}

A brief description of the index theory for arbitrary elliptic
complexes over closed manifolds is peresented in Appendix~\ref{z2-A}. There,
the general statement of the Atiyah-Singer index theorem is given
and it is argued that a result of K-theory reduces the proof
of the general theorem to that of the twisted spin index theorem.
This is a fundamental result which is used in the supersymmetric
proofs of the index theorem.

A review of the supersymmetric quantum mechanics is presented in
Appendix~\ref{z2-B}.
The idea of using supersymmetric quantum mechanics to give proofs
of index theorems was originally suggested by Witten who had discovered
the relation between the two subjects in his study of supersymmetry
breaking, \cite{witten}. Windey \cite{windey} and Alvarez-Gaume
\cite{ag1,ag2,ag3} provided the first supersymmetric proofs of the
index theorem. Few years later, WKB methods were applied to give an
alternative approach to the path integral evaluation of the index
by Ma\~nes and Zumino \cite{manes-zumino}. The same was  achieved
in a concise paper by Goodman \cite{goodman} who used the mode expansion
techniques. Among other remarkable works on the subject are the
mathematical papers of Getzler \cite{getzler} and Bismut \cite{bismut},
and a superspace approach by Friedan and Windey \cite{friedan-windey}.
Other related articles are listed in \cite{others}. \footnote{The
original proof of Atiyah and Singer was published in a series of
classical papers in 1960's, \cite{atiyah-singer}. They were rewarded
the field medal in mathematics for their achievments in the index
theory in 1966.}

In the rest of this chapter, I shall briefly review the index theorem
and give the statements of the spin and the twisted spin index theorems.
Next, I shall describe the relevance of the
supersymmetric quantum mechanics to the index theorem. In Chapter~\ref{I-2},
I shall present a new supersymmetric proof of the index theorem.
This proof  relies only on a universal definition of a Gaussian
superdeterminant. I shall follow the Peierls quantization scheme \cite{bd1}
to  quantize a particular superclassical system.
I will then demonstrate the identity of the Witten index of the corresponding
quantum system and the index of the twisted Dirac operator.
Next I shall compute the ``measure'' of the path integral exactly.
This involves a direct evaluation of $sdet\, G^+$, where $G^+$ is the
advanced Green's function for the Jacobi operator \cite{bd1}. I shall
then derive the full index formula by explicitly evaluating the Feynman
propagator of the theory. The advantage of this approach is that
unlike the earlier supersymmetric proofs, it does not involve any
complicated expansion-approximation techniques.  Moreover, it allows
to address many unresolved problems such as the problem of factor
ordering and the existence of the factor $(sdet\, G^+)^{-\frac{1}{2}}$
in the path integral ``measure.'' Another important issue is the
existence of the scalar curvature factor in the Schr\"odinger equation.
This will be addressed in Chapter~\ref{I-3}. Here, I shall present the
analysis of the path integral upto and including the two-loop order
terms. The result demonstrates a perfect match between the loop
expansion and the heat kernel expansion of the path integral. It also
serves as an important consistency check on the validity of the
supersymmetric proof of the index theorem presented in Chapter~\ref{I-2}.
Finally, the dissertation is supplemented with four appendices.
Appendix~\ref{z2-A} includes a mathematical review of the general
index theorem. Appendix~\ref{z2-B} is devoted to a review of Witten's
supersymmetric quantum mechanics. Appendices~\ref{z2-C} and~\ref{z2-D}
offer a detailed exposition of the derivation of
some of the results used in Chapter~\ref{I-2}. In particular,
Appendix~\ref{z2-D}
offers a discussion of the end point contribution to the coherent state
path integral. The technique introduced here allows for a unified
expression for the path integral by automatically including the
end point contributions in the action functional.

The Plank's constant $\hbar$ will be set to $1$ except in
Sections~\ref{I-2-1} and~\ref{I-3-1} where it is retained for
convenience.

\section{Atiyah-Singer Index Theorem}
\label{I-1-1}
The Atiyah-Singer index theorem is one of the most substantial
achievements of modern mathematics. It has been a developing subject
since its original proof by Atiyah and Singer, \cite{atiyah-singer}.
A clear exposition of the index theorem is given in the classic paper
of Atiyah, Bott, and Patodi, \cite{abp}. For the historical origins
of the index theory, see \cite{atiyah}. Several mathematical proofs of
the index theorem and its variations and generalizations are available
in the literature, \cite{palias,shanahan,gilkey,booss,berline}. A
common feature of all these proofs is the use of K-theory,
\cite{atiyah-k,nash,bundle}.
In particular, the original cobordism proof \cite{atiyah-singer}, the
celebrated heat kernal proof \cite{abp,gilkey}, and the supersymmetric
proofs, \cite{windey,ag3}, of the index theorem are based on a result
of K-theory which reduces the proof of the ``general'' index theorem
to the special case of twisted signature or alternatively twisted spin
index theorems, \cite{abp,shanahan,gilkey,booss}.

In ``general'', one has an elliptic differential operator,
\be
D: {\cal C}^{\infty}(V_{+})\longrightarrow {\cal C}^{\infty}(V_{-})
\label{eq2}
\ee
between the spaces of sections of two hermitian vector bundles
\footnote{A hermitian vector bundle is a complex vector bundle
which is endowed with a hermitian metric and a compatible
connection, see Appendix~\ref{z2-A1} for details.}
$V_{+}$ and $V_{-}$ over a closed \footnote{compact, without boundary}
Riemannian manifold, $M$. Alternatively one can speak of the  short
elliptic complex:
\[ 0 \longrightarrow {\cal C}^{\infty}(V_{+}) \stackrel{D}{\longrightarrow}
{\cal C}^{\infty}(V_{-}) \longrightarrow 0 . \]
The index theorem provides a closed formula for the analytic index of
$D$. The latter is defined by
\be
index(D) := dim[ker(D)]-dim[coker(D)]\; .
\label{eq3}
\ee
The symbols dim, ker, and coker are abbreviations of {\em dimension,
kernel}, and {\em cokernel}, respectively. Indices of elliptic operators
are of great interest in mathematics and physics because they are
topological invariants.\footnote{According to Palais, this was originally
noticed by Gel'fand, \cite{palias}.}

The ``general'' index formula computes the index as an integral
involving some characteristic classes associated with the vector
bundles $V_{\pm}$, the base manifold $M$, and the operator
$D$. A more thorough review of the general index theorem and
the related subjects is provided in Appendix~\ref{z2-A}. Here, I
shall concentrate on the special case of
the {\em twisted (or generalized) spin index theorem}.

Let $M$ be a closed $m=2l$-dimensional spin manifold, \cite{egh}. Let
${\cal S}={\cal S}^{+}\oplus {\cal S}^{-}$ denote the spin bundle
of M. The sections of ${\cal S}^{\pm}$ are called the $\pm$ chirality
spinors. Let $\not\!\! D:{\cal C}^{\infty}({\cal S}) \rightarrow {\cal C}^
{\infty}({\cal S})$ be the Dirac operator \cite{shanahan,gilkey,egh,nash},
and $\not\!\partial :=\not\!\!D \mid_{{\cal C}^{\infty}({\cal S}^{+})}$. The
short complex
\be
0 \longrightarrow {\cal C}^{\infty}({\cal S}^{+}) \stackrel{\not
\hspace{.4mm}\partial}{\longrightarrow}
{\cal C}^{\infty}({\cal S}^{-}) \longrightarrow 0
\label{eq3.1}
\ee
is called the spin complex.
\newcommand{\aroof}{\int_{M}\left[ \hat{A}(M) \right]_{\rm top}}
\newtheorem{thm}{Theorem}
\begin{thm} \label{theorem1}
Let $\not\!\partial$ be as in (\ref{eq3.1}), and  define the so-called
$\hat{A}$-genus density:
\be
\hat{A}(M) := \prod_{i=1}^{l} \left[ \frac{\frac{\Omega_{i}}{4\pi}}{\sinh
(\frac{\Omega_{i}}{4\pi})} \right]\; ,
\label{eq3.2}
\ee
where  $\Omega_{i}$ are the 2-forms defined by block-diagonalizing
\footnote{The block diagonalization is always possible since
$\Omega$ is antisymmetric.} the curvature 2-form:
\be
\Omega :=(\mbox{\fs$\frac{1}{2}$}
R_{\mu \nu \gamma \lambda} dx^{\gamma}\wedge dx^{\lambda})
=: diag \left( \left[ \begin{array}{cc}
0 & \Omega_{i} \\
-\Omega_{i} & 0
\end{array} \right] : i=1 \cdots l \right).
\label{eq5}
\ee
Then,
\be
index(\not\!\partial) = \aroof .
\label{eq4}
\ee
\end{thm}
In (\ref{eq4}), ``top'' means that the highest rank form in the power series
expansion of (\ref{eq3.2}) is integrated.

Let $V$ be a hermitian vector bundle with fiber dimension $n$, base
manifold M, and connection 1-form $A$. Then, the operator:
\[\not\!\! D_{V} :{\cal C}^{\infty}({\cal S}^{\pm})\otimes {\cal C}^{\infty}(V)
\longrightarrow {\cal C}^{\infty}({\cal S}^{\mp})\otimes {\cal C}^{
\infty}(V) \]
defined by
\be
\not\!\! D_{V}(\Psi \otimes v) :=\not\!\! D(\Psi)\otimes v + (-1)^{\Psi}\Psi
\otimes D_{A}(v),
\label{eq6}
\ee
is an elliptic operator, called the {\em twisted Dirac } operator.
In (\ref{eq6}), \\ $\Psi \in{\cal C}^{\infty}({\cal S}^{\pm})
\Leftrightarrow (-1)^{\Psi}=\pm 1$ , $v\in {\cal C}^{\infty}(V)$ ,
and $D_{A}$ is the covariant derivative operator defined by $A$,
\cite{gilkey}. Also define
\be
\not\!\partial_{V} :=\not\!\! D_{V}\mid_{{\cal C}^{\infty}({\cal S}^{+})
\otimes{\cal C}^{\infty}(V)}.
\label{eq6.1}
\ee
The twisted spin complex is the following short complex:
\be
0 \longrightarrow {\cal C}^{\infty}({\cal S}^{+}\otimes V) \stackrel{
\not\hspace{.4mm}\partial_{V}}{\longrightarrow} {\cal C}^{\infty}({\cal S}^{-}
\otimes V)\longrightarrow 0.
\label{eq7}
\ee
\begin{thm} \label{theorem2}
Let $\not\!\partial_{V}$ be as in (\ref{eq6.1}), and
\be
ch(V):=tr\left[ \exp\frac{iF}{2\pi} \right]
\label{eq9}
\ee
be the Chern character of $V$. Then,
\be
index(\not\!\partial_{V})= \int_{M}\left[ ch(V).\hat{A}(M) \right]_{top}.
\label{eq8}
\ee
\end{thm}
In (\ref{eq9}), $F$ is the curvature 2-form of the connection 1-form
$A$ (written in a basis of the structure Lie algebra of $V$):
\be
F:=\left(\mbox{\fs$\frac{1}{2}$} F_{\lambda \gamma}^{ab}
\, dx^{\lambda}\wedge dx^{\gamma}
\right)\; .
\label{eq9.1}
\ee

Throughout this dissertation, the Greek indices refer to the coordinates
of $M$, they run through $1,...,m=dim(M)$, and the indices from
the beginning of the Latin alphabet refer to the fibre coordinates
of $V$ and, hence, run through $1,...,n$.

\section{Supersymmetric Quantum Mechanics and the Index Theorem}
\label{I-1-2}
A review of the supersymmetric quantum mechanics is provided in
Appendix~\ref{z2-B}. The relevance of the index theory to supersymmetry has
been  discussed in almost every article written in this subject in the past ten
years. The idea is to realize the parallelism between the constructions
involved in the index theory, (\ref{eq2}), and the supersymmetric
quantum mechanics. In the latter, the Hilbert (Fock) space is the
direct sum of the spaces of the bosonic and the fermionic state vectors.
These correspond to the spaces of sections of $V_{\pm}$ in (\ref{eq2}).
Moreover, the supersymmetric charge $\cal Q$ plays the role of the
elliptic operator $D$, and the Hamiltonian $H$ is the analog of the
Laplacian $\Delta$ of $D$. One has:
\[ \Delta :=\{ D,D^{\dagger}\} , \]
and
\be
H=\frac{1}{2}\{ {\cal Q},{\cal Q}^{\dagger}\}.
\label{eq10}
\ee
Here, ``$\dagger$'' denotes the adjoint of the corresponding operator.
(\ref{eq10}) is also known as the superalgebra condition. Moreover, one
defines selfadjoint supersymmetric charges \footnote{
$\alpha =1,...,2N$, where N is the number of nonselfadjoint
charges (type N-SUSY). See Appendix~\ref{z2-B} for details.},
$Q_{\alpha}$ \cite{susy},
in terms of which (~\ref{eq10}) becomes:
\be
H=Q_{\alpha}^{2}\; , \;\;\; \mbox{for all}\;\; \a=1,\cdots ,2N \; .
\label{eq10.1}
\ee
The next step is to recall
\[ coker(D) = ker(D^{\dagger})\; , \]
\[ ker(\Delta) = ker(D) \oplus ker(D^{\dagger}) \]
and use (\ref{eq3}), and (\ref{eq9}) to define:
\[ index_{W} := n_{b,0}-n_{f,0}. \]
Here``$W$'' refers to Witten \cite{witten} and $n_{b,0}$ and $n_{f,0}$
denote the number of the zero-energy bosonic and fermionic
eigenstates of $H$. Realizing that due to supersymmetry any excited
energy eigenstate has a superpartner, one has the following set
of equalities:
\bea
index_{W} & = & n_{b}-n_{f}  \nonumber \\
& = & tr[(-1)^{f}]  \label{eq11} \\
& = & tr[(-1)^{f}e^{-i\beta H}] \nonumber \\
& =: & str[e^{-i\beta H}]. \nn
\eea
In (\ref{eq11}), $n_{b}$ and $n_{f}$ denote the number of the bosonic
and fermionic energy eigenstates, and $f$ is the fermion number
operator. See Appendix~\ref{z2-B} and
\cite{witten,susy,ag3} for more detailed discussions
of (\ref{eq11}).

The appearance of $e^{-i\beta H}$ in (\ref{eq11}) is quite interesting.
It had, however, been noticed long before supersymmetry was introduced
in physics. The heat kernel proof of the index theorem is essentially
based on (\ref{eq11}), \cite{abp,booss,gilkey} \footnote{In the heat
kernel proof, one computes the $index$ using the formula:
\[ index = tr[ e^{-\b \Delta_{+}}] - tr[ e^{-\b \Delta_{-}}] \; ,\]
where $\Delta_{+} := D^{\dagger}D$ and $\Delta_{-} := D\,
D^{\dagger}$. See Appendix~\ref{z2-A12} for further details.}.
The major advantage of
the supersymmetric proofs is that one can compute $index_{W}$ using
its path integral representation. In particular, since $index_{W}$
is independent of $\beta$, the WKB approximation, i.e. the first
term in the loop expansion, yields the index immediately.

\section{Remarks and Discussion}
\label{I-1-3}
In principle, one must be able to construct a supersymmetric quantum
system associated with any given elliptic operator $D$ such that the
corresponding Witten
index equals $index(D)$. However, in practice this is proven to be
extremely difficult. The main reason for this difficulty is that
in general $D$ may be of arbitrary order. This means that the
quantum mechanical Hamiltonian, if it can be constructed, would involve fourth
and higher order terms in momenta. This in turn makes the benefits of the
ordinary path integration techniques\footnote{By this I mean the
path integral techniques developed for the quadratic systems in bosonic
momenta.} unaccessible.
Nonetheless, one can
still attack the general index problem by specializing to the
twisted spin index theorem. According to  Appendix~\ref{z2-A11},
this leads to a proof of the general index theorem via K-theory.
In this respect, the supersymmetric proofs are as indirect as
the mathematical  proofs of the index theorem. The distinguishing
feature of the supersymmetric proofs is that they
are considerably shorter and, at the same time, conceptually simpler
than the other proofs.

The purpose of this dissertation is to provide a
rigorous supersymmetric proof of the index theorem and more
importantly address the fundamental issues regarding the physical techniques
that are developed and used in this proof.

The approach adopted here is a generalization of the supersymmetric proof of
the
Gauss-Bonnet theorem due to DeWitt, \cite{bd1}. In this work, the Peierls
bracket quantization is explained and applied to many interesting examples,
in particular, to the supersymmetric system described by the Lagrangian:
	\be
	L=\oot g_{\m\n}\dot{x}^{\m}\dot{x}^{\nu}+\iot g_{\m\n}\p^{\m}_{\a}
	\frac{D}{dt}(\p_{\a}^{\n})+\mbox{\small$\frac{1}{8}$}R_{\m\n\s\t}
	\p_{\a}^{\m}\p_{\a}^{\n}\p_{\b}^{\s}\p_{\b}^{\t}
	\label{eq1}
	\ee
	\[ \a,\b=1,2\hspace{1.5cm}\m,\n,\s,\t = 1,\cdots ,m\; .\]
The configuration space of (\ref{eq1}) is an $(m,2m)$ dimensional
supermanifold \cite{bd1} with $x$ and $\p$ denoting its bosonic
(commuting) and fermionic (anticommuting) coordinates, respectively.
The quantization of this system leads to a rigorous supersymmetric
proof of the Gauss-Bonnet theorem, \cite[\S 6]{bd1}.

The main difference of the analysis of the system of (\ref{eq1})
with the systems studied in Chapter~\ref{I-2} and~\ref{I-3} is in
the path integral evaluation of the index. The mathematical generality
and significance of the twisted spin index theorem
reveals its presence in the form of different nontrivial aspects of
path integration techniques used in its proof.
Some of these are discussed in Section~\ref{I-2-6} and Appendix~\ref{z2-D}.

The earlier supersymmetric proofs of the index theorem followed the
canonical quantization program. One of the difficulties of the canonical
quantization of a superclassical system is that the fermionic momentum
and coordinate variables are not independent. Thus, one needs to
consider the appropriate (fermionic) first class constraints \cite{goodman}.
In the Peierls quantization scheme this problem is completely avoided.
This is because, in the Peierls program there is no need to define
momentum variables to carry out the quantization. This also has practical
advantages in handling the bosonic variables, e.g., see Section~\ref{I-2-1}.
\chapter{Supersymmetric Proof of the Index Theorem}
\label{I-2}
\vspace{1cm}
\bcc
{\bf Abstract}
\ecc
\bcc
\parbox[b]{4.5in}{\small
The Peierls brackect quantization scheme is applied to the supersymmetric
system corresponding to the twisted spin index theorem. A detailed
study of the quantum system is presented, and the Feynman propagator
is exactly computed. The Green's function and functional integral
methods provide a direct derivation of the index as a single universal
superdeterminant.}
\ecc\vspace{.5in}
\section{The Superclassical System and Its  Quantization}
\label{I-2-1}
Consider a superclassical system described by the action functional
$S=S[\Phi ]$, with $\Phi = (\Phi^{i})$ denoting the coordinate (field)
variables. The dynamical equations are given by $\delta S = 0$, i.e.
\be
S,_{i} := S[\Phi ]\frac{\stackrel{\leftarrow}{\delta}}{\delta
\Phi^{i}(t)} = 0.
\label{eq12}
\ee
Throughout this dissertation the condensed notation of \cite{bd1}
is used where appropriate. The following example demonstrates most
of the conventions:
\[ _{i,}S_{,j'}G^{j'k''}\equiv \int dt'\left(
\frac{\stackrel{\rightarrow}{\delta}}{\delta \Phi^{i}(t)} S[\Phi ]
\frac{\stackrel{\leftarrow}{\delta}}{\delta \Phi^{j}(t')} \right)
G^{jk}(t',t'') . \]
In general, the second functional derivatives of $S$ ``are'' second
order differential operators \footnote{These are called the Jacobi
operators.}, e.g. see (\ref{eq24}) below. Let $G^{\pm i'k''}$
denote the corresponding advanced and retarded Green's functions:
\be
_{i,}S_{,j'}G^{\pm j'k''} := - \delta(t-t'')\delta_{i}^{k}
\label{eq13}
\ee
where $\delta(t-t'')$ and $\delta_{i}^{k}$ are the Dirac and Kronecker
delta functions, respectively. Furthermore, define:
\be
\tilde{G}^{jk'} := G^{+jk'} - G^{-jk'}\; .
\label{eq14}
\ee
Then, the Peierls bracket \cite{peierls,bd1} of
any two scalar fields, ${\cal A}$  and ${\cal B}$, of $\Phi$ is defined
by:
\be
({\cal A},{\cal B}') := {\cal A}_{,i}\hspace{1mm}\tilde{G}^{ij'}\hspace{.1mm}
_{j',}{\cal B}.
\label{eq14.9}
\ee
In particular, one has
\be
(\Phi^{i},\Phi^{j'}) := \tilde{G}^{ij'}.
\label{eq15}
\ee

The Green's functions $G^{\pm ij'}$ satisfy the following reciprocity
relation \cite{bd1},
\be
G^{-ij'} = (-1)^{ij'} G^{+j'i}.
\label{eq15.1}
\ee
In (\ref{eq15.1}), the indices $i$ and $j'$ in $(-1)^{ij'}$ are either
0 or 1 depending on whether $\Phi^{i}$($\Phi^{j'}$) is a bosonic
or fermionic variable, respectively.

The quantization is performed by promoting the superclassical quantities
to the operators acting on a Hilbert space and forming a superalgebra
defined by the following supercommutator:
\be
[{\cal A},{\cal B}]_{\rm super} := i \hbar ({\cal A},{\cal B})\; ,
\label{eq16}
\ee
where the right hand side is defined up to factor ordering.

The superclassical system of interest is represented by the
Lagrangian \cite{windey,ag3}:
\bea
L&=&\left[ \mbox{\fs$\frac{1}{2}$}g_{\mu \nu}(x) \dot{x}^{\mu}\dot{x}^{\nu}+
\mbox{\fs $\frac{i}{2}$}
g_{\lambda \gamma}(x)\psi^{\lambda}\frac{D}{dt}\psi^{\gamma}
\right]_{1} + \label{eq17} \\
&&+\kappa \left[ i\eta^{a*}\left( \dot{\eta}^{a}+\dot{x}^{\sigma}
A_{\sigma}^{ab}(x)\eta^{b}\right) +\mbox{\fs$\frac{1}{2}$}
F_{\lambda \gamma}^{ab}(x) \psi^{\lambda}\psi^{\gamma}\eta^{a*}\eta^{b}
\right]_{2} + \nonumber \\
&& \hspace{.5cm}+\left[ \mbox{\fs$\frac{\alpha}{\beta}$}
\eta^{a*}\eta^{a}\right]_{3}. \nn
\eea
In (\ref{eq17}), $x^{\mu}$ are the bosonic variables corresponding
to the coordinates of $M$. $g_{\mu \nu}$ are components of the
metric tensor on $M$. $\psi^{\lambda}$ and $\eta^{a}$ are fermionic
real and complex variables associated with the bundles ${\cal S}$
and $V$, respectively. $A_{\mu}^{ab}$ and $F_{\lambda \gamma}^{ab}$
are the components of the connection 1-form and the curvature
2-form on $V$ (written in an orthonormal basis of the structure Lie algebra).
One has the following well-known relation in the Lie algebra:
\be
F_{\mu \nu}=A_{\nu},_{\mu}-A_{\mu},_{\nu} + [A_{\mu},A_{\nu}].
\label{eq19}
\ee
Both $A_{\mu}$ and $F_{\mu \nu}$ in (\ref{eq19}) are antihermitian
matrices. This makes $L$ real up to total time derivatives. In
(\ref{eq17}), $\kappa =0,1$ correspond to switching off and on
of the twisting, respectively. $\alpha$ is a scalar parameter whose
utility will be discussed in Section \ref{I-2-2}. $\beta$ is the time
parameter:
$t\in [0,\beta ]$. Finally, ``dot'' means ordinary time derivative
$\frac{d}{dt}$, and $\frac{D}{dt}$ denotes the covariant time derivative
defined by the Levi Civita connection, e.g.
\[ \frac{D}{dt}\psi^{\gamma}:=\dot{\psi}^{\gamma}+\dot{x}^{\mu}
\Gamma^{\gamma}_{\mu \theta}\psi^{\theta}. \]

The following set of infinitesimal supersymmetric transformations,
\bea
\delta x^{\mu} & = & i \psi^{\mu}\delta \xi  \nonumber \\
\delta \psi^{\gamma} & = &\dot{x}^{\gamma}\delta \xi \label{eq20} \\
\delta \eta^{a} & = & i A_{\gamma}^{ab}\psi^{\gamma}\eta^{b}
\delta \xi \nonumber \\
\delta \eta^{a*} & = & -i A_{\gamma}^{ba}\psi^{\gamma}\eta^{b*}
\delta \xi \nn
\eea
\[ \delta \xi := \mbox{an infinitesimal fermionic variable} \]
leaves the action
\be
S:=\int_{0}^{\beta}L dt
\label{eq21}
\ee
invariant. Hence, the system is supersymmetric.
It is easy to see that the first two equations
in (\ref{eq20}) leave $L_{1}:=[...]_{1}$ in (\ref{eq17}) invariant.
Thus, for $\kappa = \alpha =0$, one has a supersymmetric subsystem.
This subsystem will be  used to compute the index of spin complex
(the Dirac $\hat{A}$ genus) in Section~\ref{I-2-4}. $L_{1}$ can also be
obtained from (\ref{eq1}) by reducing the system of (\ref{eq1}) by
setting $\psi_{1}^{\mu}=\psi_{2}^{\mu}$, for all $\mu$.

% Pages 12 - 14.9
The Dynamical equations (\ref{eq12}) are:
% **********************   Dynamical Equations   ********
\bea
S_{,\t } & = & -g_{\m \t}\frac{D}{dt}\dot{x}^{\m}+i\G_{\l \g \t}\p^{\g}
\frac{D}{dt}\p^{\l}+\mbox{\fs$\frac{i}{2}$}R_{\l \g \s \t}
\dot{x}^{\s}\p^{\g}\p^{\l} + \nn \\
& &  +\k \left[ -iA_{\t}^{ab}(\dot{\e}^{a*}\e^{b}+\e^{a*}\dot{\e}^{b})
+i\dot{x}^{\s}(A_{\s ,\t}^{ab}-A_{\t ,\s}^{ab})\e^{a*}\e^{b}+ \right. \nn \\
& & \left. \hspace{1in} +\mbox{\fs$\frac{1}{2}$}F_{\g \l ,\t}^{ab}
\p^{\g}\p^{\l}\e^{a*}\e^{b} \right] =0 \label{eq22} \\
S_{,\l} & = & g_{\l \g}\frac{D}{dt}\p^{\g}+\k [iF_{\g \l}^{ab}\p^{\g}
\e^{a*}\e^{b}]=0 \nn \\
S_{,a} & = & \k [i\dot{\e}^{a}+i\dot{x}^{\s}A_{\s}^{ab}\e^{b}+
\mbox{\fs$\frac{1}{2}$}F_{\g \l}^{ab}
\p^{\g}\p^{\l}\e^{b}]+\frac{\a}{\b}\e^{a}=0 \nn \\
S_{,b^{*}} & = & \k [-i\dot{\e}^{b*}+i\dot{x}^{\s}A_{\s}^{ab}\e^{a*}+
\mbox{\fs$\frac{1}{2}$}F_{\g \l}^{ab}
\p^{\g}\p^{\l}\e^{a*} ] +\frac{\a}{\b}\e^{b*}=0 \nn
\eea
% **********************************************************
where the indices from the beginning of the Greek alphabet
($\g ,\d ,\l ,\th ,\e$) label
$\p$'s and those of the middle of the Greek alphabet ($\k ,\cdots$)
label $x$'s, e.g.
\[ S_{,\t} := S\frac{\stackrel{\leftarrow}{\d}}{\d x^{\t}} \hspace{.3in}
\mbox{and}\hspace{.3in}S_{,\l} :=S\frac{\stackrel{\leftarrow}{\d}}{\d
\p^{\l}}. \]
Similarly,
\[ S_{,a} :=S\frac{\stackrel{\leftarrow}{\d}}{\d \e^{a}}\hspace{.3in}
\mbox{and}\hspace{.3in}S_{,a^{*}} := S\frac{\stackrel{\leftarrow}{\d}}{
\d \e^{a*}}. \]

The supersymmetric charge $Q$ corresponding to (\ref{eq20}) is given by
\be
Q \propto g_{\g \m}\p^{\g}\dot{x}^{\m}.
\label{eq23}
\ee
The second functional derivatives of the action are listed in the
following:
% *********************** Jacobi Operators  *******************
\bea
_{\t ,}S_{,\z '} & = & \left[ -g_{\t \z}\frac{\partial^2}{\partial t^2}+
\left( -2\G_{\t \z \m}\dot{x}^{\m}+
\mbox{\fs$\frac{i}{2}$}R_{\l \g \z \t}\p^{\g}\p^{\l}+
\right. \right. \nn  \\
 & & \left. \left. +i\k (A_{\z ,\t}^{ab}-A_{\t ,\z}^{ab})\e^{a*}\e^{b}+
i\G_{\l \g \t}\G^{\l}_{\d \z}\p^{\g}\p^{\d}\right) \pt +C_{\t \z}
\right] \d (t-t') \nn \\
_{\t ,}S_{,\g '} & = &\left[ i\G_{\g \t \d}\p^{\d}\pt +C_{\t \g}\right]
\d (t-t') \nn \\
_{\t ,}S_{,a'} & = &\k \left[ -iA_{\t}^{ca}\e^{c*}\pt +C_{\t a}\right]
\d (t-t') \nn \\
_{\t ,}S_{,a^{*'}} & = &\k \left[ iA_{\t}^{ac}\e^{c}\pt + C_{\t a^{*}}
\right] \d (t-t') \nn \\
_{\l ,}S_{,\z '} & = &\left[ i\G_{\l \z \d}\p^{\d}\pt + C_{\l \z} \right]
\d (t-t') \nn \\
_{\l ,}S_{,\g '} & = &\left[ ig_{\l \g}\pt + C_{\l \g}\right] \d (t-t') \nn \\
_{\l ,}S_{,a'} & = &\k \left[ C_{\l a}\right] \d (t-t') \label{eq24} \\
_{\l ,}S_{,a^{*'}} & = &\k \left[ C_{\l a^{*}}\right] \d (t-t') \nn \\
_{b,}S_{,\z '} & = &\k \left[ -iA_{\z}^{cb}\e^{c*}\pt +C_{b\z}\right]
\d (t-t') \nn \\
_{b,}S_{,\g '} & = &\k \left[ C_{b\g}\right] \d (t-t') \nn \\
_{b,}S_{,a'} & = &\k \left[ C_{ba}\right] \d (t-t') \nn \\
_{b,}S_{,a^{*'}} & = &\k \left[ i\d_{ba}\pt + C_{ba^{*}} \right]
\d (t-t') \nn \\
_{b^{*},}S_{,\z '} & = &\k \left[ iA_{\z}^{bc}\e^{c}\pt + C_{b^{*}\z}
\right] \d (t-t') \nn \\
_{b^{*},}S_{,\g '} & = &\k \left[ C_{b^{*}\g}\right] \d (t-t') \nn \\
_{b^{*},}S_{,a'} & = &\k \left[ i\d_{ba}\pt + C_{b^{*}a}\right]
\d (t-t') \nn \\
_{b^{*},}S_{,a^{*'}} & = &\k \left[ C_{b^{*}a^{*}}\right] \d (t-t')\; . \nn
\eea
% *************************************************************

In (\ref{eq24}) $C_{ij}$'s \footnote{$\Phi^{i}\in (x^{\m},\p^{\g},
\e^{a},\e^{b*})$ are generic variables.} are terms which do not involve
any time derivative. These terms do not actually contribute to the
equal-time commutation relations of interest, (\ref{eq33}). However, they
will contribute in part to $sdet[G^{+ij}]$ in Section~\ref{I-2-3}. One has:
\bea
C_{\t \z}&:=&-g_{\m \t ,\z}\ddot{x}^{\m}-\G_{\t \m \n ,\z}\dot{x}^{\n}
\dot{x}^{\m} + i\G_{\l \g \t ,\z}\p^{\g}\dot{\p}^{\l} +
i(\G_{\l \g \t}\G^{\l}_{\d \s})_{,\z} \dot{x}^{\s}\p^{\g}\p^{\d}
+ \nn \\
& & +{\mbox \fs \frac{i}{2}}R_{\l \g \s \t ,\z}\dot{x}^{\s}\p^{\g}\p^{\l}
+ \k \ll -iA_{\t ,\z}^{cd}(\dot{\e}^{c*}\e^{d}+\e^{c*}\dot{\e}^{d})+
\right. \nn \\ & & \left.
+i(A_{\s,\t \z}^{cd}-A_{\t ,\s \z}^{cd})\dot{x}^{\s}\e^{c*}\e^{d} +
\mbox{\fs$\frac{1}{2}$}F_{\g \l ,\t \z}^{cd}\p^{\g}\p^{\l}
\e^{c*}\e^{d} \rr \nn \\
C_{\t \g}&:=&-i\G_{\l \g \t}\dot{\p}^{\l}+i(\G_{\l \d \t}\G^{\l}_{\g \s}
-\G_{\l \g \t}\G^{\l}_{\d \s})\dot{x}^{\s}\p^{\d}+iR_{\g \l \d \t}
\dot{x}^{\s}\p^{\l} + \nn \\
& &  +\k \ll F_{\l \g ,\t}^{cd}\p^{\l}\e^{c*}\e^{d} \rr \nn \\
C_{\t a}&:=&-iA_{\t}^{ca}\dot{\e}^{c*}+i(A_{\s ,\t}^{ca}-A_{\t ,\s}^{ca})
\dot{x}^{\s}\e^{c*}+ \mbox{\fs$\frac{1}{2}$}F_{\g \l ,\t}^{ca}\p^{\g}\p^{\l}
\e^{c*} \nn \\
C_{\t a^{*}}&:=&iA_{\t}^{ac}\dot{\e}^{c}-i(A_{\s ,\t}^{ac}-
A_{\t ,\s}^{ac})\dot{x}^{\s}\e^{c} + \mbox{\fs$\frac{1}{2}$}
F_{\g \l}^{a^{*}c}\p^{\g}\p^{\l}\e^{c} \nn \\
C_{\l \z}&:=&i\G_{\l \d \m ,\z}\dot{x}^{\m}\p^{\d}+ig_{\l \d ,\z}
\dot{\p}^{\d}-\k \ll F_{\d \l ,\z}^{cd}\p^{\d}\e^{c*}\e^{d} \rr \nn \\
C_{\l \g}&:=&i\G_{\l \g \m}\dot{x}^{\m}+\k F_{\l \g}^{ab}\e^{a*}\e^{b}
\nn \\
C_{\l a}&:=&-F_{\d \l}^{ca}\p^{\d}\e^{c*} \label{eq25} \\
C_{\l a^{*}}&:=&F_{\d \l}^{ac}\p^{\d}\e^{c} \nn \\
C_{b\z}&:=&-iA_{\s ,\z}^{cb}\dot{x}^{\s}\e^{c*}-\mbox{\fs$\frac{1}{2}$}
F_{\g \l}^{cb}\p^{\g}\p^{\l}\e^{c*} \nn \\
C_{b\g}&:=&F_{\d \g}^{cb}\p^{\d}\e^{c*} \nn \\
C_{ba}&:=&0 \nn \\
C_{ba^{*}}&:=&-iA_{\s}^{ab}\dot{x}^{\s}-\mbox{\fs$\frac{1}{2}$}
F_{\d \th}^{ab}\p^{\d}\p^{\th}-\mbox{\fs$\frac{\a}{\k \b}$}\d^{ab} \nn \\
C_{b^{*}\z}&:=&iA_{\s ,\z}^{bc}\dot{x}^{\s}\e^{c}+
\mbox{\fs$\frac{1}{2}$}F_{\g \l ,\z}^{bc}\p^{\g}\p^{\l}\e^{c} \nn \\
C_{b^{*}\g}&:=&-F_{\d \g}^{bc}\p^{\d}\e^{c} \nn \\
C_{b^{*}a}&:=&iA_{\s}^{ba}\dot{x}^{\s}+\mbox{\fs$\frac{1}{2}$}
F_{\d \th}^{ba}\p^{\d}\p^{\th}+\mbox{\fs$\frac{\a}{\k \b}$}\d^{ba} \nn \\
C_{b^{*}a^{*}}&:=&0 .\nn
\eea

% Pages 14.1 - 16.1
The advanced Green's functions $G^{+ij'}$ are calculated using
(\ref{eq13}) and (\ref{eq24}). The results are listed below:
% ******************* Advanced Green's functions *****************
\renewcommand{\tt}{\theta (t'-t)}
% ------------------------------
\bea
G^{+\xi \z '} & = &\tt \ll -g^{\xi '\z'}(t-t')+
\mbox{\fs$\frac{1}{2}$}g^{\xi '\t '}
\left( 2\G_{\t '\n '\m '}\dot{x}^{\m '}- \mbox{\fs$\frac{i}{2}$}
R_{\n '\t '\l '\g '}\p^{\g '}\p^{\l '}-  \right. \right.  \nn \\
& &  \left. \left. +i\k F_{\t '\n '}^{a'b'}\e^{a*'}\e^{b'} \right) g^{\n '\z '}
(t-t')^{2}+O(t-t')^{3} \rr \nn \\
G^{+\xi \g '} & = &\tt \ll g^{\xi '\t '}\G^{\g '}_{\t '\l '}\p^{\l '}
(t-t')+O(t-t')^{2} \rr \nn \\
G^{+\xi a'} & = &\tt \ll g^{\xi '\t '}A_{\t '}^{a'c'}\e^{c'}(t-t')+
O(t-t')^{2}\rr \nn \\
G^{+\xi a^{*'}} & = &\tt \ll g^{\xi '\t '}A_{\t'}^{c'a^{*'}}\e^{c*'}
(t-t')+O(t-t')^{2} \rr \nn \\
G^{+\l \z '} & = &\tt \ll g^{\z '\t '}\G^{\l '}_{\t '\d '}\p^{\d '}
(t-t')+O(t-t')^{2} \rr \nn \\
G^{+\l \g '} & = &\tt \ll -ig^{\l '\g'}+O(t-t') \rr \nn \\
G^{+\l a'} & = &\tt \ll O(t-t') \rr \label{eq26} \\
G^{+\l a^{*'}} & = &\tt \ll O(t-t') \rr \nn \\
G^{+b\z '} & = &\tt \ll g^{\z '\t '}A_{\t '}^{b'c'}\e^{c'}(t-t')+O(t-t')\rr
\nn \\
G^{+b\g '} & = &\tt \ll O(t-t')\rr \nn \\
G^{+ba'} & = &\tt \ll O(t-t') \rr \nn \\
G^{+ba^{*'}} & = &\tt \ll - \mbox{\fs$\frac{i}{\k}$}\d^{b'a'}+
O(t-t')\rr \nn \\
G^{+b^{*}\z '} & = &\tt \ll -g^{\z '\t '}A_{\t '}^{c'b'}\e^{c*'}
(t-t')+O(t-t')^{2}\rr \nn \\
G^{+b^{*}\g '} & = &\tt \ll O(t-t')\rr  \nn \\
G^{+b^{*}a'} & = &\tt \ll - \mbox{\fs$\frac{i}{\k}$}\d^{b'a'}+O(t-t')\rr \nn \\
\nopagebreak[3]  G^{+b^{*}a^{*'}} & = &\tt \ll O(t-t')\rr \; .\nn
\eea
% **********************************************************************

The Green's functions $G^{-ij'}$ and $\tilde{G}^{ij'}$ are then obtained
using (\ref{eq15.1}) and (\ref{eq14}), respectively. Substituting the latter
in (\ref{eq15}) leads to the following Peierls brackets:
% Pages 16.1 - 16.6
%******************** Peierls Brackets ***********************
\bea
(x^{\xi},x^{\z '}) & = &-g^{\xi '\z '}(t-t')+
\mbox{\fs$\frac{1}{2}$}g^{\xi '\t'} \left(
2\G_{\t '\n '\m '}\dot{x}^{\m '}-
\mbox{\fs$\frac{i}{2}$}R_{\n '\t '\l '\g '}
\p^{\g '}\p^{\l '}+ \right. \nn \\
& & \left. -i\k F_{\t '\n '}^{a'b'}\e^{a^{*'}}\e^{b'} \right)
g^{\n '\z '}(t-t')^{2}+O(t-t')^{3} \nn \\
(x^{\xi},\p^{\g '}) & = &g^{\xi '\t '}\G^{\g '}_{\t '\d '}\p^{\d '}
(t-t')+O(t-t')^{2} \nn \\
(x^{\xi},\e^{a'}) & = &g^{\xi '\t '}A_{\t '}^{a'c'}\e^{c'}(t-t')+
O(t-t')^{2} \label{eq28} \\
(x^{\xi},\e^{a*'}) & = &-g^{\xi '\t '}A_{\t '}^{c'a'}\e^{c*'}(t-t')
+O(t-t')^{2} \nn \\
(\p^{\l},\p^{\g '}) & = &-ig^{\l '\g '}+O(t-t') \nn \\
(\e^{b},\e^{a*'}) & = &-\mbox{\fs$\frac{i}{\k}$}\d^{b'a'}+O(t-t')\; . \nn
\eea
% *******************************************************************
Other possible Peierls brackets are all of the order $(t-t')$ or
higher. Differentiating the necessary expressions in (\ref{eq28})
with respect to $t$ and $t'$, one arrives at the Peierls brackets
among the coordinates $(x,\p ,\e ,\e^{*})$ and their time derivatives.
The interesting equal-time Peierls brackets are the following:
% Pages 16.5 - 17.5
% ********************* Equal-time Peierls Brackets *****************
\bea
(x^{\xi},x^{\z}) &=&(x^{\xi},\p^{\g}) = (x^{\xi},\e^{a}) =
(x^{\xi},\e^{a*}) = 0 \nn \\
(\p^{\g},\e^{a}) &=&(\p^{\g},\e^{a*}) = (\e^{a},\e^{b}) = (\e^{a*},\e^{b*})
 = 0 \nn \\
(\p^{\l},\p^{\g}) &=&-ig^{\l \g} \label{eq29} \\
(\e^{a},\e^{b*}) &=&-\mbox{\fs$\frac{i}{\k}$}\d^{ab} \nn \\
(\dot{x}^{\xi},x^{\z}) &=&-g^{\xi \z} \nn \\
(\dot{x}^{\xi},\dot{x}^{\z}) &=&g^{\xi \t}\left[ (\G_{\n \t \m}-
\G_{\t \n \m})\dot{x}^{\m}+
\mbox{\fs$\frac{i}{2}$}R_{\t \n \l \g}\p^{\l}\p^{\g}
+i\k F_{\t \n}^{ab}\e^{a*}\e^{b}\right] g^{\n \z} \nn \\
(\dot{x}^{\xi},\p^{\g}) &=&g^{\xi \t}\G^{\g}_{\t \l}\p^{\l} \nn \\
(\dot{x}^{\xi},\e^{a}) &=&g^{\xi \t}A_{\t}^{ac}\e^{c} \nn \\ \nopagebreak[3]
(\dot{x}^{\xi},\e^{a*}) &=&-g^{\xi \t}A_{\t}^{ca}\e^{c*} \; .\nn
\eea
% *******************************************************************
The next step is to define the appropriate momenta conjugate to $x^{\n}$.
The canonical momenta are given by:
\be
p_{\n}^{\rm (canonical)} :=L\frac{\stackrel{\leftarrow}{\partial}}{\partial
\dot{x}^{\n}}=g_{\n \m}\dot{x}^{\m}+
\mbox{\fs$\frac{i}{2}$}\G_{\l \g \n}\p^{\l}\p^{\g}+
i\k A_{\n}^{ab}\e^{a*}\e^{b}.
\label{eq30}
\ee
A more practical choice is provided by:
\be
p_{\n}:=g_{\n \m}\dot{x}^{\m} \; .
\label{eq31}
\ee
This choice together with the use of (\ref{eq14.9}) and (\ref{eq29})
lead to:
\bea
(p_{\n},x^{\m}) & = &-\d_{\n}^{\m}  \nn \\
(p_{\n},\p^{\g}) & = &\G^{\g}_{\n \l}\p^{\l}  \nn  \\
(p_{\n},\e^{a}) & = &A_{\n}^{ac}\e^{c} \label{eq32} \\
(p_{\n},\e^{a*}) & = &-A_{\n}^{ca}\e^{c*} \nn \\
(p_{\n},p_{\m}) & = &\mbox{\fs$\frac{i}{2}$}R_{\n \m \l \g}\p^{\l}\p^{\g}
+\k \ll iF_{\n \m}^{ab}\e^{a*}\e^{b} \rr.  \nn
\eea
The quantization is performed via (\ref{eq16}). Enforcing (\ref{eq16})
and using (\ref{eq29}) and (\ref{eq32}), one has the following
supercommutation relations. For convenience, the commutators,
$[ . , .]$, and the anticommutators, $\{ . , .\}$, are distinguished.
% Pages 17.5 - 19
%******************** Supercommutation relations ****************
\bea
[x^{\m},x^{\n}] &=&[x^{\m},\p^{\g}] = [x^{\m},\e^{\a}] = [x^{\m},\e^{a*}]
= 0 \nn \\
\{\p^{\l},\e^{a}\} &=&\{\p^{\l},\e^{a*}\} = \{\e^{a},\e^{b}\} =
\{\e^{a*},\e^{b*}\} = 0 \nn \\
\{\p^{\l},\p^{\g}\} &=& \hbar g^{\l \g} \nn \\
\{\e^{a},\e^{b*}\} &=& \frac{\hbar}{\k}\d^{ab^{*}} \nn \\
\ll x^{\m},p_{\n}\rr &=& i\hbar \d^{\m}_{\n} \label{eq33} \\
\ll \p^{\g},p_{\n} \rr &=& -i\hbar \G^{\g}_{\n \l}\p^{\l} \nn \\
\ll \e^{a},p_{\n} \rr &=& -i\hbar A_{\n}^{ac}\e^{c} \nn \\
\ll \e^{a*},p_{\n} \rr &=& i\hbar A_{\n}^{ca}\e^{c*} \nn \\
\ll p_{\m},p_{\n} \rr &=& - \mbox{\fs$\frac{\hbar}{2}$}R_{\m \n \l \g}
\p^{\l}\p^{\g}  +\k \ll - \mbox{\fs$\hbar$}F_{\m \n}^{ab}\e^{a*}\e^{b} \rr .
\nn
\eea
One must note that in general there may be  factor ordering ambiguities in
the right hand side of (\ref{eq16}). Indeed, for the example considered
here there are  three inequivalent
choices for the last equation in (\ref{eq33}). These correspond to
the following choices of ordering $\e^{a*}\e^{b}$ in (\ref{eq32}):
\be
\e^{a*}\e^{b} \hspace{.15in},\hspace{.15in} -\e^{b}\e^{a*}
\hspace{.15in},\:{\rm and} \hspace{.15in} \mbox{\fs$\frac{1}{2}$}
(\e^{a*}\e^{b}-\e^{b}\e^{a*}).
\label{eq33.1}
\ee
The first choice is selected in
(\ref{eq33}) because, as will be seen in Section~\ref{I-2-2}, it leads to the
identification of the supersymmetric charge with the twisted
Dirac operator.
%********************************************************************
The quantum mechanical supersymmetric charge corresponding to (\ref{eq23}),
which is also hermitian, is given by:
\be
Q=\mbox{\fs$\frac{1}{\sqrt{\hbar}}$}
\p^{\n}g^{\frac{1}{4}} p_{\n} g^{-\frac{1}{4}}.
\label{eq34}
\ee
Hermiticity is ensured in view of the identity:
 \[ \p^{\n}g^{\frac{1}{4}}p_{\n}g^{-\frac{1}{4}}=g^{-\frac{1}{4}}
p_{\n}g^{\frac{1}{4}}\p^{\n}. \]
Here, $g$ is the determinant of $(g_{\m \n})$ and the proportionality
constant, $1/\sqrt{\hbar}$, is fixed by comparing the reduced form
of (\ref{eq35}) to the case: $\p =\e =0$ and $M=\R^{n}$. Equation
(\ref{eq34}) together with (\ref{eq10.1}) yield the Hamiltonian:
%************************ The Hamiltonian *************************
\be
H = Q^{2} =
\mbox{\fs$\frac{1}{2}$}g^{-\frac{1}{4}}p_{\m}g^{\frac{1}{2}}g^{\m \n}
p_{\n}g^{-\frac{1}{4}} + \mbox{\fs$\frac{\hbar^{2}}{8}$}R +\k \ll
-\mbox{\fs$\frac{1}{2}$}F_{\l \g}^{ab}\p^{\l}\p^{\g}\e^{a*}\e^{b} \rr \; .
\label{eq35}
\ee
% ********************************************************************
The derivation of (\ref{eq35}) involves repeated use of (\ref{eq33}). In
particular, the appearence of
the scalar curvature term, $\frac{\hbar^{2}}{8}R$, is a
consequence of the third and the last equations in (\ref{eq33})
and the symmetries of the Riemann curvature tensor.
A detailed derivation of (\ref{eq35}) is presented in Appendix~\ref{z2-C}.
Reducing the system of
(\ref{eq17}) to a purely bosonic one, i.e. setting $\p =\e =0$, leads
to the problem of the dynamics of a free particle moving on a Riemannian
manifold. Equation (\ref{eq35}) is in complete agreement with the analysis of
the latter problem by DeWitt \cite{bd1}. One has to emphasize,
however, that here the Hamiltonian is obtained as a result of the
superalgebra condition (\ref{eq10.1}). One can also ckeck that
(\ref{eq35}) reduces to the classical Hamiltonian, i.e. $L\frac{
\stackrel{\leftarrow}{\partial}}{\partial \dot{\Phi}^{i}}\dot{\Phi}^{i}-L$,
as $\hbar \rightarrow 0$, for the Lagrangian (\ref{eq17}) with
$\alpha = 0$. It turns out that this is precisely what one needs
for the proof of the twisted spin index theorem. See Section~\ref{I-2-2},
for a more detailed discussion of this point.

\section{The Quantum System}
\label{I-2-2}
In the rest of this chapter $\hbar$ will be set to 1.
\subsection*{The Case of Spin Complex ($\k =\a = 0$)}

Let $\{e^{i}_{\m}dx^{\m} \}$ be a local orthonormal frame for the
cotangent bundle, $TM^{*}$, i.e. $e^{i}_{\m}e^{j}_{\n}\d_{ij}=g_{\m
\n}$ and $\{e^{\m}_{i}\frac{\partial}{\partial x^{\m}}\}$ be its
dual in $TM$. Consider,
\be
\g^{i}:=i\sqrt{2} e^{i}_{\m}\p^{\m}.
\label{eq36}
\ee
Then, (\ref{eq33}) implies:
\be
\{ \g^{i},\g^{j}\}=-2\d_{ij}
\label{eq37}
\ee
In the mathematical language one says that $\g^{i}$'s are the
generators of the Clifford algebra ${\cal C}(TM_{x}^{*})\otimes \C$,
\cite[Part I\hspace{-.4mm}I, p. 6]{cd1}. Furthermore, for $i=1,...,l:=m/2$
define \cite{windey,manes-zumino}:
\bea
\xi^{i}& := & \frac{1}{2}(\g^{2i-1}+i\g^{2i}) \nn \\
\xi^{i\dagger}& := & -\frac{1}{2}(\g^{2i-1}-i\g^{2i}).
\label{eq38}
\eea
Equations (\ref{eq37}) and (\ref{eq38}) yield the following anticommutation
relations:
\bea
\{\xi^{i},\xi^{j\dagger}\}&=&\d^{ij} \nn \\
\{\xi^{i},\xi^{j}\}&=&\{\xi^{i\dagger},\xi^{j\dagger}\} = 0.
\label{eq39}
\eea
(\ref{eq39}) indicates that $\xi^{i}$ and $\xi^{i\dagger}$ behave as
fermionic annihilation and creation operators. The basic kets of the
corresponding Fock space are of the form:
\be
\mid i_{r},...,i_{1},x,t \kt :=\xi^{i_{r}\dagger}...\xi^{i_{1}\dagger}
\mid\! x,t \kt \; .
\label{eq40}
\ee
The wavefunctions are given by:
\be
\Psi_{i_{1},...,i_{r}}(x,t)=\br x,t,i_{1},...,i_{r}\mid\! \Psi \kt
\label{eq41}
\ee
where
\[ \br x,t,i_{1},...,i_{r}\mid := \mid\! i_{r},...,i_{1},x,t
\kt^{\dagger}. \]

The chirality operator $(-1)^{f}$ of (\ref{eq11}) is defined by
\be
\g^{m+1}:=i^{l}\g^{1}...\g^{m}=\prod_{i=1}^{l}(1-2\xi^{i\dagger}
\xi^{i})\; .
\label{eq42}
\ee
Equation (\ref{eq42}) is a clear indication of the relevance of the
system to the spin complex. In fact, in terms of $\g$'s the
supersymmetric charge, (\ref{eq34}), is written as
\be
Q=\mbox{\fs$\frac{-i}{\sqrt{2}}$}g^{\frac{1}{4}}\g^{\n}p_{\n}g^{-\frac{1}{4}},
\label{eq43}
\ee
where
\be
\g^{\n}:=e^{\n}_{i}\g^{i}.
\label{eq44}
\ee
It is not difficult to see that indeed $Q$ is represented by the Dirac
operator $\not\!\! D$ in the coordinate representation, i.e.
\be
\br x,t,i_{1},...,i_{r}\mid p_{\m}=-i\not\!\partial_{\m}
\br x,t,i_{1},...,i_{r}\mid
\label{eq45}
\ee
with
\be
\not\!\partial_{\m}:=\frac{\partial}{\partial x^{\m}}-
\mbox{\fs$\frac{1}{8}$}\omega_{\m}.
\label{eq45.1}
\ee
The following commutation relations can be easily computed:
\bea
\ll x^{\m},-i\not\!\partial_{\n}\rr & = & i\d^{\m}_{\n} \label{eq46} \\
\ll \g^{i},-i\not\!\partial_{\n}\rr & = & -i\omega^{i}_{j\n}\g^{j}
\label{eq47} \\
\ll-i\not\!\partial_{\m},-i\not\!\partial_{\n}\rr & = &
\mbox{\fs$\frac{1}{4}$}R_{\m \n \l \d}\g^{\l}\g^{\d} \; .\label{eq48}
\eea
In (\ref{eq45.1}) and (\ref{eq47}) $\omega_{\m}$ and $\omega^{i}_{j\n}$
refer to the spin connection:
\be
\omega^{i}_{j\n}:=\G^{\m}_{\n \s}e^{i}_{\m}e^{\s}_{j}-e^{i}_{\m ,\n}
e^{\m}_{j}=:\omega_{\n ij}\; .
\label{eq49}
\ee
\be
\omega_{\m}:=\omega_{\m ij}[\g^{i},\g^{j}]=2\omega_{\m ij}\g^{i}\g^{j}
\label{eq50}
\ee
In the derivation of (\ref{eq48}) one uses the symmetries of
$\omega_{\n ij}$, especially the identity:
\[ \omega_{\n ij}=-\omega_{\n ji}. \]
Comparing (\ref{eq46}), (\ref{eq47}), (\ref{eq48}) with the last three
equations in (\ref{eq33}) justifies (\ref{eq45}) and the claim
preceding it.

Following the analysis of \cite[\S 6.7]{bd1}, the coherent state
representation can be used to give a path integral representation
of the supertrace of any operator, $\hat{O}$. The following relations
summarize this procedure. The coherent states are defined by
\bea
\mid\! x,\xi ;t\kt &:=&e^{-\frac{1}{2}\xi^{j*}\xi^{j}+\hat{\xi}^{j\dagger}(t)
\xi^{j}}\mid\! x,t\kt \nn \\
\br x,\xi^{*};t\mid & := & \mid\! x,\xi ;t\kt^{\dagger}
\hspace{.3in} , \hspace{.3in} \xi \in \C \; .
\label{eq51}
\eea
The `` $\hat{ }$ '' is used to distinguish the operators from the
scalars where necessary. Equation (\ref{eq51}) leads to
\be
str(\hat{O})=\frac{1}{(2\pi i)^{l}}\int \br x,\xi^{*};t\mid \hat{O}
\mid\! x,\xi ;t \kt d^{m}\!\!x\: d^{l}\xi^{*}\: d^{l}\xi .
\label{eq53}
\ee
In particular, one has
\be
str(e^{-i\b H})=\frac{1}{(2\pi i)^{l}}\int \br x,\xi^{*};t+\b \mid\!
x,\xi ;t \kt d^{m}\!\!x\: d^{l}\xi^{*}\: d^{l}\xi.
\label{eq54}
\ee
The following notation is occasionally used:
\be
K(x,\xi ;\b ) :=\br x, \xi^{*} ;\b \mid x,\xi ;0 \kt
\label{eq55}
\ee
(\ref{eq55}) has a well-known path integral representation. One can change
the variables $\xi$'s to $\p$'s in (\ref{eq54}) and (\ref{eq55}) to
compute the index. This will be pursued in Section~\ref{I-2-4}.

\subsection*{The Case of the Twisted Spin Complex ($\k =1$)}
The commutation relations between $\e$'s and $\e^{*}$'s in (\ref{eq33}),
with $\hbar =1$ and $\e^{\dagger}:=\e^{*}$, read
\bea
\{ \e^{a},\e^{b\dagger} \}&=&\d^{ab} \nn \\
\{ \e^{a},\e^{b} \}&=&\{ \e^{a\dagger},\e^{b\dagger} \} = 0 \; .
\label{eq56}
\eea
Thus, $\e$ and $\e^{\dagger}$ can be viewed as the annihilation and
creation operators for ``$\e$-fermions''. The total Fock space ${\cal
F}_{\rm tot.}$ is the tensor product of the Fock space ${\cal F}_{0}$
of the $\k =0$ case and the one constructed by the action of
$\e^{\dagger}$'s on the vacuum. The basic kets are:
\[ \mid\! a_{p},...,a_{1},i_{r},...,i_{1},x,t\kt :=
\mid\! a_{p},...,a_{1},x,t\kt \otimes \mid\! i_{r},...,i_{1},x,t\kt , \]
where
\[ \mid\! a_{p},...,a_{1},x,t\kt :=\e^{a_{p}\dagger}...\e^{a_{1}\dagger}
\mid\! x,t\kt . \]
The relevant Fock space for the twisted spin complex, however, is the
subspace ${\cal F}_{V}$ of ${\cal F}_{\rm tot.}$ spanned by the
1-$\e$-particle state vectors. These are represented by the following
basic kets:
\be
\mid a,i_{r},...,i_{1},x,t\kt .
\label{eq57}
\ee
In the coordinate representation one has:
\be
\br x,t,i_{1},...,i_{r},a \mid p_{\m} =
-i(\not\!\partial_{\m}+{\cal A}_{\m})
\br x,t,i_{1},...,i_{r},a \mid ,
\label{eq56.1}
\ee
where
\[{\cal A}_{\m} := A_{\m}^{ab}\e^{a*}\e^{b} .\]
This is justified by computing
the following commutation relations and comparing them with the
last five relations in (\ref{eq33}):
\bea
\ll x^{\m},-i(\not\!\partial_{\n}+{\cal A}_{\n}) \rr &=&i\d^{\m}_{\n} \nn \\
\ll \g^{i},-i(\not\!\partial_{\n}+{\cal A}_{\n}) \rr &=&
-i\omega^{i}_{j\n}\g^{j} \nn \\
\ll \e^{a},-i(\not\!\partial_{\n}+{\cal A}_{\n}) \rr &=&
-iA_{\n}^{ab}\e^{b} \nn \\
\ll \e^{a*},-i(\not\!\partial_{\n}+{\cal A}_{\n}) \rr &=&
iA_{\n}^{ca}\e^{c*} \nn \\
\ll -i(\not\!\partial_{\m}+{\cal A}_{\m}) ,
-i(\not\!\partial_{\n}+{\cal A}_{\n}) \rr &=& \mbox{\fs$\frac{1}{4}$}
R_{\m \n \l \d}\g^{\l}\g^{\d}-F_{\m \n}^{ab}\e^{a*}\e^{b} \; . \nn
\eea
Again, the supersymmetric charge $Q$ of (\ref{eq34}) is identified with
the twisted Dirac operator, $\not\!\! D_{V}$,
in the coordinate representation.
In view of (\ref{eq34}), (\ref{eq36}), and (\ref{eq56.1}), one has
\be
\br x,t,i_{1},...,i_{r},a\!\mid Q =
\mbox{\fs$\frac{-1}{\sqrt{2}}$}
g^{\frac{1}{4}}\ll \g^{\m}(\not\!\partial_{\m}+{\cal A}_{\m})\rr
g^{-\frac{1}{4}} \br x,t,i_{1},...,i_{r},a\!\mid .
\label{eq58}
\ee

Once more, $\g^{m+1}$ of (\ref{eq42}) serves as the chirality operator.
In particular, $Q$ switches the $\pm 1$-eigenspaces of $\g^{m+1}$, i.e.
$\{ \g^{m+1},Q\} = 0$.

Coherent states are defined by
\[ \mid\! x,\xi ,\e ;t\kt :=e^{-\frac{1}{2}\e^{a*}\e^{a}+
\hat{\e}^{a\dagger}\e^{a}}\mid\! x,\xi ;t\kt . \]
The supertrace formula, the analog of (\ref{eq53}), is given by:
\[ str(\hat{O})=\frac{1}{(2\pi i)^{l+n}}\int \br x,\xi^{*},\e^{*};t\mid
\hat{O} \mid\! x,\xi ,\e ;t\kt\, d^{m}\!x\, d^{l}\!\xi^{*} d^{l}\!\xi
\, d^{n}\!\e^{*} d^{n}\!\e .\]
The application of the latter equation to the time evolution operator
leads to
\be
str(e^{-i\b H})=\frac{1}{(2\pi i)^{l+n}}\int \br x,\xi^{*},\e^{*};t+\b
\mid\! x,\xi ,\e ;t\kt d^{m}\!x\, d^{l}\!\xi^{*} d^{l}\!\xi\, d^{n}\!\e^{*}
d^{n}\!\e .
\label{eq59}
\ee
Equation (\ref{eq59}) does not, however, provide the index. This is
because in (\ref{eq59}) the supertrace is taken over
${\cal F}_{\rm  tot.}$, rather than ${\cal F}_{V}$. This is remedied
by including a term of the form
% ____________________________________________________________
\newcommand{\eno}{e^{i\a \hat{\e}^{a\dagger}\hat{\e}^{a}}}
% ------------------------------------------------------------
\[ \eno , \]
in (\ref{eq59}), and considering
\be
str \ll e^{-i\b H} \eno \rr .
\label{eq61}
\ee
The linear term in
\[ \l := e^{i\a} \]
in (\ref{eq61}) is precisely the index of
$\not\!\!\partial_{V}$.\footnote{Note that (\ref{eq61}) is a polynomial
in $\l$} The term $[...]_{3}$ in the original
Lagrangian (\ref{eq17}) is  added to fulfill this objective,
\cite{windey,manes-zumino}. In Section~\ref{I-2-5},
\be
H_{\rm eff.}:= H - \mbox{\fs$\frac{\a}{\b}$}
\hat{\e}^{a\dagger}\hat{\e}^{a}
\label{eq61.1}
\ee
will be used in the path integral evaluation of the kernel:
\be
K(x,\xi ,\e ;\b ):=\br x,\xi^{*},\e^{*};\b \mid\! x,\xi ,\e ;0\kt .
\label{eq62}
\ee
The index of $\not\!\partial_{V}$ is then given by
\be
index(\not\!\partial_{V})=\left. \frac{\partial}{\partial \l}\right|_{\l =0}
 str(e^{-i\b H_{\rm eff.}}) \hspace{.1in} .
\label{eq63}
\ee

\section{The Path Integral Evaluation of the Kernel, the Loop Expansion
and the Green's Function Methods}
\label{I-2-3}
The path integral evaluation of the kernel, (\ref{eq64}) below, is
discussed in \cite[\S 5]{bd1}. In general, for a quadratic Lagrangian the
following relation holds:
\be
K(\Phi '',t''\mid \Phi ',t'):=\br \Phi '',t''\mid \Phi ',t'\kt
=Z\int_{(\Phi ',t')}^{(\Phi '',t'')} e^{iS[\Phi ]}\left( sdet
G^{+}[\Phi ]\right)^{-\frac{1}{2}}{\cal D}\Phi .
\label{eq64}
\ee
Here , $Z$ is a (possibly infinite) constant of functional integration.
In the loop expansion of (\ref{eq64}), one expands the field
(coordinate) variables around the (classical) solutions of the dynamical
equations, $\Phi_{0}(t)$:
\be
\Phi^{i}(t)=\Phi^{i}_{0}(t)+\f^{i}(t).
\label{eq65}
\ee
Substituting (\ref{eq65}) in (\ref{eq64}) and expanding in power series
around $\Phi_{0}$, one has~\footnote{ There are additional
complications if $M$ has nontrivial first homology group, \cite{bd1}.
However, this is not relevant to the computation of the index. See
Section~\ref{I-2-6} for a further discussion of this problem. }~:
\be
K(\Phi '',t''\mid \Phi ',t')=Z (sdet G_{0}^{+})^{-\frac{1}{2}}
e^{i\so }\int e^{\frac{i}{2}\f^{i}\: _{i,}\! S_{0 ,j} \f^{j}} \{ 1+\cdots \}
{\cal D}\f \; .
\label{eq66}
\ee
Here,``$\cdots$'' denotes the higher order terms starting with the 2-loop
terms \footnote{The 2-loop terms will be analyzed in Chapter~\ref{I-3}.
They are of order $\b$ or higher. }.
The subscript ``$_{0}$'' means that the corresponding quantity is
evaluated at the classical solution $\Phi_{0}$, e.g.
\[ _{i,}S_{0 ,j}:=\: _{i,}\! S_{,j}[\Phi_{0}] . \]
The lowest order approximation
of (\ref{eq64}), which is explicitly shown in (\ref{eq66}), is the
well-known WKB approximation. This term can be further simplified
if the surviving Gaussian functional integral in (\ref{eq66}) is
evaluated. Generalizing the ordinary Gaussian integral formula,
one has \footnote{In (\ref{eq67}) $c$ is a constant of functional integration,
it may be identified with 1 if the action is appropriately rescaled.
However, this does not play any role in the application of (\ref{eq67})
and thus is not pursued here.}:
\be
\int e^{\frac{i}{2}\f^{i}\: _{i,}\! S_{0 ,j}\f^{j}} {\cal D}\f =
c [sdet(G)]^{\frac{1}{2}} \; .
\label{eq67}
\ee
The Green's function $G=(G^{ij})$, which appears in (\ref{eq67}), is the
celebrated Feynman propagator. It is the inverse of $(  _{j,}S_{0 ,i} )$, :
\be
_{i,}S_{0 ,j'}G^{j'k''} = -\d_{i}^{k}\d (t-t'') ,
\label{eq68}
\ee
defined by the boundary conditions which fix the end points:
\be
\Phi (t')=\Phi ' \hspace{.2in}\mbox{and}\hspace{.2in} \Phi (t'')=\Phi '' .
\label{eq68.1}
\ee
An important property of $G^{ij}$ is the following \cite{bd1}:
\be
G^{ij'}=(-1)^{ij'}G^{j'i} .
\label{eq68.2}
\ee
The emergence of $sdet(G^{+})$ in (\ref{eq64}) is quite important and
must not be underestimated. An explicit computation of $sdet(G^{+})$
for (\ref{eq17}) is in order. The main tool is the definition of the
$sdet$ as the solution of the variational equation \cite[\S 1]{bd1}:
\be
\d \ln [sdet(G^{+})] := str[(G^{+})^{-1} \d G^{+}] \; .
\label{eq69}
\ee
Using the definition of $G^{+}$ (\ref{eq13}):
\be
(G^{+ij})^{-1} := -( _{i,}S_{,j})
\label{eq70}
\ee
one has
% -----------------------------------------------------------
\newcommand{\gr}{G^{+}\left|_{\Phi ',t'}^{\Phi '',t''}\right.}
\newcommand{\ddt}{\d \ln\ll sdet\left( \gr \right)\rr}
% -----------------------------------------------------------
\be
\ddt =\int_{t'}^{t''}d\t\int_{t'}^{t''}d\t'
(-1)^{i}\: _{i,}\d S_{,j'}G^{+j'i} .
\label{eq71}
\ee
In (\ref{eq71}), the variation in the action is with respect to
the functional variation of the metric and the connection fields, i.e.
$\d g_{\m \n}$ and $\d A_{\m}^{ab}$, respectively.
In view of (\ref{eq24}) and (\ref{eq26}), equation (\ref{eq71}) becomes:
% ---------------------------------------------------------------------
\newcommand{\dt}{sdet\left( G^{+}\mid_{\Phi',t_{0}'}^{\Phi'',t_{0}''}\right)}
% ---------------------------------------------------------------------
\bea
\d \ln \dt &=& \int_{t'}^{t''}dt\ll \frac{d}{dt}(\d \ln g)
-iG^{+\g \l}_{1}\d g_{\l \g}+ \right. \label{eq72} \\
& & \hspace{.5cm} \left. i\d^{ab}(\d C_{ba^{*}}+ \d C_{b^{*}a})+
ig^{\g \l}\d C_{\l \g}\rr \th (0) .\nn
\eea
In (\ref{eq72}), $G^{+\g \l}_{1}$ are the coefficients of the linear terms
in the expansion of $G^{+\g \l}$ in (\ref{eq26}), i.e.
\be
G^{+\g \l '} =: \th (t-t')\ll -ig^{\g '\l '}+G^{+\g '\l '}_{1}
(t-t')+O(t-t')^{2}\rr ,
\label{eq73}
\ee
and $\d C$'s are the variations of the corresponding terms in (\ref{eq25}).
It is quite remarkable that although other higher order terms in
(\ref{eq26}) originally enter in (\ref{eq71}), their contributions
cancel and one is finally left with (\ref{eq72}). The fortunate
cancellations seem to be  primarily due to supersymmetry. Incidentally, the
calculation of $G^{+\g \l}_{1}$ is quite straightforward. The 16 coupled
equations, (\ref{eq13}), which give the next order terms in (\ref{eq26}),
decouple miraculously to yield:
\be
G^{+\g \l}_{1}=g^{\d \g}g^{\th \l}C_{\d \th}-g^{\z \t}\G^{\g}_{\z \d}
\G^{\l}_{\t \th}\p^{\d}\p^{\th} .
\label{eq74}
\ee
Substitituing (\ref{eq74}) in (\ref{eq72}), the second term in (\ref{eq74})
drops after contracting with $\d g_{\l \g}$. The surviving term combines
with the last term in (\ref{eq72}) to produce a factor of
\[ i\d (g^{\g \l}C_{\l \g}) . \]
Finally, using (\ref{eq25}) and the identity
\[ \G^{\g}_{\g \m}=\mbox{\fs$\frac{1}{2}$}\partial_{\m}(\ln g) \]
one obtains:
\be
\d \ln \dt =\int_{t'}^{t''}\d \ll \mbox{\fs$\frac{1}{2}$}
(\frac{d}{dt}\ln g) \rr \th (0) dt .
\label{eq75}
\ee
Adapting $\th (0)=1/2$, \cite[\S 6.4]{bd1}, and integrating (\ref{eq75}),
one has:
\be
\dt = \mbox{const.}g^{\frac{1}{4}}(x'') g^{-\frac{1}{4}}(x')\; .
\label{eq76}
\ee
Specializing to the Euclidean case, i.e. $g_{\m \n}=\d_{\m \n}$,
the ``const.'' is identified with ``1''. Furthermore, for the
{\em periodic boundary conditions} (\ref{eq80}), equation
(\ref{eq76}) reduces to
\be
sdet(G^{+})=1 \hspace{.3in}(\mbox{for:}\hspace{3mm} x''=x') \; .
\label{eq77}
\ee
Equation (\ref{eq77}) is a direct consequence of supersymmetry. It results in
a great deal of simplifications in (\ref{eq66}), particularly in
the higher-loop calculations of Chapter~\ref{I-3}.
\section{The Derivation of the Index of Dirac Operator}
\label{I-2-4}
Combinning (\ref{eq11}),(\ref{eq54}),(\ref{eq64}),(\ref{eq66}),(\ref{eq67}) and
(\ref{eq77}), one obtains the index of $\not\!\partial$ in the form:
% --------------------------------
\newcommand{\bo}{\b \rightarrow 0}
% --------------------------------
\be
index(\not\!\partial )=\frac{1}{(2\pi i)^{l}}\int K(x,\p ;\bo )\: \Theta
\: d^{m}\!x\, d^{m}\!\p ,
\label{eq78}
\ee
where
\be
K(x,\p ;\bo ):=\br x,\p ;\bo \mid x,\p ;0\kt \stackrel{
\mbox{\footnotesize WKB}}{=} Z c
e^{i\so } \ll sdet(G)\rr^{\frac{1}{2}} ,
\label{eq79}
\ee
and $\Theta $ is the superjacobian associated with the change of variables
of integration from $\xi ,\xi^{*}$'s to $\p$'s.

In (\ref{eq79}), the periodic boundary conditions must be adapted,
i.e.
\be
\begin{array}{ccccc}
x(0)&=&x(\b )&=:&\xo \\
\p (0)&=&\p (\b )&=:&\po \; .
\end{array}
\label{eq80}
\ee
This is consistent with supersymmetry (\ref{eq20}),
\cite{cecotti,ag3,goodman}.

Requiring (\ref{eq80}) and considering $\b \rightarrow 0$, the only
solution of the classical dynamical equations (\ref{eq22}), with
$\k =0$, is the constant configuration:
\be
\xo (t) = \xo \hspace{.4in} \po (t) = \po \; .
\label{eq81}
\ee
Substituting (\ref{eq81}) in (\ref{eq17}), one has \footnote{One must
take note of the end point contribution to $\so$. For details see
Appendix~D.}:
\be
\so = 0 \; .
\label{eq81.1}
\ee
Thus, the factor $e^{i\so}$ in (\ref{eq79}) drops. Another important
consequence of (\ref{eq81}) is that unlike $ _{i,}S_{,j}$, (\ref{eq24}),
 $ _{i,}S_{0,j}$ are tensorial quantitities.
This allows one to work with normal coordinates, \cite{cd1}, centered at
$\xo$, in which
\bea
g_{0\m \n}& :=& g_{\m \n}(\xo ) = \d_{\m \n} \label{eq83} \\
g_{\m \n ,\s}(\xo )&=&\G^{\s}_{\m \n}(\xo ) = 0 \label{eq84} .
\eea
In the rest of this section, all the fields are evaluated in such a
coordinate system. The final results hold true for arbitrary coordinates
since the quantities of interest are tensorial.

Substituting (\ref{eq81}) in  (\ref{eq24}), using (\ref{eq84}),
and noting that all the $C$'s vanish one obtains:
% --------------------------
\newcommand{\cur}{{\cal R}}
% --------------------------
\bea
_{\t ,}S_{0,\z '} & = & \ll -g_{0\t \z }\ppt + \cur_{\t \z}\pt \rr
\d (t-t')  \nn  \\
_{\t ,}S_{0,\g '} & = & _{\l ,}S_{0,\z '}  = 0  \label{eq87}  \\
_{\l ,}S_{0,\g '} & = & \ll ig_{0 \l \g}\pt \rr \d (t-t') .\nn
\eea
Here, $g_{0 \t \z}=\d_{\t \z}$ are retained for convenience, and
\be
\cur_{\t \z} := \mbox{\fs$\frac{i}{2}$}R_{\d \th \t \z}(\xo )\po^{\d}\po^{\th}.
\label{eq88}
\ee
The Green's functions (\ref{eq68}) are also tensorial quantities.
They can be explicitly computed:
\bea
G^{\z \xi '} & = & g_{0}^{\z \m} \ll \th (t-t')\frac{(e^{\cur (t-\b )}-1)
(1-e^{-\cur t'})}{\cur (1-e^{-\cur \b} )} + \right. \label{eq90} \\
& & \hspace{.5in}\left. + \th (t'-t)\frac{(e^{\cur t}-1)
(1-e^{-\cur (t'-\b )})}{\cur (e^{\cur \b}-1)} \rr_{\m}^{\xi} \nn \\
G^{\z \d '} & = & 0   \label{eq91} \\
G^{\g \xi '} & = & 0 \label{eq91.1} \\
G^{\g \d '} & = & \frac{i}{2}g_{0}^{\g \d}\ll \th (t-t') - \th (t'-t) \rr\; .
\label{eq92}
\eea
In (\ref{eq90}), the expression inside the bracket is to be interpreted
as a power series in:
\be
\cur := (\cur_{\t}^{\m}) := ( g_{0}^{\m \n}\cur_{\t \n}).
\label{eq89}
\ee
This is a finite series due to the presence of $\po$'s in (\ref{eq88})
\footnote{One can use all the properties of the ``$\exp$'' and other
analytic functions with arguments such as $\cur $. The only
rule is that the power series expansion must be postponed until
all other operations are performed.}.
Equation (\ref{eq90}) is obtained starting from the following
ansatz:
% ---------------------------------
\newcommand{\kp}{{\cal X}_{+}}
\newcommand{\km}{{\cal X}_{-}}
\newcommand{\kpm}{{\cal X}_{+\m}^{\xi}}
\newcommand{\kmm}{{\cal X}_{-\m}^{\xi}}
% ---------------------------------
\be
G^{\z \xi '} = \ll \th (t-t') (\frac{t}{\b}-1) t' g^{\z \m}
\kpm (t,t') + \th (t'-t) (\frac{t'}{\b}-1) t\, g^{\z \m}
\kmm (t,t') \rr.
\label{eq93}
\ee
(\ref{eq93}) satisfies the boundary conditions (\ref{eq68.1}), i.e.
\be
G^{\z \xi '} = 0 \hspace{.5in}\mbox{if $t$ or $t'$ $=$ $0$ or $\b$} .
\label{eq90.1}
\ee
Moreover, the reduction to $\p =0$ case is equivalent to choosing
${\cal X}_{\pm}=\I $. Imposing (\ref{eq68}) on (\ref{eq93})
and using (\ref{eq87}), one obtaines the following equations:
\bea
t>t' & : & (t-\b )\ll \ppt \kp -\cur \pt \kp \rr + 2\ll \pt \kp
-\mbox{\fs$\frac{1}{2}$} \cur \kp \rr =  0 \label{eq94} \\
t<t' & : & t \ll \ppt\km -\cur \pt \km \rr + 2\ll \pt \km
-\mbox{\fs$\frac{1}{2}$} \cur \km \rr =  0 \label{eq95} \\
t=t' & : & \ll \frac{t}{\b}\kp -(\frac{t}{\b}-1)\km + \right.
\label{eq96}  \\
& & \left.+ t(\frac{t}{\b}
-1) \left( \pt \kp - \pt \km - \cur \kp + \cur \km \right) \rr_{t=t'}
= \I . \nn
\eea
(\ref{eq94}) and (\ref{eq95}) are easily solved by the power series method.
The final result, (\ref{eq90}), follows using (\ref{eq68.2}) and (\ref{eq96}).
The derivation of (\ref{eq91}),(\ref{eq91.1}), and (\ref{eq92}) is
straightforward.

The next step is to compute $sdet(G)$. This is accomplished by
considering a functional variation in the metric tensor, $\d g_{\m \n}$,
and using the definition of $sdet$, (\ref{eq69}). After considerable
amount of algebra and repeated use of the symmetries of $g_{\m \n}$,
and $\cur_{\m}^{\n}$, one arrives at the following expression:
\be
\d \ln [sdet(G)] = \th (0) tr\ll 2\d \ln (\b \cur ) + \b \d \cur
(\frac{1+e^{\b \cur}}{1-e^{\b \cur}}) \rr\; .
\label{eq97}
\ee
Setting $\th (0)=1/2$ and integrating the right hand side of (\ref{eq97}),
one has:
\be
\d \ln [sdet(G)] = \d tr \left( \ln \ll \frac{-\frac{\b \cur}{2}}{\sinh
(\frac{\b \cur}{2})} \rr \right)\; .
\label{eq98}
\ee
Since $\cur$ is  antisymmetric in its indices, it can be put in the
following block-diagonal form:
\be
\cur = diag \left( \ll
\begin{array}{cc}
0 & \cur_{i} \\
-\cur_{i} & 0
\end{array}
\rr : i=1,\cdots ,l \right) \; .
\label{eq99}
\ee
% -------------------------------
\newcommand{\AO}{\frac{\frac{\b \cur}{2}}{\sinh (\frac{\b \cur}{2})} }
% -------------------------------
Another important observation is that $\AO$ and hence $\ln [-\AO ]$
are polynomials in $(\frac{\b \cur}{2})^{2}$ . In view of (\ref{eq99}),
it is easy to see that $\cur^{2}$ is diagonal, namley
\be
\cur^{2}=diag(-\cur_{1}^{2},-\cur_{1}^{2},\cdots,-\cur_{l}^{2},-\cur_{l}^{2}).
\label{eq100}
\ee
Implementing (\ref{eq100}) in (\ref{eq98}), one has:
% --------------------------------
\renewcommand{\cur}{i{\cal R}_{j}}
\newcommand{\pl}{\prod_{j=1}^{l}}
% --------------------------------
 \[ \d \ln [sdet(G)] = 2 \d \ln \pl \ll -\AO \rr  , \]
and finally
\be
sdet(G)^{\frac{1}{2}} = \tilde{c} \pl \ll \AO \rr ,
\label{eq101}
\ee
where $\tilde{c}$ is a constant of functional integration.
Substituting (\ref{eq101}) and (\ref{eq81}) in (\ref{eq79}), one obtains
the kernel in the form:
\be
K(x,\p ;\bo ) =  Zc\,\tilde{c}\pl \ll \AO \rr .
\label{eq102}
\ee
% ---------------------------------
\renewcommand{\cur}{{\cal R}}
\newcommand{\crj}{\cur_{j}}
\newcommand{\oj}{\Omega_{j\d \th }\po^{\d}\po^{\th}}
\newcommand{\boj}{\frac{\b \oj}{2}}
% ---------------------------------
To write $\crj$ as functions of $\xo$ and $\po$, one needs to also
block-diagonalize:
\be
\left( \mbox{\fs$\frac{1}{2}$}R_{\m \n \d \th}
\mbox{\fs$(\xo)$}\,\po^{\d}\po^{\th} \right) = diag
\left( \ll
\begin{array}{cc}
0 & \oj \\
-\oj & 0
\end{array} \rr : \mbox{\fs$j=1,\cdots ,l$} \right) .
\label{eq103}
\ee
Combinning (\ref{eq88}),(\ref{eq89}),(\ref{eq99}),(\ref{eq102}) and
(\ref{eq103}), one is led to
\be
K(\xo ,\po ;\bo )= Zc\tilde{c} \pl \ll \frac{\boj}{\sinh (\boj )}
\rr .
\label{eq104}
\ee
The proportionality constant $Zc\,\tilde{c}=:\tilde{Z}$ is determined
by specializing to the $M=\R^{m}$ case. Using the results of
\cite[\S 5,\S 6]{bd1}, one has
\be
\tilde{Z} = (2\pi i dt)^{-\frac{\b m}{2 dt}}\times (2\pi i)^{
-\frac{\b l}{dt}+l}\; .
\label{eq105}
\ee
The limit $\b \rightarrow 0$ is taken by setting
\be
\b =dt\; .
\label{eq106}
\ee
Finally, combinning (\ref{eq78},\ref{eq104}) and (\ref{eq106}), and realizing
that $e^{\m}_{i}(\xo)=\d^{\m}_{i}$ so that $\Theta =i^{-l}$, one has
\be
index(\not\!\partial )=\frac{1}{(2\pi )^{l}}\int \pl \ll
\frac{\boj}{\sinh (\boj )}\rr \frac{d^{m}\po \, d^{m}\!\xo}{(2\pi i\b )^{l}}.
\label{eq108}
\ee
The $\po$-integration rules, \cite{bd1}:
\be
\begin{array}{ccc}
\int d\po &=& 0 \\
\int \po \, d\!\po &=& \sqrt{2\pi i}
\end{array}
\label{eq109}
\ee
allow only the highest degree term in the integrand to survive. This
implies the cancellation of $\b $'s. Performing the $\po$-integrations
yields an expression for (\ref{eq108}) which is identical with the
following:
\bea
index(\not\!\partial )&=&\frac{1}{(2\pi )^{l}}\int_{M}\pl \ll
\frac{\frac{\Omega_{j}}{2}}{\sinh (\frac{\Omega_{j}}{2})}
\rr_{\mbox{top}} \nn \\
& = & \int_{M}\pl \ll \frac{\frac{\Omega_{j}}{4\pi}}{\sinh
(\frac{\Omega_{j}}{4\pi})} \rr_{\mbox{top}}\; .
\label{eq110}
\eea
In (\ref{eq110}), $\Omega_{j}$ are defined by (\ref{eq5}) and the identity
\[ \Omega_{j} = \Omega_{j\d \th}\: dx^{\d}\wedge dx^{\th}. \]
(\ref{eq110}) is precisely the statement of Theorem~\ref{theorem1},
i.e. (\ref{eq4}).
The fact that the integrand in (\ref{eq110})
is an even polynomial in $\frac{\Omega{_j}}{4\pi}$ implies that only
for $l=2k$, i.e. $m=4k$, is the index nonvanishing.

\section{The Derivation of the Index of the Twisted Dirac Operator}
\label{I-2-5}
Equations (\ref{eq59}),(\ref{eq62}),(\ref{eq63}),(\ref{eq64}),
(\ref{eq66}),(\ref{eq67}), and (\ref{eq77}) provide the following formula
for the index
\be
index(\not\!\partial_{V})=\left. \frac{\partial}{\partial \l}
\right|_{\l =0} \ll
\frac{1}{(2\pi i)^{l+n}}\int K(x,\p ,\e ;\b\rightarrow 0)\,\Theta\,
d^{m}\! x\, d^{m}\!\p\, d^{n}\!\e^{*}\, d^{n}\!\e \rr
\label{eq111}
\ee
where
\be
K(x,\p,\e ;\b\rightarrow 0):=\br x,\p,\e^{*};\b\rightarrow 0 \mid
x,\p ,\e ;0 \kt\stackrel{\mbox{\footnotesize WKB}}{=\hspace{-1.5mm}=}
Z'c'e^{i\so}[sdet(G)]^{\frac{1}{2}} .
\label{eq112}
\ee
The first step in the calculation of the kernel is to obtain the
classical solution of the dynamical equations, (\ref{eq22}), in the
$\b\rightarrow 0$ limit.

As one can see from the analysis of Section~\ref{I-2-4}, absorbing a factor
of $\sqrt{\b}$ in $\p$'s can take care of the limiting process
automatically. Furthermore, following \cite{manes-zumino}, define
the parameter:
\[ s:=\frac{t}{\b}, \]
and expand the coordinate variables in powers of $\b$:
\bea
x(t)&=&\tilde{x}_{0}(s)+\tilde{x}_{1}(s)\b +O(\b^{2}) \nn \\
\p (t)&=&\frac{1}{\sqrt{\b}}\ll \tilde{\p}_{0}(s)+\tilde{\p}_{1}(s)
\b +O(\b^{2})\rr \label{eq113} \\
\e (t)&=&\tilde{\e}_{0}(s)+\tilde{\e}_{1}(s)\b +O(\b^{2}). \nn \\
\e^{*}(t)&=&\tilde{\e}^{*}_{0}(s)+\tilde{\e}^{*}_{1}(s)\b+O(\b^{2}) \; .\nn
\eea
The appropriate boundary conditions for (\ref{eq112}) are the periodic
boundary conditions \footnote{In fact, one can adopt antiperiodic boundary
conditions for $\e$'s and $\e^{*}$'s, \cite{ag3}. Since the relevant
trace is taken over the 1-$\e$-particle state vectors, this would
introduce an extra minus sign.}, i.e., (\ref{eq80}) and:
\bea
\e & = &\e ':=\e (t=0) =\e (\b ) =: \e '' \nn \\ \nopagebreak[3]
\e^{*'}& := &\e^{*}(t=0)=\e^{*}(t=\b )=:\e^{*''}=\e^{*} \; .
\label{eq114}
\eea

It follows from (\ref{eq80}),(\ref{eq114}), and (\ref{eq113}) that the
dynamical
equations are solved by:
\bea
\xo (t)&=&\xo + O(\b ) \nn \\
\po (t)&=&\frac{1}{\sqrt{\b}}\ll \tilde{\p}_{0} + \tilde{\p}_{1}(s) \b
+O(\b^{2}) \rr \label{eq115} \\
\e_{0} (t)&=&\tilde{\e}_{0}(s)+O(\b ) \nn \\
\e_{0}^{*}(t)&=&\tilde{\e}^{*}_{0}(s)+O(\b ), \nn
\eea
where $\xo$ and $\po =:\frac{1}{\mbox{\fs$\sqrt{\b}$}}\tilde{\p}_{0}$
are constants.
Adopting a normal coordinate frame centered at $\xo$ and using (\ref{eq83}),
(\ref{eq84}) and
\[ A_{\m}^{ab}(\xo)=0 ,\]
one has:
%_______________________________________________________
\newcommand{\te}{\tilde{\e}_{0}}
%\newcommand{\cuf}{{\cal F}}
%-------------------------------------------------------
\bea
\frac{d}{ds}\tilde{\p}_{1}^{\g}-ig_{0}^{\g \l}F_{\l \d}^{ab}(\xo )
\tilde{\p}_{0}^{\d}\tilde{\e}_{0}^{a*}\tilde{\e}_{0}^{b}&=&0 \nn \\
\frac{d}{ds}\tilde{\e}_{0}^{a}-i(\tcuf^{ab}+\a \d^{ab})\te^{b}&=&0
\label{eq116} \\
\frac{d}{ds}\te^{a*}+i(\tcuf^{ba}+\a \d^{ba})\te^{b*}&=&0 .\nn
\eea
Here $\tcuf$ is defined by
\be
\tcuf = \left( \tcuf_{ab} \right) := \left( \mbox{\fs$\frac{1}{2}$}
F_{\g \l}^{ab}(\xo)\, \tilde{\p}_{0}^{\g}\tilde{\p}_{0}^{\l} \right)\; .
\label{eq117}
\ee
The quantity:
\be
\cuf =\left( \cuf_{ab} \right) := \left( \mbox{\fs$\frac{1}{2}$}
F_{\g \l}^{ab}(\xo )\, \po^{\g}\po^{\l} \right) = \mbox{\fs$\frac{1}{\b}$}
\tcuf
\label{eq118}
\ee
will also be used.

The next step is to compute $\so$. This is done in Appendix~\ref{z2-D}. The
final result is :
\be
\so = -i\e^{a*}\ll e^{i(\b \cuf +\a \In )}-\In \rr^{ab}\e^{b}\; .
\label{eq119}
\ee

Next, one needs to calculate the Feynman propagator $G$. This is
done in two steps. First, consider the special case of
\bea
\xo (t)&=&\xo \nn \\
\po (t)&=&\po = \mbox{\fs$\frac{1}{\sqrt{\b}}$}\tilde{\p}_{0} \label{eq120} \\
\e_{0}(t)&=&0 \nn \\
\e^{*}_{0}(t)&=&0 \; . \nn
\eea
It is clear that (\ref{eq120}) satisfies the dynamical equations,
(\ref{eq22}). Substituting (\ref{eq120}) in (\ref{eq24}) and (\ref{eq25}),
one recovers (\ref{eq87}). The other nonvanishing $_{i,}S_{0,j}$'s are:
\be
\begin{array}{ccc}
_{b,}S_{0,a^{*'}}&=&\ll i\d_{ba}\pt -(\cuf_{ab}+\frac{\a}{\b}\d_{ba})\rr
\d (t-t') \\
_{b^{*},}S_{0,a'}&=&\ll i\d_{ba}\pt +(\cuf_{ba}+\frac{\a}{\b}\d_{ba})\rr
\d (t-t')\; .
\end{array}
\label{eq121}
\ee
In other words, one has:
\be
\left( \, _{i,}S_{0,j'} \right)=\ll
\begin{array}{cccc}
_{\t ,}S_{0,\z '} & 0 & 0 & 0 \\
0 & _{\l ,}S_{0,\g '} & 0 & 0 \\
0 & 0 & 0 & _{b,}S_{0,a^{*'}} \\
0 & 0 & _{b^{*},}S_{0,a'} & 0
\end{array} \rr\; .
\label{eq122}
\ee
(\ref{eq122}) suggests:
\be
\left( G^{ij'} \right) = \ll
\begin{array}{cccc}
G^{\z \xi '} & 0 & 0 & 0 \\
0 & G^{\g \e '} & 0 & 0 \\
0 & 0 & 0 & G^{ac^{*'}} \\
0 & 0 & G^{a^{*}c'} & 0
\end{array} \rr\; .
\label{eq123}
\ee
In view of (\ref{eq68}), it is clear that $G^{\z \xi '}$ and $G^{\g \e '}$
are given by equations (\ref{eq90}) and (\ref{eq92}), respectively. Defining
\[ G_{1}(t,t') = \left( G_{1}^{ac}(t,t') \right) := \left( G^{a^{*}c'}\right)\]
\[ G_{2}(t,t') = \left( G_{2}^{ac}(t,t') \right) := \left(
G^{ac^{*'}}\right),\]
and using (\ref{eq123}),(\ref{eq87}),(\ref{eq121}),(\ref{eq90}),(\ref{eq92}),
and (\ref{eq68}), one obtains:
\be
\begin{array}{ccc}
\ll i\pt - (\cuf^{*}+\frac{\a}{\b}) \rr G_{1}(t,t')&=&-\d (t-t') \\
\ll i\pt + (\cuf +\frac{\a}{\b}) \rr G_{2}(t,t')&=&-\d (t-t') \; .
\end{array}
\label{eq124}
\ee
In (\ref{eq124}), use has been made of the fact that $\cuf$ is hermitian
and hence
\[ \cuf^{\rm transpose}=\cuf^{*}. \]
To compute $G_{1}$ and $G_{2}$, consider the ansatz:
%------------------------------
\newcommand{\xx}{{\cal X}}
\newcommand{\yy}{{\cal Y}}
%------------------------------
\be
G_{k}(t,t')=\th (t-t')e^{i\xx_{k}(t,t')}-\th (t'-t)e^{i\yy_{k}(t,t')}
\hspace{12mm}(k=1,2) .
\label{eq125}
\ee
Substituting (\ref{eq125}) in (\ref{eq124}), one obtains:
\bea
t>t' &:&\pt\xx_{1}=-(\cuf^{*}+\frac{\a}{\b})\hspace{2mm},\hspace{2mm}
\pt\xx_{2}=\cuf + \frac{\a}{\b} \nn \\
t<t' &:&\pt\yy_{1}=-(\cuf^{*}+\frac{\a}{\b})\hspace{2mm},\hspace{2mm}
\pt\yy_{2}=\cuf + \frac{\a}{\b} \nn \\
t=t' &:&\ll e^{i\xx_{k}}+e^{i\yy_{k}}\rr_{t=t'}=i \hspace{5mm}k=1,2. \nn
\eea
Finally, using (\ref{eq68.2}):
\[ G_{1}(t,t')=-G_{2}^{\rm transpose}(t',t) ,\]
one arrives at the following expressions:
\be
\begin{array}{ccccc}
G_{1}(t,t')&=&\left( G^{a^{*}c'} \right)&=&\frac{i}{2}
e^{-i(\cuf^{*}+\frac{\a}{\b})(t-t')}\ll \th (t-t')-\th (t'-t) \rr \\
G_{2}(t',t)&=&\left( G^{ac^{*'}} \right)&=&\frac{i}{2}
e^{i(\cuf +\frac{\a}{\b})(t-t')}\ll \th (t-t')-\th (t'-t) \rr\; .
\end{array}
\label{eq126}
\ee
Next step is to observe that, in the limit: $\b \rightarrow 0$, (\ref{eq123})
with (\ref{eq90}),(\ref{eq92}), and (\ref{eq126}) actually satisfies equation
(\ref{eq68}) even in the general case of (\ref{eq115}). To see this, it is
sufficient to examine:
\bea
_{\t ,}S_{0,\z '}&=&\b^{-3}\ll -g_{0\t \z}\frac{\partial^{2}}{\partial s^{2}}
+\frac{i}{2}R_{\l \g \z \t}(\xo )\, \tilde{\p}_{0}^{\g}\tilde{\p}_{0}^{\l}\pss
\rpt \nn \\
&&\hspace{.8cm}\lpt  +O(\b )\rr \d (s-s') \label{eq127} \\
_{\l ,}S_{0,\g '}&=&\b^{-2}\ll ig_{0\l \g}\pss +O(\b )\rr\d (s-s')
\label{eq128}
\eea
and
\bea
_{b,}S_{0,a^{*'}} & = &\b^{-2}\ll i\d_{ba}\pss -(\tcuf_{ab}+\a\d_{ab})+O(\b
)\rr
\d (s-s') \nn \\
_{b^{*},}S_{0,a'} & = & \b^{-2}\ll i\d_{ba}\pss +(\tcuf_{ba}+\a\d_{ba})+O(\b
)\rr
\d (s-s') .
\label{eq129}
\eea
The next step is to compute  $sdet(G)$. Since $G$ is block-diagonal, one
has:
\be
sdet(G)=sdet(G_{0}).sdet(G_{\e})
\label{eq130}
\ee
where
\[ G_{0} := \left( \begin{array}{cc}
G^{\z\xi '} & 0 \\
0 & G^{\g\l '} \end{array} \right) \mbox{\hspace{2mm}and\hspace{2mm}}
G_{\e} := \left( \begin{array}{cc}
0 & G^{ac^{*'}} \\
G^{a^{*}c'} & 0 \end{array} \right) . \]
Clearly, $sdet(G_{0})$ is given by (\ref{eq101}). $sdet(G_{\e})$ is
computed following the procedure of Section~\ref{I-2-4}. Applying (\ref{eq69})
and (\ref{eq71}) to $G_{\e}$ and writing only the nonvanishing terms,
one has:
\be
\d \ln sdet(G_{\e}) = \int_{0}^{\b}dt\int_{0}^{\b}dt' \ll
-\, _{c,}\d\! S_{0,a^{*'}}G^{a^{*'}c}-\, _{c^{*},}\d\! S_{0,a'}G^{a'c^{*}} \rr
{}.
\label{eq131}
\ee
Let us define:
% -----------------------------
\newcommand{\zz}{{\cal Z}}
% -----------------------------
\be
\zz (t,t'') := \int_{0}^{\b}dt'\ll -\, _{c,}\d\! S_{0,a^{*'}}G^{a^{*'}c''}
-\, _{c^{*},}\d\! S_{0,a'}G^{a'c^{*''}} \rr .
\label{eq132}
\ee
Taking the functional variation of (\ref{eq129}) with respect to
$\d\cuf$ and substituting the result in (\ref{eq132}), one has:
\bea
\zz (t,t'') & = & -\d\cuf_{ac}G^{a^{*}c''}+\d\cuf_{ca}G^{ac^{*''}}
\label{eq133} \\
& = & tr\ll -\d\cuf^{*}\, G_{1}(t,t'')+\d\cuf\, G_{2}(t,t'')\rr \nn \\
& = & tr \ll \d e^{-i(\cuf^{*}+\frac{\a}{\b})(t-t'')}+
             \d e^{i(\cuf +\frac{\a}{\b})(t-t'')} \rr \mbox{\fs$
\ll \frac{\th (t-t'') -\th (t''-t)}{2(t''-t)} \rr$} \nn \\
& = & tr \ll \d \left( \cos [ (\cuf +\frac{\a}{\b})(t-t'')] \right) \rr
\mbox{\fs$\ll \frac{\th (t-t'')-\th (t''-t)}{t''-t} \rr$} \nn \\
& = & \d tr \ll \frac{1}{2}(\cuf +\frac{\a}{\b})^{2}(t''-t)+O(t-t'')^{3} \rr
\ll \th (t-t'') -\th (t''-t) \rr. \nn
\eea
Combinning equations (\ref{eq131}),(\ref{eq132}), and (\ref{eq133}), one
obtains:
\[ \d \ln sdet(G_{\e}) = \int_{0}^{\b}dt \:\zz (t,t) = 0 ,\]
and hence:
\be
sdet(G_{\e}) = const.
\label{eq134}
\ee
Equations (\ref{eq101}),(\ref{eq130}), and (\ref{eq134}) yeild:
% ----------------------------------------
\renewcommand{\cur}{i{\cal R}_{j}}
% ----------------------------------------
\be
\ll sdet(G)\rr^{\frac{1}{2}}=\tilde{c}'\pl \ll \AO \rr\; .
\label{eq135}
\ee

Substituting (\ref{eq119}) and (\ref{eq135}) in (\ref{eq112}), one
finds:
\be
K(\xo ,\po ,\e ;\b\rightarrow 0) = Z'c'\tilde{c}'\pl \ll \AO \rr
\mbox{\fs$\exp \left( \e^{a*}\ll e^{i(\b\cuf +\a\In)}-\In\rr^{ab}\e^{b}
\right)$}\; .
\label{eq136}
\ee
The constant $\tilde{Z}':=Z'c'\tilde{c}'$ is fixed by specializing to
the case of $\cuf =\a =0$. In this case (\ref{eq136}) must reduce to
(\ref{eq104}). Thus,
\[ \tilde{Z}'=\tilde{Z}=(2\pi i\b )^{-l}. \]
Substituting (\ref{eq136}) in (\ref{eq111}) and performing the $\e^{*}$
and $\e$ integrations, one has:
\be
index(\not\!\partial_{V})=\left. \frac{\partial}{\partial \l}
\right|_{\l =0} \left( \frac{1}{(2\pi )^{l}}\int \pl \ll \AO \rr
\mbox{\fs$ det \ll e^{i(\b\cuf +\a\In )} - \In \rr $}\right)  \; .
\label{eq137}
\ee
At this stage, one can take the $\l$-derivative. Since $\cuf$ is hermitian
it can be diagonalized, i.e.
\[ \cuf =: diag ( \cuf_{1},\cdots ,\cuf_{n}). \]
Then,
% ------------------------------------------------
\newcommand{\plo}{\left. \frac{\partial}{\partial \l}\right|_{\l =0}}
% ------------------------------------------------
\bea
\plo det\ll e^{i(\b\cuf +\a\In)}-\In\rr & = & \plo \ll \prod_{a=1}^{n}
(\l e^{i\b\cuf_{a}}-1) \rr \nn \\
&=& (-1)^{n}\plo \ln \ll \prod_{a=1}^{n}(\l e^{i\b\cuf_{a}}-1) \rr \nn \\
& = & \sum_{a=1}^{n}e^{i\b\cuf_{a}} \nn \\
& = & tr\, (e^{i\b\cuf}) \; . \label{eq138}
\eea
Using (\ref{eq138}) and performing $\po$-integrations, one finally
arrives at:
\bea
index(\not\!\partial_{V})&=& \frac{1}{(2\pi )^{l}}\int_{M}
\ll \pl \left( \frac{\frac{\Omega_{j}}{2}}{\sinh (\frac{\Omega_{j}}{2})}
\right) tr\, (e^{iF})\rr_{\rm top.} \nn \\
& = & \int_{M}\ll \pl \left( \frac{\frac{\Omega_{j}}{4\pi}}{
\sinh (\frac{\Omega_{j}}{4\pi})} \right) tr\, (e^{\frac{iF}{2\pi}})
\rr_{\rm top.} .
\label{eq139}
\eea
In (\ref{eq139}), $\Omega_{j}$ and $F$ are defined by (\ref{eq5}) and
(\ref{eq9.1}), respectively. This completes the proof of
Theorem~\ref{theorem2}.

\section{Remarks and Discussion}
\label{I-2-6}
The following is a list of remarks concerning some of the aspects of
the present work:
\begin{itemize}
\begin{enumerate}
\item The reality condition on the Lagrangian (\ref{eq17}), requires
$F_{\m\n}$ to be antihermitian matrices. This is equivalent to
requiring the structure group of the bundle $V$ to be (a subgroup)
of $U(n)$. This is consistent with the existence of a hermitian
structure on $V$. For, any hermitian vector bundle can be reduced
to one with $U(n)$ as its structure group.
\item The term $i\e^{a*}\dot{\e}^{a}$ in the Lagrangian (\ref{eq17})
can be replaced with
\[ \mbox{\fs$\frac{i}{2}$} (\e^{a*}\dot{\e}^{a}-\dot{\e}^{a*}\e^{a}) \]
with no consequences for the action. This is because the boundary
conditions on $\e$'s and $\e^{*}$'s are periodic.
\item In the Peierls quantization scheme, the momenta, ``$p_{\m}$'',
are not {\em a~priori} fixed. They may be chosen in a way that facilitates
the analysis of the problem. In particular, the choice: (\ref{eq31})
has obvious advantages. The factor ordering problem may arise in
general. In the case of th index problem, however, the requirement that
the supersymmetry generator be identified with the elliptic
operator in question, enables one to choose the appropriate ordering.
It is also remarkable that in the special case that the structure
group of $V$ can be chosen $SU(n)$, the factor ordering ambiguities,
(\ref{eq33.1}), disappear. It seems that the obstruction for having
a unique quantum system is the first Chern class.
\item The problem of nonuniqueness of the momentum operators was
not addressed in Section~\ref{I-2-2}. In general,
\[ {\bf p}:=p_{\m}\, dx^{\m} \]
is unique up to closed forms, $\omega\in H^{1}(M,\R\! )$, \cite{bd1}.
This is reflected in the path integral formulation as the
necessity of concidering different homotopy meshes of paths. This
is, however, not relevant to the index problem. In general, the
index of an elliptic operator is a map:
\[ index\, :\:K(TM)\,\rightarrow \Z \; , \]
where $K(TM)\cong K(M)\cong\bigoplus_{i} H^{2i}(M,\!\Z )$ is the
Grothendieck's K-group (ring), \cite{shanahan}. Thus the index does
not detect the first cohomology group of M. For more details
of K-theory see Appendix~\ref{z2-A11}.
\item The procedure used in the evaluation of the index in this paper
are essentially based on a single basic assumption. This is the
Gaussian functional integral formula, (\ref{eq67}). The other
important ingredient is the definition of the superdeterminant,
(\ref{eq69}),
which is assumed to hold even for infinite dimensional matrices.
The latter can be easly shown to be a consequence of (\ref{eq67}).
\item The factor $[ sdet\, G^{+}]^{-\frac{1}{2}}$, in (\ref{eq64}), is
usually constant for the case of flat spaces. However, it contributes
in an essential way in the case of curved spaces. For the system
considered in this paper, it was explicitely calculated and shown
to be 1. It is not difficult to see that the amazing cancellations
are due to supersymmetry. Considering the complexity of the system,
(\ref{eq17}), (\ref{eq24}), and (\ref{eq25}), this might be a
general pattern for a large class of supersymmetric systems.
\item The appearence of $\frac{\hbar^{2}}{8}R$ in the Hamiltonian,
(\ref{eq35}), can be verified independently by comparing the heat
kernel and loop expansions of the path integral, (\ref{eq64}).
This  term contributes to the linear term in
the heat kernel expansion, \cite{bd1}. On the other hand, the
presence of $\hbar^{2}$ is reminiscent of the contribution
to the path integral in the 2-loop order. The complete
2-loop analysis of this problem is the subject of the next chapter.
\end{enumerate}
\end{itemize}
\chapter{The Scalar Curvature Term in the Schr\"odinger Equation}
\label{I-3}
\renewcommand{\cur}{{\cal R}}
\vspace{1cm}
\bcc
{\bf Abstarct}
\ecc
\bcc
\parbox[b]{4.5in}{\small
The quantization of the superclassical system used in the proof of the
index theorem results in a factor of $\frac{\hbar^2}{8}R$ in this
Hamiltonian. The path integral expression for the kernel is analyzed
up to and including 2-loop order. The existence of the scalar curvature
term is confirmed by comparing the linear term in the heat kernel
expansion with the 2-loop order terms in the loop expansion. In the
operator formalism this term arises from the fermionic sector whereas
in the path integral formulation it comes from the bosonic sector.}
\ecc
\vspace{.5in}
\section{Introduction}
\label{I-3-1}
In the preceding chapter, a supersymmetric proof of the
twisted spin index theorem was presented.
There, the Peierls bracket quantization was applied to the
following supersymmetric Lagrangian:
\bea
L&=&\ll \oot g_{\m\n}\, \dot{x}^{\m}\,\dot{x}^{\n}+\iot g_{\l\g}\,\p^{\l}
\left( \dot{\p}^{\g}+\dot{x}^{\m}\G_{\m\th}^{\g}\p^{\th}\right) \rr +
\label{e1} \\
&&+\k\ll i\e^{a*}\left( \dot{\e}^{a}+\dot{x}^{\s}A_{\s}^{ab}\e^{\b}\right)
+\oot F_{\l\g}^{ab}\,\p^{\l}\p^{\g}\e^{a*}\e^{b}\rr + \nn \\
&&+\mbox{\fs$\frac{\a}{\b}$}\e^{a*}\e^{a} \; .\nn
\eea
The classical ``momenta'' were defined by:
\be
p_{\m}:=g_{\m\n}\dot{x}^{\n}\; .
\label{e2}
\ee
The Peierls bracket quantization led to the quantization of the
supersymmetric charge:
\be
Q=\mbox{\fs$\frac{1}{\sqrt{\hbar}}$}\p^{\n}\, g^{\frac{1}{4}}\, p_{\n}
g^{-\frac{1}{4}} \; .
\label{e4}
\ee
The quantum mechanical Hamiltonian was then given by:
\be
H=Q^{2}=\oot g^{-\frac{1}{4}}p_{\m}g^{\oot}g^{\m\n}p_{\n}g^{-\frac{1}{4}}
+\mbox{\fs$\frac{\hbar^{2}}{8}$}R-\mbox{\fs$\frac{\k}{2}$}
F_{\l\g}^{ab}\p^{\l}\p^{\g}
\e^{a*}\e^{b} \; .
\label{e6}
\ee
The fact that $Q$ is identified with the (twisted) Dirac operator,
\be
Q \equiv \not\!\! D \; ,
\label{e7}
\ee
defined $H$ uniquely.

Reducing (\ref{e1}) to a purely bosonic system, i.e. setting $\p =\e =\e^{*}
=0$, one arrives at the Lagrangian for a free particle moving on a Riemannian
manifold \footnote{In this dissertation the case of closed and simply
connected Riemannian manifolds is considered. The reason is explained
in Section~\ref{I-2-6} .}.
Equation~(\ref{e6}) is in complete agreement with the analysis of
the reduced system presented in \cite{bd1}. There, the Hamiltonian was defined
by requiring a particular factor ordering, namely by the time ordering
\cite[\S 6.5]{bd1}. Furthermore, it was shown in \cite[\S 6.6]{bd1} that
the term $\frac{\hbar^{2}}{8}R$ in the Hamiltonian contributed to the
kernel:
\be
K(x';t''|x';t') := \br x''=x';t''|x';t'\kt \; ,
\label{e8}
\ee
a factor of
\be
-\mbox{\fs$\frac{i\hbar^{2}}{24}$}\,R(x')\:(t''-t') \; .
\label{e9}
\ee
This was obtained using the heat kernel expansion of (\ref{e8}). The
quantity: (\ref{e9}) is the linear term in the heat kernel expansion.
In \cite[\S 6.6]{bd1}, the 2-loop terms were computed.
It was shown that indeed the loop expansion validates the
existence of $\frac{\hbar^{2}}{8}R$ term. The presence of
the scalar curvature factor in the Schr\"{o}dinger equation is discussed
in \cite{bd2}. For further review see,
\cite{marinov,cecile} and references therein.\footnote{In \cite{cecile}
it is shown that for the purely bosonic system whose Hamiltonian does
not include the curvature factor, the short time and the WKB limit
of the path integral does not coincide. This implies that there are
examples for which the first term in the loop expansion is different from
the first term in the heat kernel expansion. As it will be explicitly
shown, the supersymmetric systems studied in this dissertation are
not among these examples.}

The present chapter is devoted to the 2-loop analysis of the path integral
formula for the kernel defined by the quantization of (\ref{e1}).
It is shown that indeed the path integral used in the derivation of the index
formula  corresponds to the Hamiltonian given by (\ref{e4}).
This serves as an important consistency check for the  supersymmetric
proof of the index theorem presented in Chapter~\ref{I-2}. The curvature term
in the Hamiltonian that was derived in Section~\ref{I-2-2} via operator
methods, arises from the fermionic sector. This chapter displays the
astonishing fact that in the path integral derivation it arises purely
from the bosonic sector.\footnote{This is a virtue of supersymmetry.
One must keep in mind that one cannot extend the results of the
analysis of a supersymmetric system to a purely bosonic or purely
fermionic system by simply reducing the supersymmetric system to its bosonic
and fermionic parts. This would clearly destroy the supersymmetry.}

In Section~\ref{I-3-2}, the loop expansion
is reviewed and the relevant 2-loop terms for the system
of (\ref{e1}) are identified. In Sections~\ref{I-3-3} and~\ref{I-3-4},
the 2-loop
calculations are presented for the spin ($\k =\a =0$) and twisted
spin ($\k =1$) cases, respectively.

Again the
Latin indices label $\e$'s, the indices from the first and
the second halves of the Greek alphabet label $\p$'s and $x$'s, respectively
\footnote{$\p$'s are labelled by $\g ,\d ,\l ,\th ,\e$.}.
The following conventional choices are also made:
\[ \hbar =1 \; \; ,\; \; \b :=t'' \;\;,\; \mbox{and}\; t'=0 \; .\]
\section{The Loop Expansion}
\label{I-3-2}
Let $\Phi^{i}$ denote the coordinate (field) variables of a superclassical
system. Then, if the Lagrangian is quadratic in $\dot{\Phi}$'s, one has
\cite[\S 5]{bd1}:
\be
K(\Phi '',t''|\Phi ',t'):=\br \Phi '',t''|\Phi ',t'\kt = Z
\int_{\Phi ',t'}^{\Phi '',t''}e^{iS[\Phi ]}(sdet\, G^{+}[\Phi ])^{
-\frac{1}{2}}{\cal D}\Phi \; .
\label{e11}
\ee
In the loop expansion of (\ref{e11}), one expands $\Phi$ around the
classical paths $\Phi_{0}$. Defining $\f$ by:
\[ \Phi (t)=:\Phi_{0}(t)+\f (t)\; , \]
one obtains:
\bea
\lefteqn{K(\Phi '',t''|\Phi ',t')\, =\, Z\, (sdet\, G_{0}^{+})^{-\frac{1}{2}}
\, e^{i\so}\, \int e^{\frac{i}{2}\f^{i}\, _{i,}S_{0,j}\f^{j}}
\{ \: 1 \: + }  \label{e12} \\
&& \; +\mbox{\fs$\frac{i}{24}$}S_{0,ijkl}\,\f^{l}\f^{k}\f^{j}\f^{i}
-\mbox{\fs$\frac{1}{72}$}S_{0,ijk}\, S_{0,lmn}\, \f^{n}\f^{m}\f^{l}
\f^{k}\f^{j}\f^{i} + \nn \\
&& \;\ll  -\mbox{\fs$\frac{i}{24}$}\left(\epsilon_{1}\, S_{0,ijk}S_{0,lmn}
\, G_{0}^{+nm} + \epsilon_{2}\,S_{0,mni}G_{0}^{+nm}S_{0,jkl}\right)
\f^{l}\f^{k}\f^{j}\f^{i} + \right. \nn \\
&& \;  \left. +\mbox{\fs$\frac{1}{8}$}\epsilon_{3}\, S_{0,jki}G_{0}^{+kj}
S_{0,mnl}G_{0}^{+nm}\, \f^{l}\f^{i} \rr + \cdots \} {\cal D}\f \; ,\nn
\eea
where,
\[ \epsilon_{1}:=(-1)^{m(l+1)+ln}\, ,\;
\epsilon_{2}:=(-1)^{m(i+1)+in}\, ,\mbox{ and}\;
\epsilon_{3}:=(-1)^{j(i+1)+m(l+1)+k+ln}\; . \]
The subscript ``$0$'' indicates that the corresponding quantity is
evaluated at the classical path, e.g. $G_{0}^{+ij}:=G^{+ij}[\Phi_{0}]$.
Finally, ``$\cdots$'' are 3 and higher loop order terms.

To evaluate the right hand side of (\ref{e12}), one needs to perform
the following functional Gaussian integrals:
% ------------------------------------------
\newcommand{\fgi}{\int e^{\frac{i}{2}\, \f^{i}\, _{i,}S_{0,j}\f^{j}}}
\newcommand{\cud}{{\cal D}\f}
% -------------------------------------------
\bea
\fgi \cud &=& c\, (sdet\, G)^{\frac{1}{2}} \nn \\
\fgi \f^{k}\f^{l}\, \cud &=& -ic\, (sdet\, G)^{\frac{1}{2}}\, G^{kl}
\label{e14} \\
\fgi \f^{k}\f^{l}\f^{m}\f^{n}\, \cud &=& (-i)^{2}c\, (sdet\, G)^{\frac{1}{2}}
\, \left( G^{kl}G^{mn}\pm \mbox{permu.} \right) \nn \\
\fgi \f^{k}\f^{l}\f^{m}\f^{n}\f^{p}\f^{q}\, \cud &=& (-i)^{3}c (sdet\, G)^{
\frac{1}{2}}\, \left( G^{kl}G^{mn}G^{pq}\pm \mbox{permu.} \right) \; .\nn
\eea
In (\ref{e14}), ``$G$'' is the Feynman propagator, ``permu.'' are terms
obtained by some permutations of the indices of the previous term,
``$\pm$'' depends on the ``parity'' of $\f$'s appearing on the left
hand side, and ``$c$'' is a possibly infinite constant of functional
integration. The functional integrals in (\ref{e11}) and (\ref{e14})
are taken over all paths with fixed end points:
\[ \f '':=\f (t'')=0 \;\;\mbox{and}\;\;\f ':=\f (t')=0 \; .\]
This justifies the appearence of $G$ in (\ref{e14}).

One must realize that the terms in square bracket in (\ref{e12}) originate
from the expansion of $(sdet\, G^{+})^{-\frac{1}{2}}$ in (\ref{e11}).
For the system under consideration (\ref{e1}), it was shown in
Section~\ref{I-2-3} that
\be
sdet(G^{+}) = 1 \hspace{2cm}\mbox{for: $x''=x'$}.
\label{e15}
\ee
This simplifies the computations of Sections~\ref{I-3-3} and~\ref{I-3-4}
considerably.
In view of (\ref{e15}), the square bracket in (\ref{e12}) drops
and (\ref{e12}) reduces to:
\bea
K&=&K_{\rm WKB}\left\{ 1 -\mbox{\fs$\frac{i}{24}$}\, S_{0,ijkl}\left(
G^{lk}G^{ji}\pm \mbox{permu.}\right) + \right. \label{e16} \\
&& \left. +\mbox{\fs$\frac{i}{72}$}\, S_{0,ijk}S_{0,lmn}\left(
G^{nm}G^{lk}G^{ji}\pm \mbox{permu.}\right) +\cdots \right\} \; , \nn
\eea
where,
\be
K_{\rm WKB}:= Zc\, e^{i\so}(sdet\, G)^{\frac{1}{2}}
\label{e17}
\ee
is the ``WKB'' approximation of the kernel. In Sections~\ref{I-3-3}
and~\ref{I-3-4}, the terms:
\bea
I&:=&S_{0,ijkl}\left( G^{lk}G^{ji}\pm \mbox{permu.}\right) \label{e18} \\
J&:=&S_{0,ijk}S_{0,lmn}\left( G^{nm}G^{lk}G^{ji}\pm\mbox{permu.}\right) \nn
\eea
are computed explicitly. They correspond to the following Feynman
diagrams:
\[ I \equiv \bigcirc \hspace{-1.5mm}\bigcirc \;\;\; \mbox{and}\;\;\;
J \equiv \mbox{\large $\ominus$}\; . \]
\section{2-Loop Calculations for the Case: $\k=0$}
\label{I-3-3}
For $\k =\a =0$, the dynamical equations (\ref{eq22}) are
solved by (\ref{eq81}):
\be
\xo (t) = \xo \;\;\; , \;\;\; \po (t) = \po \; .
\label{e20}
\ee
As in Chapter~\ref{I-2}, all the computations will be performed
in a normal coordinate system centered at $\xo$. Since $K$ and $K_{\rm WKB}$
in (\ref{e16}) have the same tensorial properties, the curly bracket in
(\ref{e16}) must be a scalar. This justifies the use of the normal
coordinates.

The Feynman propagator was computed in Section~\ref{I-2-4}:
\bea
G^{\z\xi '}&=&g_{0}^{\z\m}\ll \mbox{\fs$\th (t-t')$} \frac{(e^{\cur (t-\b )}-1)
(1-e^{-\cur t'})}{\cur (1-e^{-\cur \b })}+ \right. \label{e21} \\
&&\hspace{1cm}\left. \mbox{\fs$\th (t'-t)$}
\frac{(e^{\cur t}-1)(1-e^{-\cur (t'-\b)})}{\cur (e^{\cur \b}-1)}\rr_{\m}^{\xi}
\nn \\
G^{\z\d '}&=&G^{\g\xi '}\;=\; 0 \label{e22} \\
G^{\g\d '}&=&\iot g_{0}^{\g\d}\ll \th (t-t')-\th (t'-t) \rr \label{e23}
\eea
where,
\be
\cur = \left( \cur_{\t}^{\m} \right) =\left( g_{0}^{\m\n}\cur_{\t\n}\right)
:= \left( \iot g_{0}^{\m\n}R_{0\t\n\d\th} \po^{\d}\po^{\th}\right)\; .
\label{e24}
\ee

The functional derivatives of the action which appear in (\ref{e18}) are
listed below:
\bea
S_{0,\t\z '\rho ''}&=&\cur_{\t\z ,\rho}\, \mbox{\fs$
\ll\pt \d (t-t')\rr\d (t-t'')$}
+\cur_{\rho\t ,\z}\, \mbox{\fs$\d (t-t')\ll\pt \d (t-t'')\rr$} \nn \\
S_{0,\t\z '\l ''}&=&iR_{0\t\z\g\l}\po^{\g}\, \mbox{\fs$
\ll\pt\d (t-t')\rr \d (t-t'')$}
+i\G_{0\l\g\t,\z}\po^{\g}\, \mbox{\fs$
\d (t-t')\ll\pt\d (t-t'')\rr$} \nn \\
S_{0,\t\g '\l ''}&=& S_{0,\l\g '\e ''}\: =\: 0 \nn \\
S_{0,\t\z '\rho ''\k '''}&=&\left\{ -g_{0\t\z ,\rho\k}\,
\mbox{\fs$\ll\ppt \d (t-t')\rr \d (t-t'')\d (t-t''')$}+ \right.
\label{e25} \\
&&\hspace{1cm}  -g_{0\t\rho ,\k\z}\, \mbox{\fs$\d (t-t')
\ll\ppt \d (t-t'')\rr \d (t-t''')$} + \nn \\
&&\hspace{2cm} \left. -g_{0\t\k ,\z\rho}\, \mbox{\fs$\d (t-t')
\d (t-t'') \ll\ppt  \d (t-t''')\rr$} \right\}_{1} + \nn \\
&&\left\{ -2\G_{0\t\z\rho ,\k}\, \mbox{\fs$\ll\pt  \d (t-t')\rr
\ll\pt \d (t-t'') \rr \d (t-t''')$} +\right. \nn \\
&&\hspace{1cm} -2\G_{0\t\rho\k ,\z}\, \mbox{\fs$\d (t-t')
\ll\pt  \d (t-t'')\rr \ll\pt \d (t-t''')\rr$} + \nn \\
&&\hspace{2cm}\left. -2\G_{0\t\k\z ,\rho}\,
\mbox{\fs$\ll\pt\d (t-t')\rr\d (t-t'')\ll\pt \d (t-t''')\rr$}\right\}_{2}
+ \nn  \\
&&\left\{ \cur_{\t\z ,\rho\k}\, \mbox{\fs$\ll \pt \d (t-t')\rr \d (t-t'')
\d (t-t''')$} + \right. \nn \\
&&\hspace{1cm}+\cur_{\t\rho ,\k\z}\, \mbox{\fs$\d (t-t')\ll\pt\d (t-t'')\rr
\d (t-t''')$} + \nn \\
&&\hspace{2cm}\left. +\cur_{\t\k ,\z\rho}\, \mbox{\fs$\d (t-t')\d (t-t'')
\ll\pt  \d (t-t''')\rr$}\right\}_{3} + \nn \\
&& \left\{ i( \G_{0\m\g\t}\G^{\m}_{0\l\z})_{,\rho\k}\po^{\g}\po^{\l}\,
\mbox{\fs$\ll\pt\d (t-t')\rr\d (t-t'') \d (t-t''')$} + \right. \nn \\
&&\hspace{.4cm} +i(\G_{0\m\g\t}\G^{\m}_{0\l\rho})_{,\k\z}\po^{\g}\po^{\l}\,
\mbox{\fs$\d (t-t')\ll\pt\d (t-t'')\rr\d (t-t''')$} + \nn \\
&&\hspace{.8cm}\left. +
i(\G_{0\m\g\t}\G^{\m}_{0\l\k})_{,\z\rho}\po^{\g}\po^{\l}\,
\mbox{\fs$\d (t-t')\d (t-t'')\ll\pt\d(t-t''')\rr$}\right\}_{4} \nn \\
\pagebreak[4]
S_{0,\t\z '\l ''\g '''}&=&iR_{0\t\z\g\l}\,
\mbox{\fs$\ll\pt\d (t-t')\rr \d (t-t'')\d (t-t''')$}+ \nn \\ \nopagebreak[3]
&&\hspace{1cm}+i\G_{0\l\g\t ,\z}\, \mbox{\fs$\d (t-t')
\ll\pt\d (t-t'')\rr\d 9t-t''')$} + \nn \\ \nopagebreak[3]
&&\hspace{2cm}-i\G_{0\g\l\t ,\z}\, \mbox{\fs$
\d (t-t')\d (t-t'')\ll\pt\d (t-t''')\rr$} \nn \\
S_{0,\l\g'\e ''\d '''}&=& 0 \nn
\eea
where the indices are placed on some of the curly brackets for
identification purposes.
$S_{0,\t\z '\rho ''\l '''}$ and $S_{0,\t\l '\g ''\e '''}$ are omitted
because, as will be seen shortly, they do not contribute to (\ref{e18}).

In view of (\ref{e22}), one needs to consider only the following terms:
\bea
I_{1}&:=&S_{0,\t\z'\rho''\k'''}\ll G^{\t\z'}G^{\rho''\k'''}+
G^{\t\rho''}G^{\z'\k'''}+G^{\t\k'''}G^{\z'\rho''} \rr \nn \\
&=&3S_{0,\t\z'\rho''\k'''}G^{\t\z'}G^{\rho''\k'''} \nn \\
I_{2}&:=&S_{0,\t\z'\l''\g'''}\, G^{\t\z'}G^{\l''\g'''} \label{e25.1} \\
J_{1}&:=&S_{0,\t\z'\rho''}\, S_{0,\m'''\n^{\iv}\s^{v}}\, G^{\t\z'}
G^{\rho''\m'''}G^{\n^{\iv}\s^{v}} \nn \\
J_{2}&:=&S_{0,\t\z'\l''}\, S_{0,\m'''\n^{\iv}\g^{v}}\, G^{\t\z'}
G^{\l''\g^{v}}G^{\m'''\n^{\iv}} \; ,\nn
\eea
where,
\be
I = I_{1}+6I_{2}\; .
\label{e26}
\ee
Here, $6$ is a combinatorial factor and $J$ is a linear combination
of $J_{1}$ and $J_{2}$.

\subsection*{Calculation of $I_{1}$ and $I_{2}$}
Let us denote by $I_{1.\a}$ the terms in $I_{1}$ which correspond
to $\{\; \}_{\a}$ in $S_{0,\t\z'\rho''\k'''}$, with $\a =1,2,3,4$.
The computation of $I_{1.\a}$ is in order. One has:
\bea
I_{1.1}&:=& 3\ib dt\ib dt' \lll -g_{0\t\z ,\r\k}\,\ll\ppt\d (t-t')\rr
\, G^{\t\z'}G^{\r\k}\rrr +\nn \\
&&3\ib dt\ib dt'' \lll -g_{0\t\r ,\k\z}\,\ll\ppt\d (t-t'')\rr\,
G^{\t\z}G^{\r''\k}\rrr+\nn \\
&&3\ib dt\ib dt'''\lll -g_{0\t\k ,\z\r}\,
\ll\ppt\d (t-t''')\rr\, G^{\t\z}G^{\r\k'''} \rrr \nn \\
&=&-3\ib dt \lll \ll g_{0\t\z ,\r\k}+2g_{0\z\r ,\t\k}\rr\, G^{\r\k}I^{\t\z}
\rrr \label{e27} \\
&=&3\ib dt \lll R_{0\z\t\r\k}\, G^{\r\k}I^{\t\z} \rrr \nn \\
&=&0\; . \nn
\eea
In (\ref{e27}),
\be
I^{\t\z}:=\ib dt'\,\ll\frac{\partial^{2}}{\partial t^{'2}} \d (t-t')
G^{\t\z'}\rr \; ,
\label{e28}
\ee
and the third and forth equalities are established using
\cite[p.~56]{stephani}:
\be
g_{0\m\n ,\s\t}=-\mbox{\fs$\frac{1}{3}$}\lll R_{0\m\s\n\t}+R_{0\n\s\m\t}\rrr
\; ,
\label{e30}
\ee
and
\be
G^{\r\k}=G^{\k\r}\; .
\label{e29}
\ee
Next, we compute:
\bea
I_{1.2}&:=& -6\G_{0\t\z\r ,\k}\ib dt\ib dt'\ib dt'' \lll
\mbox{\fs$\ll\pt\d (t-t')\rr\pt\ll\d (t-t'')\rr$}
\, G^{\t\z'}G^{\r''\k} \rrr + \nn \\
&&-6\G_{0\t\r\k ,\z}\ib dt\ib dt''\ib dt''' \lll
\mbox{\fs$\ll\pt\d (t-t')\rr\pt\ll\d (t-t''')\rr$}
\, G^{\t\z}G^{\r''\k'''} \rrr + \nn \\
&&-6\G_{0\t\k\z ,\r}\ib dt\ib dt'\ib dt''' \lll
\mbox{\fs$\ll\pt\d (t-t')\rr\pt\ll\d (t-t''')\rr$}
 \, G^{\t\z'}G^{\r\k'''}\rrr \nn \\
&=&-12\G_{0\t\z\r ,\k}\ib dt\lll \lpt \frac{\partial}{\partial t'} G^{\t\z'}
\right|_{t'=t}\lpt \frac{\partial}{\partial t''}G^{\r''\k}\right|_{t''=t}
\rrr + \label{e31} \\
&&-6\G_{0\t\r\k ,\z}\ib dt\lll G^{\t\z}\lpt \frac{\partial^{2}}{\partial t''
\partial t'''}G^{\r''\k'''}\right|_{t''=t'''=t} \rrr \; . \nn
\eea
To evaluate the right hand side of (\ref{e31}), one needs the following
relations:
\bea
\lpt \frac{\partial}{\partial t''}G^{\r''\k}\right|_{t''=t}&=&
\lpt \frac{\partial}{\partial t'}G^{\r\k'}\right|_{t'=t}\: =\:
\oot g_{0}^{\r\n}\ll \frac{1+e^{\cur \b}-2e^{\cur t}}{1-e^{\cur\b}}
\rr_{\n}^{\k} \label{e32} \\
\lpt\frac{\partial^{2}}{\partial t'\partial t}G^{\k\r'}\right|_{t'=t}
&=&-\oot g_{0}^{\k\m}\ll\frac{\cur (1+e^{\cur\b})}{
1-e^{\cur\b}}\rr_{\m}^{\r} \; ,\label{e33}
\eea
and \cite[p.~55]{stephani}:
\be
\G_{0\t\z\r ,\k}=-\mbox{\fs$\frac{1}{3}$}\lll R_{0\t\z\r\k}+
R_{0\t\r\z\k}\rrr \; .
\label{e34}
\ee
Equations (\ref{e32}) and (\ref{e33}) are obtained by differentiating
(\ref{e21}), using symmetries of $\cur$ and:
\be
\th (0) := \oot \; .
\label{e35}
\ee
Combinning equations (\ref{e31})-(\ref{e34}) and using (\ref{e21}),
one obtains:
\bea
\lefteqn{I_{1.2}=\lll R_{0\t\z\r\k}+R_{0\t\r\z\k}\rrr \times } \nn \\
\nopagebreak
&&\ib dt \lll
g_{0}^{\z\m}\ll\frac{1+e^{\cur\b}-2e^{\cur t}}{1-e^{\cur\b}}\rr_{\m}^{\t}
g_{0}^{\r\n}\ll\frac{1+e^{\cur\b}-2e^{\cur t}}{1-e^{\cur\b}}\rr_{\n}^{\k}
\rrr + \label{e36} \\ \nopagebreak
&&-\lll R_{0\t\r\k\z}+R_{0\t\k\r\z} \rrr \times  \nn \\ \nopagebreak
&&\ib dt \lll
g_{0}^{\t\n}\ll \frac{e^{\cur\b}-e^{\cur t}-e^{-\cur (t-\b )}+1}{
\cur (1-e^{\cur\b})}\rr_{\n}^{\z}
g_{0}^{\k\m}\ll\frac{\cur (1+e^{\cur\b})}{1-e^{\cur\b}}\rr_{\m}^{\r}
\rrr  \nn
\eea
To identify the terms in (\ref{e36}) which are linear in $\b$, one
may recall that for every integral
\[ {\cal I}(\b ):=\ib dt\, f(\b ,t) \; ,\]
with an analytic integrand in both $t$ and $\b$, the linear term in
$\b$ is given by:
\be
\ll \lpt \frac{\partial}{\partial \b}\right|_{\b =0}{\cal I}(\b )\rr \b
= f(\b =0,t=0)\, \b \; . \label{e37}
\ee
Thus, one needs to examine the integrands in (\ref{e36}). This
leads to
\be
I_{1.2}=\lll R_{0\t\z\r\k}+R_{0\t\r\z\k}\rrr\, g_{0}^{\z\t}g_{0}^{\r\k}\,
\b +O(\b^{2})= R_{0} \b +O(\b^{2})\; ,
\label{e39}
\ee
where, $R_{0}$ is the Ricci scalar curvature evaluated at $\xo$.

Next step is to compute:
\bea
I_{1.3}&:=&3\mbox{\fs$ \lll \cur_{\t\z ,\r\k}\ib dt \, G^{\r\k} \lpt
\frac{\partial}{\partial t'}G^{\t\z'}\right|_{t'=t} +
\cur_{\t\r ,\k\z}\ib dt\, G^{\t\z} \lpt
\frac{\partial}{\partial t''}G^{\r''\k}\right|_{t''=t} + \rpt $}\nn \\
&&\hspace{1cm}\mbox{\fs$\lpt \cur_{\t\k ,\z\r}\ib dt \, G^{\t\z} \lpt
\frac{\partial}{\partial t'''}G^{\r\k'''}\right|_{t'''=t}\rrr$} \; . \nn
\eea
This is done by using (\ref{e32}) and (\ref{e37}). The result is:
\be
I_{1.3}=O(\b^{2}) \; .
\label{e40}
\ee
The computation of $I_{1.4}$ is similar. Again, one obtains:
\be
I_{1.4}=O(\b^{2}) \; .
\label{e41}
\ee
This completes the calculation of $I_{1}$. Combinning (\ref{e27}),
(\ref{e39}), (\ref{e40}), and (\ref{e41}), one has:
\be
I_{1}=R_{0}\b +O(\b^{2}) \; .
\label{e42}
\ee

The computation of $I_{2}$ is straightforward. Substituting (\ref{e23})
and (\ref{e25}) in (\ref{e25.1}), one has:
\[ I_{2}=\ib dt\ib dt'\ib dt''\ib dt''' \ll {\cal X}_{\t\z\g\l}(t,t',t'',t''')
\: {\cal Y}^{\t\z\g\l}(t,t',t'',t''')\rr \; , \]
where,
\bea
{\cal X}_{\t\z\g\l}&:=& iR_{0\t\z\g\l}
\mbox{\fs$\ll\pt\d (t-t')\rr\d (t-t'')\d (t-t''')$}+ \nn \\
&& +i\G_{0\l\g\t ,\z}\mbox{\fs$\d (t-t')\ll\pt\d (t-t'')\rr\d (t-t''')$}+ \nn
\\
&& -i\G_{0\g\l\t ,\z}\mbox{\fs$\d (t-t')\d (t-t'')\ll\pt\d (t-t''')\rr$} \nn \\
{\cal Y}^{\t\z\g\l}&:=&\mbox{\fs$-\frac{1}{4}$} G^{\t\z'}g_{0}^{\l\g}
\mbox{\fs$\ll \th (t''-t''') -\th (t'''-t'')\rr$} \; . \nn
\eea
Since ${\cal X}$ is antisymmetric in ($\g \leftrightarrow \l$) and
${\cal Y}$ is symmetric in ($\g\leftrightarrow\l$), $I_{2}$ vanishes.
This together with (\ref{e26}) and (\ref{e42}) lead to:
\be
I=R_{0}\b +O(\b^{2}) \; .
\label{e43}
\ee

\subsection*{Calculation of $J_{1}$ and $J_{2}$}
Substituting (\ref{e25}) in (\ref{e25.1}) and peforming the integrations
which involve $\d$-functions, one obtains:
\bea
J_{1}&=&\ib dt\ib dt'''\lll \cur_{\t\z ,\r}\cur_{\m\n ,\s} \,
\lpt \frac{\partial}{\partial t'}G^{\t\z'}\right|_{t'=t}\,
\lpt   \frac{\partial}{\partial t^{\iv}}G^{\n^{\iv}\s^{v}}\right|_{t^{\iv}
=t^{v}=t'''}\, G^{\r\m'''} \rpt +\nn \\
&&+\cur_{\t\z ,\r}\cur_{\s\m ,\n}\,
\lpt \frac{\partial}{\partial t'}G^{\t\z'}\right|_{t'=t}\,
\lpt \frac{\partial}{\partial t^{v}}G^{\n^{\iv}\s^{v}}\right|_{t^{\iv}=
t^{v}=t'''}\, G^{\r\m'''}  +\nn \\
&&+\cur_{\r\t ,\z}\cur_{\m\n ,\s}\,
\lpt \frac{\partial}{\partial t''}G^{\r''\m'''}\right|_{t''=t}\,
\lpt \frac{\partial}{\partial t^{\iv}}G^{\n^{\iv}\s^{v}}\right|_{t^{\iv}=
t^{v}=t'''}\, G^{\t\z}  +\label{e44} \\
&&\lpt +\cur_{\r\t ,\z}\cur_{\s\m ,\n}\,
\lpt \frac{\partial}{\partial t''}G^{\r''\m'''}\right|_{t''=t}\,
\lpt \frac{\partial}{\partial t^{v}}G^{\n^{\iv}\s^{v}}\right|_{t^{\iv}=
t^{v}=t'''}\, G^{\t\z} \rrr \nn \\
&=& \ib dt\ib dt'''\lll \cur_{\t\z ,\r}(\cur_{\m\n ,\s}+\cur_{\n\m ,\s})
\ll \cdots \rr +\cur_{\r\t ,\z}(\cur_{\m\n ,\s}+\cur_{\n\m ,\s})
\ll \cdots \rr \rrr \nn \\
&=& 0 \nn
\eea
In (\ref{e44}), the second equality is obtained by rearranging the indices
and using (\ref{e32}). The terms $\ll \cdots\rr$ involve $G$'s and their
time derivatives. The last equality is established using the antisymmetry
of $\cur$:
\be
\cur_{\m\n}=-\cur_{\n\m}\; .
\label{e45}
\ee

The computation of $J_{2}$ is a little more involved. Carrying out the
integrations involving $\d$-functions, one can write $J_{2}$ in the following
form:
\be
J_{2}=\sum_{\a=1}^{4} J_{2.\a} \; ,
\label{e46}
\ee
where,
\bea
J_{2.1}&:=&-R_{0\t\z\d\l} R_{0\m\n\e\g}\, \po^{\d}\po^{\e} \: \times \nn \\
&&\ib dt \ib dt''' \lll \lpt \lpt
\frac{\partial}{\partial t'}G^{\t\z'}\right|_{t'=t}
\frac{\partial}{\partial t^{\iv}}G^{\m'''\n^{\iv}}\right|_{t^{\iv}=t'''}
G^{\l\g'''} \rrr \label{e47} \\
J_{2.2}&:=&\mbox{\fs$\frac{1}{3}$} R_{0\t\z\d\l}( R_{0\g\e\m\n}+
R_{0\g\m\e\n} ) \po^{\d}\po^{\e} \: \times \nn \\
&&\ib dt \ib dt''' \lll \lpt \lpt
\frac{\partial}{\partial t'}G^{\t\z'}\right|_{t'=t}
\frac{\partial}{\partial t^{v}}G^{\l\g^{v}}\right|_{t^{v}=t'''}
G^{\m'''\n'''} \rrr  \label{e48} \\
J_{2.3}&:=&\mbox{\fs$\frac{1}{3}$} R_{0\m\n\e\g}( R_{0\l\d\t\z}+
R_{0\l\t\d\z} ) \po^{\d}\po^{\e}\: \times \nn \\
&&\ib dt \ib dt''' \lll \lpt \lpt
\frac{\partial}{\partial t''}G^{\l''\g'''}\right|_{t''=t}
\frac{\partial}{\partial t^{\iv}}G^{\m'''\n^{\iv}}\right|_{t^{\iv}=t'''}
G^{\t\z} \rrr \label{e49} \\
J_{2.4}&:=&\mbox{\fs$-\frac{1}{9}$}\ll (R_{0\l\d\t\z}+R_{0\l\t\d\z})
(R_{0\g\e\m\n}+R_{0\g\m\e\n})\rr\, \po^{\d}\po^{\e}\: \times \nn \\
&&\ib dt\ib dt''' \lll
G^{\t\z}G^{\m'''\n'''}\lpt \frac{\partial^{2}}{\partial t^{v}\,
\partial t''}G^{\l''\g^{v}}\right|_{t''=t,t^{v}=t'''} \rrr \; .
\label{e50}
\eea
In (\ref{e48})-(\ref{e50}), use has been made of (\ref{e34}).

Consider the integrals ($:=\int \int$) in (\ref{e47}). $\int \int$ is symmetric
under the exchange of the pairs $(\t ,\z)\leftrightarrow (\m ,\n)$. The
term  $\po^{\d}\po^{\e}$ is antisymmetric in $\d\leftrightarrow\e$.
Thus, the term $(R.R)$ is antisymmetrized in $\l\leftrightarrow\g $.
However, according to (\ref{e23}), $G^{\l\g'''}$ involves $g_{0}^{\l\g}$
which is symmetric in $\l\leftrightarrow\g$. This makes $J_{2.1}$
vanish. Furthermore, using (\ref{e37}) one finds out that
$J_{2.2}$ and $J_{2.3}$ are at least of order $\b^{2}$.
$\int \int$ in (\ref{e50}) is symmetric under
$\t\leftrightarrow\z$, $\m\leftrightarrow\n$, and
$(\m ,\n )\leftrightarrow (\t ,\z )$. This allows only the term
$R_{0\l\t\d\z}R_{0\g\m\e\n}$ to survive in the square bracket in
(\ref{e50}). Moreover, this term is symmetrized in
$(\m ,\n )\leftrightarrow (\t ,\z )$, or alternatively in
$(\l ,\d )\leftrightarrow (\g ,\e )$. Since $\po^{\d}\po^{\e}$ is
antisymmetric in $\d\leftrightarrow\e$,  the surviving
term which is multiplied by $\int\int$ can
be antisymmetrized in $\l\leftrightarrow\g $. However, due to
(\ref{e23}) $\int\int$ is symmetric in these indices. Hence, $J_{2.4}$
vanishes too.

This concludes the computation of the 2-loop terms in the case
$\k =\a =0$. Combinning (\ref{e16}), (\ref{e18}), (\ref{e26}),
(\ref{e43}), (\ref{e44}), and (\ref{e46}),
one finally obtains:
\be
K=K_{\rm WKB} \llc 1 - \mbox{\fs$\frac{i}{24}$} R_{0}\b +O(\b^{2})\rrc\; .
\label{e51}
\ee

\section{2-Loop Calculations for the Case: $\kappa =1$}
\label{I-3-4}
First, the following special case will be considered:
\be
\tilde{K}:=\br x ,\p ,\e^{*}=0 |x,\p ,\e =0 \kt\; .
\label{e52}
\ee
The dynamical equations (\ref{eq22}), are solved by (\ref{eq120}):
\bea
\xo (t)&=&\xo \nn \\
\po (t)&=&\po\:=\:\frac{1}{\sqrt{\b}}\tilde{\po} \label{e53} \\
\eo (t)&=&0 \nn\\
\eo^{*}(t)&=&0 \; .\nn
\eea
Following Section~\ref{I-2-5}, one chooses a normal coordinate system
centered at $\xo$.

The Feynman propagator is given by:
\be
\lll G^{ij'}\rrr=\lll\begin{array}{cccc}
G^{\z\xi'}& 0 & 0 & 0 \\
0 & G^{\g\d'} & 0 & 0 \\
0 & 0 & 0 & G^{ac^{*'}} \\
0 & 0 & G^{a^{*}c'} & 0
\end{array} \rrr\; ,
\label{e54}
\ee
where $G^{\z\xi'}$ and $G^{\g\d'}$ are given by (\ref{e21}) and (\ref{e23}),
respectively. Equations~(\ref{eq126}) provide the remaining components
of the Feynman propagator:
\be
\begin{array}{ccc}
G^{ac^{*'}}&=&\ll \iot e^{i(\cuf +\frac{\a}{\b}\In )(t-t')}\rr^{ac}
\mbox{\fs$\ll\th (t-t')-\th (t'-t)\rr$} \\
G^{a^{*}c'}&=&\ll \iot e^{i(\cuf^{*}+\frac{\a}{\b}\In )(t-t')}\rr^{ac}
\mbox{\fs$\ll\th (t-t')-\th (t'-t)\rr$} \; .\\
\end{array}
\label{e55}
\ee
Here,
\[ \cuf =\lll \cuf^{ab}\rrr := \lll \oot F_{0\l\g}^{ab}\po^{\l}\po^{\g}
\rrr\; , \]
and $\In$ is the $n\times n$ unit matrix.

The functional derivatives of the action which enter into the computation
of $I$ and $J$, (\ref{e18}), are listed below \footnote{The other terms
are obtained from this list using the rules of changing the order of
differentiation.}:
\bea
\ts_{0,\t\z'\xi''}&=&\lpt S_{0,\t\z'\xi''}\right|_{\k =\a =0}
\label{e56} \\
\ts_{0,\t\z'\l''}&=&\lpt S_{0,\t\z'\l''}\right|_{\k =\a =0}
\label{e57} \\
\ts_{0,\l\g'\th''}&=&0\: =\: \ts_{0,\l\g'\th''\d'''}
\label{e58} \\
\ts_{0,\t\z'\xi''\r'''}&=&\lpt S_{0,\t\z'\xi''\r'''}\right|_{\k =\a =0}
\label{e59} \\
\ts_{0,\t\z'\g''\l'''}&=&\lpt S_{0,\t\z'\g''\l'''}\right|_{\k =\a =0}
\label{e60} \\
\ts_{0,\t\z' c''}&=&0\: =\:\ts_{0,\t\z' c{*''}}
\label{e61} \\
\ts_{0,\t\g' c''}&=&0\: =\:\ts_{0.\t\g' c^{*''}}
\label{e62} \\
S_{,\t a' c''}&=&0\: =\:S_{,\t a^{*'}c^{*''}}
\label{e63} \\
\ts_{0,\t a'c^{*''}}&=&\cuf_{,\t}^{ca}\,\d (t-t')\,\d (t-t'')
\label{e64} \\
\ts_{0,\l\g' c''}&=&0\: =\:\ts_{0,\l\g' c^{*''}}
\label{e66} \\
S_{,\l c'd''}&=&0\: =\: S_{,\l c^{*'}d^{*''}}
\label{e67} \\
\ts_{0,\l c'd^{*''}}&=&F_{0\d\l}^{dc}\po^{\d}\,\d (t-t')\,\d (t-t'')
\label{e68} \\
\ts_{0,\t\z'\xi'' c'''}&=&0\: =\:\ts_{0,\t\z'\xi'' c^{*'''}}
\label{e70} \\
\ts_{0,\t\z'\g'' c'''}&=&0\: =\:\ts_{0,\t\z'\g'' c^{*'''}}
\label{e71} \\
S_{,\t\z' c''d'''}&=&0\: =\:S_{,\t\z' c^{*''}d^{*'''}}
\label{e73} \\
\ts_{0,\t\z' c''d^{*'''}}&=&iF_{0\t\z}^{dc}\,\mbox{\fs$
\ll\pt\d (t-t')\rr\d (t-t'')\d (t-t''')$}+\label{e74} \\
&&\hspace{-1.5cm}
-iA_{0\t ,\z}^{dc}\,\mbox{\fs$\d (t-t')\pt\ll\d (t-t'')\d (t-t''')\rr$}
+\cuf_{,\t\z}^{dc}\,\mbox{\fs$\d (t-t')\d (t-t'')\d (t-t''')$}
\nn \\
\ts_{0,\t\g'\l'' c'''}&=&0\: =\:\ts_{0,\t\g'\l'' c^{*'''}}
\label{e72} \\
\ts_{0,\t\g' c''d^{*'''}}&=&F_{0\l\g,\t}^{dc}\,\d (t-t')\d (t-t'')
\d (t-t''') \label{e75} \\
\ts_{0,\l\g' c''d^{*'''}}&=&F_{0\l\g}^{dc}\, \d (t-t')\d (t-t'')
\d (t-t''')\; . \label{e76}
\eea
Here, `` $\tilde{\;} $ '' 's are placed to indicate that the special case
of (\ref{e52}) is under consideration.

Equations (\ref{e56})-(\ref{e60}), indicate that the 2-loop contributions
due to the terms which involve only the Greek indices are the same as
the case of $\k =\a =0$, i.e. these terms contribute a factor of
$-\frac{i}{24}R_{0}\b$ to the kernel. Consequently, it is sufficient
to show that the remaining 2-loop terms vanish.

\subsection*{Computation of the Terms of Type $I$}
In view of (\ref{e54}) and (\ref{e56})-(\ref{e76}), the following
terms must be considered:
\bea
\tilde{I}_{1}&:=&\ts_{0,\t\z' c''d^{*'''}}\, G^{\t\z'}G^{c''d^{*'''}}
\label{e76.1} \\
\tilde{I}_{2}&:=&\ts_{0,\l\g' c''d^{*'''}}\, G^{\l\g'}G^{c''d^{*'''}}\; .
\label{e76.2}
\eea
Performing the integration overs $\d$-functions, one has:
\bea
\tilde{I}_{1}&=&\ib dt\lll iF_{0\t\z}^{dc}\,\lpt
\frac{\partial}{\partial t'}G^{\t\z'}\right|_{t=t'}G^{cd^{*}} + \rpt
\label{e77} \\
&&\lpt -iA_{0\t ,\z}^{dc}\, G^{\t\z}\ll \lpt
\frac{\partial}{\partial t''}G^{c''d^{*}}\right|_{t''=t} +
\lpt \frac{\partial}{\partial t'''}G^{cd^{*'''}}\right|_{t'''=t}\rr
+\cuf_{,\t\z}^{dc}\, G^{\t\z}G^{cd^{*}} \rrr \; .\nn
\eea
The first and the last terms in the integrand of (\ref{e77}) vanish
because according to (\ref{e55}):
\be
G^{cd^{*}}:=\lpt G^{cd^{*'}}\right|_{t'=t}=0 \; .
\label{e78}
\ee
Moreover, one has:
\be
\lpt \frac{\partial}{\partial t'}G^{c'd^{*}}\right|_{t'=t}=i\d (0)\d^{cd}
=\lpt -\frac{\partial}{\partial t'}G^{cd^{*'}}\right|_{t'=t} \; .
\label{e79}
\ee
Thus, the remaining terms cancel and one obtains:
\[ \tilde{I}_{1}=0 \; .\]
The computation of $\tilde{I}_{2}$ is quite simple. Substituting
(\ref{e23}), (\ref{e55}), (\ref{e76}) in (\ref{e76.2}) and using
(\ref{e78}), one finds:
\be
\tilde{I}_{2}=\ib dt\lll F_{0\l\g}^{dc}\, G^{\l\g}G^{cd^{*}}\rrr =0 \; .
\label{e80}
\ee
The other terms of type $I$ which involve Latin indices are proportional
to $\tilde{I}_{1}$ or $\tilde{I}_{2}$ and hence vanish.

\subsection*{Computation of the Terms of Type $J$}
There are six different terms of type $J$ which must be considered.
These are:
\bea
\tilde{J}_{1}&:=&\ts_{0,\t\z'\xi''}\ts_{0,\k''' c^{\iv}d^{*v}}\,
G^{\t\z'}G^{\xi''\k'''}G^{c^{\iv}d^{*v}} \nn \\
\tilde{J}_{2}&:=&\ts_{0,\t\z'\l''}\ts_{0,\g''' c^{\iv}d^{*v}}\,
G^{\t\z'}G^{\l''\g'''}G^{c^{\iv}d^{*v}} \nn \\
\tilde{J}_{3}&:=&\ts_{0,\t a'b^{*''}}\ts_{0,\z'''c^{\iv}d^{*v}}\,
G^{\t\z'''}G^{a'b^{*''}}G^{c^{\iv}d^{*v}} \nn \\
\tilde{J}_{4}&:=&\ts_{0,\t ab^{*'}}\ts_{0,\z'''c^{\iv}d^{*v}}\,
G^{\t\z'''}G^{a'd^{*v}}G^{c^{\iv}b^{*''}} \nn \\
\tilde{J}_{5}&:=&\ts_{0,\l a'b^{*''}}\ts_{0,\g''' c^{\iv}d^{*v}}\,
G^{\l\g'''}G^{a'b^{*''}}G^{c^{\iv}d^{*v}} \nn \\
\tilde{J}_{6}&:=&\ts_{0,\l a'b^{*''}}\ts_{0,\g'''c^{\iv}d^{*v}}\,
G^{\l\g'''}G^{a'd^{*v}}G^{c^{\iv}b^{*''}} \; .\nn
\eea
The following relations are useful in the computation of $\tilde{J}$'s.
Using (\ref{e78}), one has:
\bea
\ts_{0,\k c'd^{*''}}\, G^{c'd^{*''}}&=&\cuf_{,\k}^{dc}\, G^{cd^{*}}\: = \:
0 \label{e81} \\
\ts_{0,\g c'd^{*''}}\, G^{c'd^{*''}}&=&F_{0\d\g}^{dc}\po^{\d}\, G^{cd^{*}}
\: = \; 0  \; .\label{e82}
\eea
Equations (\ref{e81}) and (\ref{e82}) lead immidiately to:
\be
\tilde{J}_{\a}=0 \;\;\;\;{\rm for:}\; \a =1,2,3,5\; .
\label{e83}
\ee
It remains to calculate $\tilde{J}_{4}$ and $\tilde{J}_{6}$. In view
of (\ref{e64}) and (\ref{e68}), one has:
\bea
\tilde{J}_{4}&=&\cuf_{,\t}^{ba}\cuf_{,\z}^{dc}\,\ib dt\ib dt'''\,
G^{\t\z'''}G^{ad^{*'''}}G^{c'''b^{*}}
\label{e84} \\
\tilde{J}_{6}&=&F_{0\d\l}^{ba}F_{0\th\g}^{dc}\po^{\d}\po^{\th}\,
\ib dt\ib dt'''\, G^{\l\g'''}G^{ad^{*'''}}G^{c'''b^{*}} \; ,
\label{e85}
\eea
Using (\ref{e37}) and examining the integrands in (\ref{e84})
and (\ref{e85}), one finally arrives at:
\be
\tilde{J}_{\a}=O(\b^{2}) \;\;\;\;\;{\rm for:}\; \a =4,6 \; .
\label{e86}
\ee

This concludes the 2-loop calculations for the special case of (\ref{e53}).
For this case the kernel is given by:
\be
\tilde{K}=\tilde{K}_{\rm WKB}\lll 1 -\mbox{\fs$\frac{i}{24}$}R_{0}\b
+ O(\b^{2}) \rrr \; .
\label{e87}
\ee
In the rest of this section, it is shown that the same conclusion
is reached even for the general case where $\e \neq 0 \neq \e^{*}$,
i.e. for
\[ K := \br x,\p ,\e^{*}|x,\p ,\e \kt \; .\]

In the general case, the dynamical equations~(\ref{eq22}) are
solved by (\ref{eq115}), (\ref{eqa6}), and~(\ref{eqa7}):
\bea
\xo (t)&=&\xo + O(\b ) \nn \\
\po (t)&=&\frac{1}{\sqrt{\b}}\ll \tilde{\p}_{0}+O(\b )\rr
\label{e88} \\
\eo^{a}(t)&=&O(1)\nn \\
\eo^{a*}(t)&=&O(1)\; .\nn
\eea
It is easy to check that the terms in $S_{0,\cdots}$'s which
involve $\e$'s or $\e^{*}$'s, and thus survive in the general case,
are all of higher order in $\b$ than the terms considered above.
Therefore, the contribution of these terms are of order $\b^{2}$ or
higher. For instance,
\[
S_{0,\t\z'\r''}=\ts_{0,\t\z'\r''}+\Sigma_{\t\z\r} \]
where,
\[
\ts_{0,\t\z'\r''}=O(\b^{-4})\; \]
is given by (\ref{e56}), and
\bea
\Sigma_{\t\z\r}&:=&\ll i(A_{0\z ,\t}^{ab}-A_{0\t ,\z}^{ab})_{,\r}
\eo^{a*}\eo^{b}+
-iA_{0\t ,\z\r}^{cd}(\dot{\e}_{0}^{c*}\eo^{d}+\eo^{c*}\dot{\e}_{0}^{d})+
\rpt \nn \\
&&+\lpt i(A_{0\s ,\t}^{cd}-A_{0\t ,\s}^{cd})_{,\z\r}\dot{\xo}^{\s}
\eo^{c*}\eo^{d}+\cuf_{,\t\z\r}\eo^{c*}\eo^{d}\rr\d (t-t')\d (t-t'') +
\nn \\
&&+i(A_{0\r ,\t}-A_{0\t ,\r})_{,\z}\d (t-t')\frac{\partial}{\partial t}
\ll \d (t-t'')\rr \nn \\
&=& O(\b^{-3})\; . \nn
\eea
To determine the order of the terms in $\b$, one proceeds as in
Section~\ref{I-2-5}, namely one makes the following change of time
variable:
\be
s:=\frac{t}{\b}\, \in [0,1] \; .
\label{e89}
\ee
For example, one has:
\[ \frac{\partial}{\partial t}\d (t-t') =\b^{-2}\frac{\partial}{\partial s}
\d (s-s')=O(\b^{-2}) \; . \]

The terms for which the above argument might not apply are those
that vanish identically in the special case of (\ref{e53}) but
survive otherwise. These are:
\bea
J_{7}&:=&S_{0,\t\z' c''}S_{0,\xi'''\r^{\iv}d^{*v}}\,
G^{\t\z'}G^{\xi''\r^{\iv}}G^{c''d^{*v}} \nn \\
J_{8}&:=&S_{0,\t\z' c''}S_{0,\xi'''\r^{\iv}d^{*v}}\,
G^{\t\xi'''}G^{\z'\r^{\iv}}G^{c''d^{*v}} \nn \\
J_{9}&:=&S_{0,\t\g' c''}S_{0,\z'''\l^{\iv}d^{*v}}\,
G^{\t\z'''}G^{\g'\l^{\iv}}G^{c''d^{*v}} \nn \\
J_{10}&:=&S_{0,\l\g' c''}S_{0,\d'''\th^{\iv}d^{*v}}\,
G^{\l\g'}G^{\d'''\th^{\iv}}G^{c''d^{*v}} \nn \\
J_{11}&:=&S_{0,\l\g' c''}S_{0,\d'''\th^{\iv}d^{*v}}\,
G^{\l\d'''}G^{\g'\th^{\iv}}G^{c''d^{*v}}\; . \nn
\eea
However, one can easily show that the contribution of these terms to the
kernel is of the order $\b^{2}$. The following relations are
useful:
\bea
S_{0,\t\z' c''}&=&O(\b^{-3})\: =\: S_{0,\t\z' c^{*''}} \nn \\
S_{0,\t\g' c''}&=&O(\b^{-\frac{5}{2}}) \nn \\
S_{0,\l\g' c''}&=&O(\b^{-2})\: =\: S_{0,\l\g' c^{*''}} \nn \\
G^{\t\z'}&=&O(\b ) \nn \\
G^{\g\l'}&=&O(1)\: =\:G^{cd^{*'}}\; . \nn
\eea
Furthermore, each integral:
\[
\ib dt\cdots =\b \int_{0}^{1} ds \cdots \]
contributes a factor of $\b$. Putting all this together, one finds out
that $J_{7},\cdots ,J_{11}$ are all of order $\b^{2}$.

This concludes the 2-loop calculations for the general case of
(\ref{e88}). The final result is:
\be
K=K_{\rm WKB}\ll 1-\mbox{\fs$\frac{i}{24}$}R_{0}\b +O(\b^{2})\rr\; .
\label{e90}
\ee
\section{Remarks and Discussion}
\label{I-3-5}
Equation (\ref{e90}) verifies the existence of the scalar curvature
factor in the Hamiltonian and provides a consistency check for the
supersymmetric proof of the Atiyah-Singer index theorem presented
in Chapter~\ref{I-2}. \footnote{After the completion of the
present work, the author was informed of an article by
Peeters and van Nieuwenhuizen \cite{peter} whose results confirm
the existence of the curvature factor in the Hamiltonian.}
It is remarkable that in the operator formulation the
scalar curvature factor originates from the supercommutation relations
involving the fermionic variables  whereas in the
path integral formulation it appears in the bosonic sector.

It must be emphasized that the 2-loop term in (\ref{e90})
does not contribute to the index formula (\ref{eq139}).
This is simply because the $\p$-integrations in (\ref{eq111})
eliminate such a term.

\chapter{Conclusion}
The Peierls quantization scheme provides a systematic procedure for
quantizing superclassical systems. Supersymmetry results in remarkable
simplifications in the path integral evaluation of the kernel. In
particular, the miraculous simplifications that occur in the computation
of the path integral measure are due to supersymmetry. The WKB approximation
for the path integral yields the index formula exactly. The supersymmetric
proof of the index theorem is based on the assumption that the ordinary
Gaussian integral formula holds even in the functional integral case.
This can be viewed as a definition of the Gaussian superdeterminant.
The existence of a rigorous supersymmetric proof of a substantial
mathematical result such as the index theorem, is another indication
of the validity and the power of the path integral techniques. It is
particularly remarkable to observe that even the normalization
constants agree with those chosen by mathematicians.

The scalar curvature factor in the Schr\"odinger equation yields a factor
of $-iR\b /24$ in the heat kernel expansion of the path integral. The
loop expansion provides an independent test of the validity of this
assertion. In particular, this factor is obtained in the two-loop order.
Thus, it is shown that indeed the path integral used in the derivation
of the index formula corresponds to the Hamiltonian defined by the twisted
Dirac operator.

\appendix
\chapter{A Brief Review of Index Theory}
\label{z2-A}
\vspace{1cm}
\bcc
{\bf Abstract}
\ecc
\bcc
\parbox[b]{4.5in}{\small
The mathematical concepts related to the index theory are introduced.
The Atiyah-Singer index theorem is presented and its relation to
K-theory is briefly discussed. The heat kernel proof of the index
theorem is also outlined. The latter is closely related to the
supersymmetric proofs of the index theorem.}
\ecc
\vspace{.5in}
\section{Complex of Hermitian Bundles}
\label{z2-A1}
Let $M$ be a Riemannian manifold of dimension $m$. $M$ is said to be
{\em closed} if it is compact and has no boundary. A complex
vector bundle $V$ on $M$, \cite{cd1}, is said to be a {\em
hermitian bundle} if it is endowed with a {\em (fibre) metric} and a
compatible {\em connection}. The word compatible means that the
 covariant derivative of the metric vanishes. Let $\{ \varphi_{i}\sx\}$
be a set of local basic sections over an open neighborhood $O\subset M$. Then,
every section $s$ of $V$ is locally expressed by:
\[ s(x)=s^{i}\sx\, \varphi_{i}\sx \;\;\; ;\;\;\; i=1,\cdots ,n\; ; \;
\forall x\in O\; .\]
Here $n$ denotes the fibre dimension of $V$ and $s^{i}\sx$ are the
local components of $s$ in the basis $\{\varphi_{i}\sx\}$. Furthermore, let
${\bf h}$ denote the hermitian metric  on $V$. It can be locally
determined by the $(n\times n)$ dimensional matrix $(h_{ij}\sx )$ of its
components, i.e.,
\[ {\bf h}=:h_{ij}\sx\, \varphi^{i}\sx\,\varphi^{j}\sx\; , \]
where $\{\varphi^{i}\sx \}$ is the dual basis to $\{\varphi_{i}\sx\}$
\footnote{The bases $\{ \varphi_{i}\sx\}$ and $\{\varphi^{i}\sx\}$
are the analogs of the local frames $\{ e_{\m}\sx\}$ and $\{ e^{\m}\sx\}$
of the Riemannian geometry. Persuing this analogy, one has the
correspondence between $V$ and the tangent bundle $TM$, and ${\bf h}
=(h_{ij}\sx )$ and the Riemannian metric ${\bf g}=(g_{\m\n}\sx )$.}.
${\bf h}$ defines a
hermitian inner product on $V$. For any pair of sections $s_{1}$ and
$s_{2}$ of $V$, their inner product is given by
\[ (s_{1},s_{2}):=\int_{M} s_{1}^{i*}\sx\, h_{ij}\sx\, s_{2}^{j}\sx
\, \vol\; ,\]
where $\vol$ is the volume form of $M$. For a more detailed discussion
of hermitian bundles, refer to \cite{bott-chern,abp}.

Let $V_{0},V_{1},\cdots ,V_{k}$ be a sequence of hermitian vector
bundles over $M$. The vector space of smooth ($C^{\infty}$) sections of
$V_{i}$ is denoted by $C^{\infty}(V_{i})$. Let
\[ D_{i}:\cinf~(V_{i})\rar\cinf~(V_{i+1}) \;\;\; ,\;\;\; i=0,\cdots ,k-1\]
be arbitrary differential operators. Then, the sequence
\be
{\cal C}:\; \cinf~(V_{0})\stackrel{D_{0}}{\rar}\cinf~(V_{1})
\stackrel{D_{1}}{\rar}\cdots\stackrel{D_{k-1}}{\rar}\cinf~(V_{k})
\label{m1}
\ee
is said to be a {\em complex} if
\[ im(D_{i})\subset ker(D_{i+1}) \;\; , \;\; \forall i=0,\cdots ,k-2\; , \]
or alternatively
\[ D_{i+1}\, o\, D_{i}=0 \; .\]
Here, the symbols $im$ and $ker$ stand for {\em image} and {\em kernel},
respectively.
The sequence ${\cal C}$ is said to be {\em exact} if
\[ im(D_{i})=ker(D_{i+1})\; .\]
Clearly, any exact sequence is a complex.

\section{Symbol of a Differential Operator}
\label{z2-A2}
Let $V_{+}$ and $V_{-}$ be two hermitain vector bundles
of fibre  dimension $n_{+}$ and $n_{-}$ and
\[ D:\cinf~(V_{+})\rar\cinf~(V_{-})\]
be a differential operator. Let us choose a local coordinate
patch  and denote its points by their coordinates, $x=(x^{\m})$.
The operator $D$ can be locally expressed in the form
\be
D=\sum_{l=1}^{d}{\bf a}^{\m_{1}\cdots\m_{l}} \mbox{\fs$(x)$}\, D_{\m_{1}}\cdots
D_{\m_{l}}+{\bf a\sx}\;\;\; ,\;\; \; x=(x^{\m})
\label{m2}
\ee
where $\m$'s take values in $1,\cdots ,m=dim(M)$, $D_{\m}$'s
are defined by
\[ D_{\m} := -i\frac{\partial}{\partial x^{\m}}\; ,\]
and ${\bf a}^{\m_{1}\cdots\m_{l}}$'s and ${\bf a}$ are $(n_{-}\times n_{+})$
dimensional matrix-valued functions of $x^{\m}$.
The integer $d$ is called the order of $D$.
For example a first order differential operator is locally
represented  by
\be
D^{(1)}= {\bf a}^{\m}\mbox{\fs$(x)$}\, D_{\m}+{\bf a\sx} \; .
\label{m3}
\ee
It is easy to verify that only the coefficients (${\bf a}^{\m_{1}\cdots
\m_{d}}$) of the highest order part of a differential
operator have tensorial properties. This is quite straightforward
to see in the case of a first order operator (\ref{m3}).
The {\em symbol} of $D$ is obtained by taking the highest order part
of $D$ and viewing it as a homogeneous polynomial in $D_{\m}$'s.
For convenience, one often changes the notation
\[ D_{\m}\rar\xi_{\m}\; .\]
Because of the obvious transformation property of $D_{\m}$'s under
coordinate transformations, one can actually view $\xi_{\m}$'s as
the coordinates of some point $\xi$ in the {\em cotangent space} $TM^{*}_{x}$
at $x$. Then $(x,\xi)$ denote the points of the {\em cotangent
bundle} $TM^{*}$ of $M$. The symbol of $D$ is locally given by
\be
\s_{D}(x,\xi):={\bf a}^{\m_{1}\cdots\m_{d}}\mbox{\fs$(x)$}\,\xi_{\m_{1}}
\cdots\xi_{\m_{d}}\; .
\label{m4}
\ee
As $\s_{D}$ transforms tensorially it can be globally defined on $TM^{*}$.

On the other hand, since $D$ is a linear operator, $\s_{D}(x,\xi )$
is a linear transformation (homomorphism) of the vector spaces:
\[\s_{D}(x,\xi ):V_{+x}\rar V_{-x} \; .\]
Here $V_{\pm x}$ denote the fibres of $V_{\pm}$ over $x$.
Globally, the symbol of $D$ defines the following mapping:
\be
\s_{D}:TM^{*}\rar End(V_{+},V_{-})\; ,
\label{m5}
\ee
where $End(V_{+},V_{-})$ denotes the set of all fibrewise
linear maps from $V_{+}$ to $V_{-}$.

The operator $D$ is said to be an {\em elliptic} operator if
for all $x\in M$ and all nonzero $\xi \in TM^{*}_{x}$ the
linear map (\ref{m4}) defined by the symbol is an isomorphism.
If $D$ is an elliptic operator, its symbol defines a map
\be
\s_{D}:TM^{*}-M\rar Iso(V_{+},V_{-})\; .
\label{m6}
\ee
In (\ref{m6}), I have used the canonical identification of the {\em zero
section} of $TM^{*}$ with $M$. The symbol $Iso(V_{+},V_{-})$
denotes the subset of $End(V_{+},V_{-})$ consisting of
the fibrewise bijective (one-to-one and onto) maps. In particular,
if there is an elliptic operator between two hermitian bundles
they must have the same fibre dimension.

\section{Adjoint of a Differential Operator}
\label{z2-A3}
Let $D$ be an arbitrary differential operator as given locally by
(\ref{m2}). The adjoint of $D$ is a differential operator
\be
D^{\dagger}:\cinf (V_{-})\rar\cinf(V_{+})
\label{m7}
\ee
of the same order.
Let $s_{\pm}$ be arbitrary sections of $V_{\pm}$, then $D^{\dagger}$
is defined by the following expression:
\be
\int_{M}( s_{-}\, ,\, D\, s_{+})_{-}\, \vol =:
\int_{M}( D^{\dagger}\, s_{-}\, ,s_{+})_{+}\, \vol\; ,
\label{m8}
\ee
where $(.\, ,.)_{\pm}$ are the hermitian inner products on $V_{\pm}$
defined by their hermitian metrices, and $\vol$ denotes the volume
form  of the manifold $M$. It is easy to see that the adjoint
operator $D^{\dagger}$ depends not only on the original operator $D$,
but also on the geometric structures of the base manifold and the
vector bundles.

\section{Elliptic Complexes and Their Indices}
\label{z2-A4}
Let ${\cal C}$ be a complex of $k+1$ hermitian vector bundles as in
(\ref{m1}) and denote
the symbols of the differential operators $D_{i}$ by $\s_{i}(x,\xi )$.
Then, if for all $x\in M$ and all $\xi\in TM_{x}^{*}-\{ 0\}$ the sequence
\be
0\hookrightarrow V_{0x}\stackrel{\s_{0}(x,\xi )}{\rar}V_{1x}
\stackrel{\s_{1}(x,\xi )}{
\rar}\cdots\stackrel{\s_{k-1}(x,\xi )}{\rar}V_{kx}\stackrel{0}{\rar}0
\label{m9}
\ee
is exact, the complex ${\cal C}$ is called an {\em elliptic
complex}. Note that in (\ref{m9}), the symbols are viewed as linear
maps between vector spaces, and the exactness means that
\[ Im\:\s_{i}(x,\xi )=Ker\:\s_{i+1}(x,\xi )\; .\]
In particular for $k=1$, there is a single differential operator $D:=D_{0}$
\be
 {\cal C}_{1}: 0\hookrightarrow\cinf (V_{0})\stackrel{D}{\rar}
\cinf (V_{1})\rar 0\; .
\label{m9.1}
\ee
The sequence defined by the symbol, in this case, is
\be
 0\hookrightarrow V_{0x}\stackrel{\s_{D}(x,\xi )}{\rar}V_{1x}
\stackrel{0}{\rar} 0\; .
\label{m10}
\ee
If the complex ${\cal C}_{1}$ is elliptic, then the sequence (\ref{m10})
is exact. This implies that for all $x\in M$ and  $\xi\in TM_{x}^{*}-{0}$,
$Ker\:\s_{D}(x,\xi )=\{ 0\}$ and $Im\:
\s_{D}(x,\xi )=Ker\, (\mbox{0-map})=V_{1x}$. The former condition means
that the symbol map is one-to-one and the latter indicates that
it is onto. Thus, the map $\s_{D}(x,\xi )$ is bijective and
$D$ is an elliptic operator. In this
sense the concept of ``elliptic complex'' is a generalization of the
concept of  ``elliptic operator''. I will show that indeed for any
elliptic complex there is an elliptic operator that carries all
the topological information and as far as the index problem is
concerned one can restrict to the {\em short elliptic complexes}
(\ref{m9.1}), i.e., elliptic operators.

Let us consider a linear operator $P:{\cal H}_{+}\rar{\cal H}_{-}$,
where ${\cal H}_{\pm}$ are vector spaces. The {\em kernel} and
the {\em cokernel} of $P$ are defined respectively by:
\bea
ker(P)&:=&\{v_{+}\in {\cal H}_{+}\: :\: P(v_{+})=0\:\} \nn \\
coker(P)&:=&{\cal H}_{-}/im(P)\; .\nn
\eea
If the dimensions of
${\cal H}_{\pm}$ are not finite, then the kernel and the cokernel
of $P$ might not be finite dimensional. The operators
whose kernel and cokernel are finite dimensional are called
{\em Fredholm operators}. For a Fredholm operator, one  defines
its {\em analytic index} by
\[ index(P):= dim(ker\: P)-dim(coker\: P)\; .\]
For the case of differential operators
which act on the sections of hermitian vector bundles, the spaces
${\cal H}_{\pm}$ correspond to the spaces of sections $\cinf (V_{\pm})$.
These are clearly infinite dimensional. However, one can show that
elliptic operators are Fredholm operators \cite{booss} and their
analytic index is a finite integer.
I shall denote the analytic index of an elliptic operator $D$,
by ``index''. The index theorem
gives a formula to compute this quantity in terms of the
geometric structures of the spaces involved and the symbol of the
operator (i.e., the coefficients of the highest order part of $D$).
Furthermore, it is not difficult to show that
\[ coker(D)=ker(D^{\dagger})\; .\]
Thus, one has
\be
index(D):=dim(ker\: D) - dim(ker\: D^{\dagger}) \; .
\label{m11}
\ee

The concept of the index of an elliptic operator can be generalized
to elliptic complexes. The index of an elliptic complex ${\cal C}$:
\be
{\cal C}:\; 0
\stackrel{D_{-1}}{\hookrightarrow}\cinf (V_{0})
\stackrel{D_{0}}{\rar}\cinf (V_{1})
\stackrel{D_{1}}{\rar}\cdots
\stackrel{D_{k-1}}{\rar}\cinf (V_{k}0
\stackrel{D_{k}:=0}{\rar}0
\label{m12}
\ee
is defined by
\be
index({\cal C}):=\sum_{l=0}^{k}(-1)^{l}\, dim(ker\, D_{l}/Im\, D_{l-1})
\; .
\label{m13}
\ee
Incidentally, the vector spaces
\[ H_{l}({\cal C}):=ker(D_{l})/im(D_{l-1}) \]
are called the $l$-th cohomology of the complex.
Let us also define the {\em Laplacian}'s of the complex:
\be
\Delta_{l}:=D_{l-1}D_{l-1}^{\dagger}+D_{l}^{\dagger}D_{l}:
\cinf (V_{l})\rar\cinf (V_{l})\; .
\label{m14}
\ee
The $index$ can be conveniently expressed in terms of the Laplacians:
\be
index({\cal C})=\sum_{l=0}^{k}(-1)^{l}dim(ker\:\Delta_{l})\; .
\label{m15}
\ee

For a short elliptic complex (\ref{m9.1}), i.e. $k=1$, the expression
for the index (\ref{m13}) becomes:
\bea
index({\cal C}_{1})&:=&dim(ker\: D_{0}/Im\: D_{-1})-dim(ker\: D_{1}/
Im\: D_{0})  \nn \\
&=&dim(ker\: D)-dim\lll \cinf (V_{1})/im\: D\rrr \nn \\
&=&dim(ker\: D)-dim(coker\: D)\: =:\: index(D) \; , \label{m16}
\eea
where in accordance with (\ref{m9.1}), $D_{0}$ is identified with $D$.
Thus, the concept of the index of an elliptic complex is a
generalization of the (analytic) index of an elliptic operator.

Next, I will proceed in the opposite direction, namely I will show
that the computation of the index of a (long) elliptic complex
can be reduced to that of a short complex. In view of
(\ref{m16}), the latter is equivalent
to the index problem for an elliptic operator.

\section{Rolling up a Complex}
\label{z2-A5}
Consider the complex ${\cal C}$ of (\ref{m12}). One can construct
an associated short complex ${\cal C}_{1}$ such that ${\cal C}$
and ${\cal C}_{1}$ have the same index. To show this, one starts
with taking the direct sums of $V_{l}$'s in (\ref{m12}) to define
\bea
\Omega_{+}&:=&\cinf (\bigoplus_{l}V_{2l}) \nn \\
\Omega_{-}&:=&\cinf (\bigoplus_{l}V_{2l+1}) \; , \nn
\eea
where $l=0,1,\cdots ,\ll \frac{k}{2}\rr$, and $\ll z\rr$ denotes
the largest integer part of $z\in \R$. Next, one defines
\[ D_{+}:\Omega_{+}\rar\Omega_{-} \;\;\;\mbox{and}\;\;\;
D_{-}:\Omega_{-}\rar\Omega_{+} \]
by requiring that
\bea
\lpt D_{+}\right|_{\cinf (V_{l})}&:=&\llc
\begin{array}{cc}
0&\mbox{if $l$ is odd}\\
D_{l}&\mbox{if $l$ is even}
\end{array}
\rpt \label{m17} \\
\lpt D_{-}\right|_{\cinf (V_{l})}&:=&\llc
\begin{array}{cc}
D_{l}&\mbox{if $l$ is odd}\\
0&\mbox{if $l$ is even}
\end{array}
\rpt \; . \nn
\eea
In fact, these define $D_{\pm}$ over
\[ \Omega:=\Omega_{+}\oplus\Omega_{-}\; . \]
Hence, one can view them as oprators from $\Omega$ to itself.
Also, one defines:
\[ d:=D_{+}+D_{-}\, :\, \Omega\rar\Omega\; . \]
The following identities follow trivially from (\ref{m17}):
\bea
D_{+}^{2}&=&D_{-}^{2}\: =\: D_{+}^{\dagger 2}\: =\:D_{-}^{\dagger 2}\: =\: 0
\label{m18} \\
D_{+}D_{-}^{\dagger}&=&D_{-}D_{+}^{\dagger}\: =\: D_{+}^{\dagger}D_{-}\:
=\: D_{-}^{\dagger}D_{+}\: =\: 0 \; ,\nn
\eea
where $D_{\pm}$ and $D_{\pm}^{\dagger}$ are viewed as operators defined
on $\Omega$.
The Laplacian of $d$ is given by:
\bea
\Delta&:=&d\, d^{\dagger}+d^{\dagger}\, d \nn \\
&=&\lll D_{+}^{\dagger}D_{+}+D_{-}\, D_{-}^{\dagger}\rrr +
   \lll D_{-}^{\dagger}D_{-}+D_{+}\, D_{+}^{\dagger}\rrr
\label{m19}
\eea
Let us denote the terms in the the right hand side of (\ref{m19}) by
$\Delta{\pm}$, i.e.,
\bea
\Delta_{+}&:=&D_{+}^{\dagger}D_{+}+D_{-}\, D_{-}^{\dagger}\nn \\
&=&\lll D_{+}+D_{-}^{\dagger}\rrr^{\dagger}
   \lll D_{+}+D_{-}^{\dagger}\rrr \label{m20} \\
\Delta_{-}&:=&D_{-}^{\dagger}D_{-}+D_{+}\, D_{+}^{\dagger}\nn \\
&=&\lll D_{+}+D_{-}^{\dagger}\rrr
   \lll D_{+}+D_{-}^{\dagger}\rrr^{\dagger} \; .\label{m21}
\eea
In (\ref{m20}) and (\ref{m21}), use has been made of the identities
listed in (\ref{m18}). Equations (\ref{m20}) and (\ref{m21})
suggest the definition of
\[ \hD:=D_{+}+D_{-}^{\dagger}:\Omega\rar\Omega \; .\]
$\hD$ is determined by its restriction to $\Omega_{+}$:
\[ D:=\lpt \hD\right|_{\Omega_{+}}:\Omega_{+}\rar\Omega_{-} \; .\]
In terms of $D$, $\Delta_{\pm}$ are written conveniently in the
form:
\be \Delta_{+}=D^{\dagger}D \;\;\;\mbox{and}\;\;\;
   \Delta_{-}=D\, D^{\dagger} \; .
\label{m21.1}
\ee
Another important observation is  that $\Delta_{\pm}$ are related
to the Laplacians $\Delta_{l}$, (\ref{m14}), of the complex ${\cal C}$
according to the following folmulea
\be
\Delta_{+}=\sum_{l=0}^{\ll k/2 \rr} \Delta_{2l}\;\;\;
\mbox{and}\;\;\;
\Delta_{-}=\sum_{l=0}^{\ll k/2 \rr} \Delta_{2l+1}\; .
\label{m22}
\ee
Let us consider the short complex defined by $D$
\[ {\cal C}_{1}:\; 0\hookrightarrow\Omega_{+}\stackrel{D}{\rar}
\Omega_{-}\rar 0 \; .\]
The opertor $D$ is by construction an elliptic differential
operator. This can be infered using (\ref{m22})
and realizing that since $\Delta_{l}$ are elliptic operators,
so are $\Delta_{\pm}$. In view of (\ref{m21.1}),
the ellipticity of $\Delta_{\pm}$ is
a sufficient condition for the ellipticity of $D$ and $D^{\dagger}$,
\cite{shanahan}.

One can directly check that indeed $D$ and $D^{\dagger}$ have
finite dimensional kernels. To see this, one first shows that
\be
ker(D)=ker(\Delta_{+}) \;\;\;\mbox{and}\;\;\; ker(D^{\dagger})
=ker(\Delta_{-})\; .
\label{m23}
\ee
To prove the first equaltity, one chooses
an arbitrary element $s_{+}$ of $ker(\Delta_{+})$, i.e.,
\[ D^{\dagger}D(s_{+})=\Delta_{+}(s_{+})=0 \; .\]
Then, using the definition of the adjoint of a differential operator
one shows that
\be
0 = (s_{+},D^{\dagger}D\, s_{+})_{+} = (D\, s_{+},D\, s_{+})_{-}\; .
\label{m24}
\ee
Since the inner product (metric) induced
\footnote{The hermitian metrices on $V_{l}$ induce hermitian
metrices on their direct products.} on $\Omega_{-}$ is positive
definite, (\ref{m24}) implies that $D\, s_{+}$ vanishes, i.e.,
$s_{+}\in ker(D)$. Thus, $ker(\Delta_{+})
\subset ker(D)$. The converse of the latter relation
is trivially correct. This completes the proof of the claim that
\[ ker(D)=ker(\Delta_{+}) \; .\]
The proof of the other identity uses the same argument.
Now, in view of (\ref{m22}) and the fact that $ker(\Delta_{l})$
are finite dimensional, one observes that $ker(\Delta_{\pm})$
have also finite dimensions. This together with (\ref{m23}) indicate
that both $D$ and $D^{\dagger}$ have finite dimensional kernels.

The analytic index of ${\cal C}_{1}$ is given by
\be
index({\cal C}_{1})=index(D)=dim(ker\, D)-dim(ker\, D^{\dagger})\; .
\label{m25}
\ee
It is quite easy to employ the results derived in the
preceding paragraph to prove that
\be
index({\cal C})=index({\cal C}_{1})=index(D) \; .
\label{m26}
\ee
One has the following list of identities:
\bea
index({\cal C}_{1})&:=&dim(ker\, D)-dim(ker\, D^{\dagger}) \nn \\
&=& dim(ker\,\Delta_{+})-dim(ker\,\Delta_{-}) \nn \\
&=& \sum_{l}dim(ker\, \Delta_{2l})-\sum_{l}dim(ker\,\Delta_{2l+1})\nn \\
&=& \sum_{l}(-1)^{l}dim(ker\,\Delta_{l}) \: = \: index({\cal C}) \; . \nn
\eea

Before ending this section, let us also consider the extension
of $D$ and $D^{\dagger}$ to whole $\Omega$, namely $\hD$ and $\hD^{\dagger}$:
\bea
\Omega & \stackrel{\hD}{\rar} & \Omega \nn \\
\Omega & \stackrel{\hD^{\dagger}}{\longleftarrow} & \Omega \; .\nn
\eea
In view of (\ref{m18}), one has the following obvious equalities
\bea
\hD^{2}&=&D_{+}^{2}+D_{-}^{\dagger 2}+D_{+}D_{-}^{\dagger}
+D_{-}^{\dagger}D_{+}\: =\: 0 \nn \\
\hD^{\dagger 2}&=&D_{-}^{2}+D_{+}^{\dagger 2}+D_{-}D_{+}^{\dagger}
+D_{+}^{\dagger}D_{-}\: = \: 0 \; .
\label{m27}
\eea
Moreover, the total Laplacian can be expressed as
\be
\Delta =\Delta_{+}+\Delta_{-}=\lll \hD+\hD^{\dagger}\rrr^{2} \; .
\label{m28}
\ee
Equations (\ref{m27}) and (\ref{m28}) are used in Appendix~\ref{z2-B}
in the discussion of the relation between the index theorem
and the supersymmetric quantum mechanics.
\section{Symbol Bundle}
\label{z2-A6}
The index of an elliptic operator depends not only
on the operator but also on the spaces on which it acts.
This is easily seen by recalling the definition of
the adjoint operator. Thus, the index
formula is expected to involve the coefficients of the operator
as well as the geometric quantities associeted with the
bundles and their base manifold. It is practical to
encode all these information in a single
geometric (topological) construction and then try to
use its characteristic properties to obtain the index.
In this way, one does not need to deal with both the
operator and the vector bundles at the same time.
The space that carries all the necessary information is called
the {\em symbol bundle}. It is a vector bundle over a
base manifold which is itself a fibre bundle.
This section is devoted to the
description of the symbol bundle associated to an arbitrary
short elliptic complex. Certainlly, one can apply the same
construction to long elliptic complexes by first rolling
them up.

Let
\[ {\cal C}:\; \cinf (V_{+})\stackrel{D}{\rar}\cinf (V_{-}) \]
be a short elliptc complex. The {\em unit ball bundle} $B(M)$
and the {\em unit sphere bundle} $S(M)$ associated
with the cotangent bundle $TM^{*}$ of the base manifold are defined by
\bea
B(M)&:=&\{ z=(x,\xi )\in TM^{*}\, :\, |\xi |\leq 1\} \nn \\
S(M)&:=&\{ z=(x,\xi )\in TM^{*}\, :\, |\xi |=1\}\: =: \: \partial B(M) \: .
\label{m28.1}
\eea
B(M) and S(M) are indeed fibre bundles over $M$. They will be
simply denoted by $B$ and $S$. Let us take two copies of $B$
and label them as $B^{\pm}$. These are fibre bundles
\be
 B^{\pm}\stackrel{\pi^{\pm}}{\rar}M
\label{m29}
\ee
of fibre dimension $m$.
Next, let us glue the fibres $B_{x}^{+}$ and $B_{x}^{-}$ along their
boundries, i.e., identify the points of their boundaries. The boundaries
are unit spheres sitting inside the cotangent space $TM^{*}_{x}$,
$\forall x\in M$. Let us denote the single bundary that is obtained
in this way by $S_{x}$. Clearly, the collection of all $S_{x}$ leads
to a description of the sphere bundle $S$. The object obtained by
gluing $B^{\pm}$ in this manner is denoted by
\[ \Sigma (M) = B^{+}\cup_{S} B^{-} \; ,\]
or simply $\Sigma$. This is also a fibre bundle over $M$:
\be
\Sigma \stackrel{\pi}{\rar} M \; .
\label{m30}
\ee
The (total) space $\Sigma$ as a manifold will play the role
of the base space of the symbol bundle. By its construction
$\Sigma$ depends solely on $M$.

Next step is to incorporate some information about the bundles
$V_{\pm}$. This is realized by viewing $B^{\pm}$ as seperate
submanifolds of $\Sigma$ and pulling back the vector bundles $V_{\pm}$
on $B^{\pm}$ using the obvious projection maps $\pi^{\pm}$ of
(\ref{m29}). This is demonstrated by the following commutative
diagram:
\be
\begin{array}{ccc}
\pi^{\pm *}(V_{\pm}) &\stackrel{\pi^{\pm *}}{\longleftarrow}& V_{\pm} \\
\downarrow & \bigcirc & \downarrow \\
B^{\pm}&\stackrel{\pi^{\pm}}{\rar}&M \: .
\end{array}
\label{m31}
\ee
$\pi^{\pm *}(V_{\pm})$ are  pullback bundles on $B^{\pm}$.
To obtain a fibre bundle over $\Sigma$, one needs to somehow
identify the fibres along the boundaries of $B^{\pm}$. To clarify
what the last sentence means, let us consider an arbitrary point $z=(x,\xi )$
in $\Sigma$. Either $z$ belongs to $S$, i.e., the common boundary of
$B^{\pm}$ or it belongs to their interior.
In the latter case the fibre over $z$ is uniquely defined to be
$V_{x+}$ or $V_{x-}$ depending on whether
$z$ belongs to $B^{+}$ or $B^{-}$, respectively. In the case
where $z$ is in $S=\partial B^{\pm}$, there are two different choics
of fibers, namely $V_{x\pm}$. To obtain a well-defind fibre bundle structure on
$\Sigma$, one needs to identify these two fibres. The
identification can introduce `topological twists' and thus
effects the topology of the bundle in a crucial manner. This is
precisely the way one introduces the information about the
operator in the construction.

To begin describing this in
detail, let us recall the mathematical meaning of the
word `identification'. If the two spaces which are to be
identified have only the structure of point sets, their
identification means that there
is a one-to-one and onto map that takes the points of the
first space to the other. If the two spaces have further
structures one would like to also obtain a corrspondence
between these structures, e.g., for vector spaces one
would like to also identify their linear structure. Thus, two
vector spaces are identified with a bijection which also
preserve the linear (vector space) structures. Such a map
is called an isomorphism. Therefore,
for each $z\in S$, one needs
an isomorphism between the two vector spaces $V_{x\pm}$.
This is precisely what the {\em symbol} $\sigma_{D}$, (\ref{m6}),
of the elliptic operator $D$ provides. Note that by definition
(\ref{m28.1}), for all $z=(x,\xi )\in S$, $|\xi |=1$. This means that
as an element of $TM_{x}^{*}$, $\xi $ is nonzero. Consequently,
since $D$ is an elliptic operator, the symbol
\[\s_{D}(z):V_{+x}\rar V_{-x} \;\; \; ,\; \;\;\forall z=(x,\xi )\in S\]
is an isomorphism.
The construction of the symbol bundle is then completed by
using the symbol $\s_{D}$ of $D$ to uniquely define the fibres
over $S\subset\Sigma$. The symbol bundle, which is denoted by
{\bf $\s $}, is a vector bundle of fibre dimension $n$ over
the space $\Sigma$,
\be
\s\rar\Sigma\; .
\label{symbol}
\ee
We will see that it carries all the necessary
information about the index of $D$ (${\cal C}$). In particular,
the index depends on the coefficients of the highest order part
of the operator.
\section{Characteristic Classes}
\label{z2-A7}
In this section, I shall  review the basic definitions of
some of the {\em charactristic classes} of vector bundles
that appear in the statement of the index theorem. For a
more detailed survey of characteristic classes see \cite{egh,milnor}.

Let $V$ be a complex hermitian vector bundle over the closed
Rimannian manifold $M$. Let $A$ and $F$ denote the {\em connection
one-form} and the {\em curvature two-form} of $V$, respectively.
$F$ is a two-form
with values in the Lie algebra of the structure group of $V$. Since $V$
has a hermitian structure, one can assume, without loss of generality,
that the structure group of $V$ is $U(n)$ \cite{bundle}. Then, in a
basis of the Lie algebra of $U(n)$, $F$ can be exprssed as an antihermitian
matrix. This makes it diagonalizable. The eigenvalues of $F$ define a
set of basic {\em characteristic classes} which
are called the {\em Chern roots} of $V$ and denoted by $x_{i}$, $i=1,\cdots
,n$.
These are by definition differential two-forms on $M$. One has
\[ F=diag(-2\pi i x_{1},\cdots ,-2\pi i x_{n}) \; .\]
The {\em Chern classes} \cite{cd1} of $V$ are symmetric polynomials of the
Chern roots. They are defined by
\bea
c_{1}\V &:=&\sum_{i}x_{i} \nn \\
c_{2}\V &:=&\sum_{i<j}x_{i}\wedge x_{j}
\nn \\
\cdots &:=& \cdots \nn \\
c_{n}\V &:=&x_{1}\wedge x_{2}\wedge \cdots \wedge x_{n} \; .\nn
\eea
The sum of the Chern classes is called the {\em total Chern class}.
It is given by
\[ c\V :=c_{1}+\cdots +c_{n} =\prod_{i=1}^{n} (1+x_{i}) \; .\]
The total Chern class has the following property
\be
 c(V_{1}\oplus V_{2})=c(V_{1})\wedge c(V_{2}) \; ,
\label{m36}
\ee
where $V_{1}$ and $V_{2}$ are hermitian vector bundles and $\oplus$
denotes their {\em direct sum} \cite{cd1,egh}.
Another interesting characteristic class associated with a complex
hermitian bundle $V$ is the {\em Chern character}, $ch(V)$. It is
defined by
\[ ch(V):=\sum_{i=1}^{n}e^{x_{i}}=n+c_{1}+\oot(c_{1}^{2}-2c_{2})+\cdots
\; .\]
The Chern character has the following important properties
\bea
ch(V_{1}\oplus V_{2})&=&ch(V_{1})+ch(V_{2})\nn \\
ch(V_{1}\otimes V_{2})&=&ch(V_{1})\wedge ch(V_{2}) \; ,\label{mchern}
\eea
where $\otimes$ denotes the {\em tensor product} \cite{cd1,egh} of two
vector bundles. These properties
have important implications for the index theorem, \cite{booss}.
Another useful characteristic class of a complex vector bundle is
the {\em Todd class}
\be
td(V):=\prod_{i=1}^{n}\frac{x_{i}}{1-e^{-x_{i}}}=1+\oot c_{1}+
\mbox{\fs$\frac{1}{12}$}(c_{2}+c_{1}^{2})+\cdots \; .
\label{mtodd}
\ee
The Todd class of the direct sum of two vector bundles is the
product of their Todd classes. In this respect, the Todd class
is similar to the total Chern class (\ref{m36}).

All these characteristic classes can be expressed in terms of the
traces (determinants) of the (wedge) products of the curvature two-form.
For example, one has
\bea
c\V &=&\det\ll \In +\frac{iF}{2\pi}\rr \nn \\
ch(V)&=& tr\ll \exp \frac{iF}{2\pi}\rr  \; . \label{chern}
\eea

There are also the characteristic classes which are associated with
{\em symmetric vector bundles}. A symmetric vector bundle is a  real
vector bundles which is endowed with a symmetric positive definite metric
and a compatible connection. Briefly, symmetric bundles are the real
analogs of the hermitian bundles. Undoubtedly, the most well-known
examples of symmetric vector bundles are the tengent bundle $TM$ and
its dual the cotangent bundle $TM^{*}$ of a Riemannian manifold $M$.
The structure group of a symmetric vector bundle can be chosen to be
$O(n)$. Thus, in this case the curvature two-form $\Omega$ is an antisymmetric
matrix of differential two-forms. It can be block-diagonalized
\be
\Omega= diag\lll \ll
\begin{array}{cc}
0&\Omega_{i} \\
-\Omega_{i}&0
\end{array} \rr \, : \, i=1,\cdots n \rrr \; .
\label{m37}
\ee
The analogs of the Chern roots are defined by
\[ r_{i}:=2\pi \Omega_{i} \; ,\]
and the analogs of the Chern classes are the symmetric polynomials
of $r_{i}^{2}$:
\bea
p_{1}\V &:=&\sum_{i} r_{i}^{2} \nn \\
p_{2}\V &:=&\sum_{i<j}r_{i}^{2}\wedge r_{j}^{2} \nn \\
\cdots &:= & \cdots \nn \\
p_{n}\V &:=&r_{1}^{2}\wedge\cdots\wedge r_{n}^{2} \; .\nn
\eea
These are called the {\em Pontrijagin classes}.
Once more, one can define the {\em total Pontrijagin class} and other
polynomial combinations of $p_{i}$'s. Some of the more famous
examples of these are  {\em Herzebruch's signature density}:
\be
L:= \prod_{i=1}^{n}\frac{r_{i}}{\tanh (r_{i})}=1+\mbox{\fs$\frac{1}{3}$}
p_{1}+\mbox{\fs$\frac{1}{45}$}(7p_{2}-p_{1}^{2})+\cdots \; ,
\label{signature}
\ee
and {\em Dirac's $\hat{A}$-genus density}:
\be
\hat{A}:=\prod_{i=1}^{n}\frac{r_{i}/2}{\sinh (r_{i}/2)}=
1-\mbox{\fs$\frac{1}{24}p_{1}$}+\mbox{\fs$\frac{1}{5760}$}(-4p_{2}+
7p_{1}^{2})+\cdots\; .
\label{spin}
\ee
In fact, the Pontrijagin
classes of a real vector bundle are related to the Chern classes
of its complexification, namely
\[ p_{i}\V=(-1)^{i}c_{2i}(V\otimes \C ) \; .\]
Unlike the Chern classes $c_{i}$ that are $2i$-forms, the Pontrijagin classes
$p_{i}$ are $4i$-forms. Thus, if the dimension of $M$ is less than four all
the Pontrijagin classes vanish.
\section{The Atiyah-Singer Index Theorem}
\label{z2-A8}
Let ${\cal C}: \cinf (V_{+})\stackrel{D}{\rar}\cinf (V_{-})$ be an
elliptic complex and $\s\rightarrow\Sigma$ be the symbol bundle (\ref{symbol})
of $D$.
Then, the {\em topological index} of ${\cal C}$ (alternatively of $D$)
is defined by
\be
index^{\mbox{\fs top.}}(D):= \int_{\Sigma} ch\,(\s )\wedge \pi^{*}\ll
td\,(TM^{*}\otimes \C )\rr\; ,
\label{top}
\ee
where $\pi^{*}$ denotes the pullback operation defind by the
projection map $\pi :\Sigma\rar M$, of (\ref{m30}), and $ch$ and $td$ stand
for the corresponding Chern character and Todd class
(\ref{mchern},\ref{mtodd}),
respectively.

{\em The Atiyah-Singer index theorem states that the analytic index (\ref{m11})
and the topological index (\ref{top}) are identical.} In other words,
it provides a formula for the analytic index which is given by (\ref{top}).
The latter statement is more illuminating since the motivation for
defining  the topological index is precisely to give a closed
expression for the analytic index in terms of some differential geometric
quantities.

There are alternative expressions for the topological index
which are almost as equally complicated as (\ref{top}) \cite{shanahan}.
However, in particular cases  the expression
for the topological index may be simplified. There are four
so called {\em classical complexes} whose indices give  important
topological information about the associated structures.
These are the {\em de Rham}, {\em signature},
{\em spin}, and {\em Dolbeault} complexes. They are discussed extensively in
the
most of the text books on the index theorem, \cite{shanahan,gilkey}. A rather
complete survey is provided in \cite{egh}. In the following section, I
will review the index theorem for the de Rham complex. It has all
the ingrediants of a general elliptic complex.
\section{The de Rham Complex and the Euler-Poincar\'{e} Characteristic}
\label{z2-A9}
Let $M$ be a closed, oriented smooth manifold of dimension $m$.
The de Rham complex is the long elliptic complex:
\be
{\cal C}_{\rm de Rham}: 0\hookrightarrow\La_{0}\stackrel{d}{\rar}\La_{1}
\stackrel{d}{\rar}\cdots\stackrel{d}{\rar}\La_{m}\stackrel{d}{\rar} 0\; ,
\label{derham}
\ee
where $\La_{i}$ are the spaces of $i$-th differential forms. The bundles
$V_{i}$ of (\ref{m1}) are, in this case, the $i$-the wedge products of
the cotangent bundle, i.e.,
	\[ V_{i}:=\underbrace{ TM^{*}\wedge\cdots\wedge
	TM^{*} }_{i-\mbox{times}}\; ,\]
and the operators $D_{i}$ are the restrictions
of the exterior derivative operator $d$ to the $i$-th forms.
Rolling up ${\cal C}$, one obtains the following short complex:
\[{\cal C}_{dR}: \La_{\rm even}\stackrel{d+d^{*}}{\rar}\La_{\rm odd} \; ,\]
where $d^{*}$ is the adjoint of $d$ \cite{cd1,egh}.

The expression for the topological index of ${\cal C}_{dR}$ can be
simplified as described in \cite{shanahan}. The index is called the {\em
Euler-Poincar\'{e} characteristic} of $M$. The statement of the
index theorem for the de Rham complex reads:
\be
 index({\cal C}_{dR})=\int_{M}e\, (TM^{*}) \; .
\label{euler}
\ee
Here $e\, (TM^{*})$ denotes the so called {\em Euler class} of $TM^{*}$.
This is another characteristic class of a real vector bundle. It is
an $m$-differential form on $M$ whose coefficient is determined by
the Riemann curvature tensor of $M$, \cite{cd1,egh,gilkey,bd1}.

For $m=2$, M is a Riemann surface and the Riemann curvature
tensor is determined by the Gaussian curvature $K$.
In this case, (\ref{euler}) reduces to
\be
 index({\cal C}_{dR})=\mbox{\fs$\frac{1}{2\pi}$}\int_{M}K\,ds\: =:\: \chi \; ,
\label{poincare}
\ee
where $ds$ is the {\em surface (volume) element} of $M$. Equation
(\ref{poincare}) is a classical result of the differential geometry
of surfaces. It is known as the {\em Gauss-Bonnet} theorem. A
remarkable fact about the Riemann surfaces is that
\[ \chi = 2-2g \; ,\]
where $g$ is the {\em genus}, i.e., the number of handles of $M$.
Thus, Riemann surfaces are classified by their Euler-Poincar\'{e}
characteristic.

Euler-Poincar\'{e} characteristic is also related to the zeros of
the global sections of $TM$ which have a discrete number of zeros.
This is known as {\em Pincar\'{e}-Hopf theorem},
\cite[Part I\hspace{-.4mm}I p.321]{cd1}.
Sections of $TM$ are usually called vector fields on $M$. A {\em zero}
of a vector field $s$ is a point of $M$ at which $s$ vanishes.
For instance if there exists a nowhere vanishing vector field
on $M$ then the Euler-Poincar\'{e} characteristic  must be zero. Thus the
vanishing of this topological invariant is a necessary condition
for the existence of nowhere zero vector fields.
A simple application of this is the fact that the only
two-dimensional surface that supports nowhere zero vector fields is
the torus. This is one of the main reasons why toroidal geometries
are so commonly encountered in areas such as plasma physics.

\section{~Spin Bundle and Spin Complex}
\label{z2-A10}
Let $E$ be an {\em inner product space} with an orthonormal basis
$\{ e_{i}\}$. The {\em Clifford algebra} ${\cal C}(E)$ of $E$ is
generated by $e_{i}$ according to the Clifford multiplication rule
\cite{shanahan,gilkey,cd1,spinor}:
	\bea
	\{ e_{i},e_{j}\} &=& e_{i}*e_{j}+ e_{j}*e_{i} \nn \\
		 	 &=& -2{\bf (}e_{i},e_{j}{\bf )} \nn \\
			 &=&-2\d_{ij} \; , \nn
	\eea
where $*$ and ${\bf (\: ,\, )}$ denote the Clifford multiplication and
the inner product, respectively. As a vector space the Clifford algebra
${\cal C}(E)$ can be decomposed into the following subspaces
	\[ {\cal C}(E)={\cal C}^{+}(E)\oplus {\cal C}^{-}(E)\, \]
where
	\[ {\cal C}^{\pm}(E):=\mbox{Span}\{ e_{i_{1}}*\cdots *e_{i_{p}}\}
\;\; \mbox{with} \; \pm =(-1)^{p}\; .\]
In fact, ${\cal C}^{+}(E)$ forms a Clifford subalgebra of ${\cal C}(E)$.

Let us assume that $E$ is even dimensional, dim($E$)$=m=2l$, and
denote by $t:{\cal C}(E)\to{\cal C}(E)$ the operation of transposition:
	\[ (e_{i_{1}}*\cdots *e_{i_{p}})^{t}:=e_{i_{p}}*\cdots
	e_{i_{1}}\; .\]
Furthermore, let $\{ v_{i}\} $ be a set of unit vectors in $E$. Then the
subset
	\[ Spin(m):=\{ w=v_{1}*\cdots *v_{2p} : p\leq l\} \; ,\]
of ${\cal C}^{+}(E)$ forms a group under Clifford multiplication.
The inverse of each element is given by its transpose. Clearly,
$w*w^{t}=1$.

$Spin(m)$ acts on ${\cal C}(E)$ by conjugation
	\[ x\stackrel{w}{\rar}\rho (w)(x):=w*x*w^{t} \; .\]
This action of $Spin(m)$ on $\R^{m}\cong E\subset{\cal C}(E)$ may be
used to show that indeed there is a two-to-one canonical homomorphism
from $Spin(m)$ onto $So(m)$, i.e., $Spin(m)$ is a double cover of $So(m)$,
\cite{gilkey,cd1}.
On the other hand, ${\cal C}(E)\otimes \C$ provides a $2^{l}\times 2^{l}$
dimensional complex representation $\rho$ of $Spin(m)$. This decomposes
into $2^{l}$ copies of a more basic
representation, ${\cal S}$, which itself consists
of two $2^{l-1}$ dimensional irreducible representations ${\cal S}^{\pm}$,
\cite{shanahan,gilkey,spinor}.\footnote{In fact, $\rho$ may be used as
a representation of the Clifford algebra ${\cal C}(E)$; then
${\cal S}$ is an irrducible representation of this algebra. It
is, however, reducible as a representation of $Spin(m)$, \cite{shanahan}.}
The representations ${\cal S}$ and
${\cal S}^{\pm}$ are called the {\em spinor} and {\em chirality $\pm$
spinor representations}. Briefly one has
	\bea
	{\cal C}(\R^{m})&=&2^{l}\, {\cal S} \nn \\
	{\cal S}&=&{\cal S}^{+}\oplus {\cal S}^{-} \nn \\
	{\cal S}^{\pm}&:=&{\cal S}\cap \lll {\cal C}^{\pm}(\R^{m})\otimes
	\C \rrr \; .\nn
	\eea

Let $M$ be an oriented compact Riemannian manifold of dimension $m$.
For all $x\in M$,
one can use $E=TM^{*}_{x}$ to construct the associated Clifford
algebra ${\cal C}(TM^{*}_{x})$. These may be viewed as the
fibres of  a global
bundle structure over $M$ which is called the {\em Clifford bundle}
of $M$. The transition functions of this bundle are induced from those
of $TM^*$, so the structure group is again $SO(m)$. However, we
know that there is a  (fibrewise transitive) action of $Spin(m)$ on such
a vector bundle. Thus, it is interesting to construct another bundle
structure on the same set of points  whose structure group
is  $Spin(m)$ rather than $So(m)$. This is done by lifting the
transition functions of the Clifford bundle
from $So(m)$ to $Spin(m)$.  The
associated principal bundle to this new construction (if it exists)
is called the {\em spin bundle}, ${\cal S}$.
The existence of such a structure depends on the topology of
$M$ and it is detected by yet another characteristic class of $TM^{*}$
called the second {\em Stiefel-Whitney} class,
\cite{gilkey,egh,cd1,nakahara}.

Let us consider the associated $So(m)$ principal bundle to $TM^*$, the
coframe bundle $coframe(M)$.  The transition
functions of ${\cal S}$ are lifted from those of $coframe(M)$ via the
canoniacal homomorphism from $Spin(m)$ onto
$So(m)$. If the transition functions can be consistently defined then
the manifold is called a {\em spin manifold}, \cite{shanahan,egh,gilkey}.
The associated vector bundles
to ${\cal S}$ defined by the spinor representations of $Spin(m)$
are of utmost importance in physics. In particuler, their sections are
the so called {\em spinors} which are  often used to represent the
matter fields of physical theories. For further details, see
\cite[Part I\hspace{-.4mm}I p.~134]{cd1} and references therein.

The spin and the twisted spin complexes are described in Chapter~\ref{I-1}.
For completeness I shall briefly recall the expression for the twisted spin
complex. It is given by
	\be
	 0\to \cinf ({\cal S}^{+}\otimes V)\stackrel{\not\partial_{V}}{
	\rar}\cinf ({\cal S}^{-}\otimes V)\to 0 \; ,\label{xxx}
	\ee
where $V$ is a hermitian vector bundle over $M$ and $\not\!\!\partial_{V}$
is the {\em twisted Dirac opertor}. For more details see Chapter~\ref{I-1} and
the references cited there. (\ref{xxx}) is indeed a family of elliptic
complexes
which are parametrized by the auxiliary bundle $V$. It turns out that for
any elliptic complex (operator) there exists an auxiliary bundle $V$
such that the indices of the original complex and the twisted spin
complex defined by $V$ are equal. This is a substantial result for
it reduces the proof of the index theorem to that of the twisted Dirac
complex. The body of knowledge leading to this result is called {\em
K-theory}. I shall briefly outline the utility of K-theory in relation
to the index theorem in the next section.
\section{~K-Theory and the Index Theorem}
\label{z2-A11}
Let $Vec(X)$ be the set of all complex vector bundles over a manifold $X$.
$Vec(X)$ forms an abelian semigroup under the operation of direct (Whitney)
sum, \cite{shanahan,gilkey,egh,booss}. The abelian semigroup
$(Vec(X),\oplus )$ is not a group because there is no consistent
way of defining the inverse of a vector bundle with respect to $\oplus$.

Interestingly, the situation is in perfect analogy with the set of
non-negative integers, $\Z^{+}$, and the operation of addition.
Clearly, $(\Z^{+},+)$ is an abelian
semigroup which is not a group because the negative integers are not
included in $\Z^{+}$. There is however a very simple but rather
clever way of defining the negative integers from $\Z^{+}$ which
generalizes to any abelian semigroup. The construction is as follows.

Let us consider the set of all pairs of non-negative integers
	\[ Z:=\{ (p,q)\: :\: p,q\in\Z^{+}\} \; .\]
and define an equivalence relation $\sim$ on $Z$:
	\[ (p_{1},q_{1})\sim (p_{2},q_{2})\]
if and only if there exist an $r\in\Z^{+}$ such that
	\[ p_{1}+q_{2}+r=p_{2}+q_{1}+r\; .\]
In the last equality one cannot simply cross out $r$ on the both sides
as one usually does since there is still no notion of ``$-r$'' defined for
$\Z^{+}$. The quotient set $\Z :=Z/\!\sim$ is the set of all integers, where
$\Z^{+}$ can be viewed as the subset of pairs with the second entry
being zero. The negative numbers are those with the first entry being zero.
In fact one usually uses the following familiar notation:
	\[ (p,q)=:p-q\; .\]

This simple but general construction can be applied for $Vec(X)$ to
construct a group structure. Again one considers the set of pairs
of vector bundles
	\[ \{ (E_{+},E_{-})\: :\: E_{\pm}\in Vec(X)\}\; ,\]
and considers the (quotient) set of equivalence classes of the
following equivalence relation:
	\[(E_{+},E_{-})\sim (F_{+},F_{-})\]
if and only if there is a $U\in Vec(X)$ such that
	\[ E_{+}\oplus F_{-}\oplus U\cong F_{+}\oplus E_{-}\oplus U\; .\]
Here $\cong$ denots the isomorphism of the vector bundles. The
quotient set $K(X)$ has the structure of an abelian group. It is
known as Grothendieck's K-group. \cite{booss,shanahan,gilkey,nash}.
There is another representation of $K(X)$ which makes it quite useful
in the index theory.

Let $h:E_{+}\to E_{-}$ be a homomorphism of vector bundles over $X$.
If there is a compact region in $X$ outside of which $h$ is an
isomorphism then the triplet
	\be
	k:\; E_{+}\stackrel{h}{\rar}E_{-}
	\label{triplet0}
	\ee
is said to have a {\em compact support} and the  compact region is
called the {\em support} of the triplet. Two triplets are said to
be {\em equivalent} if they are {\em homotopic}, \cite{booss}.
The following is a definition of the concept of homotopy for the
triplets of the form (\ref{triplet0}). Let $k$ and $k'$ be two
such triplets with $k$ given by (\ref{triplet0}) and $k'$ by
	\[k':\; E_{+}'\stackrel{h'}{\rar}E_{-}'\; .\]
Then, $k$ and $k'$ are said to be {\em homotopic} if there
exists a triplet
	\[{\cal K}:\; {\cal E}_{+}\stackrel{H}{\rar}{\cal E}_{-}\; ,\]
where ${\cal E}_{\pm}$ are vector bundles over the manifold
$X\times [0,1]$ such that the restrictions of ${\cal K}$ to $X\times\{ 0\}$
and $X\times\{ 1\}$ yield $k$ and $k'$, respectively.
Next, let us define the following two sets
	\bea
	C(X)&:=& \mbox{ the set of compactly supported triplets}\nn \\
	C_{0}(X)&:=&\mbox{ the set of triplets with empty support}\; .\nn
	\eea
Evidently, $C_{0}(X)\subset C(X)$. Then one can prove,
\cite{shanahan,booss}, that $C(X)-C_{0}(X)$ and $K(x)$ are in one-to-one
correspondence. The Whitney sum $\oplus$ can be also trivially
extended on this set. Thus one obtains an alternative representation
of the K-group. In this representation it can be manifestly seen that
the K-group is sensitive to the homotopic properties of $X$.

The other advantage of the second representation is in relation to
the index problem. Let $M$ be a closed Riemannian manifold and $V_{\pm}$
be two hermitian vector bundles on $M$. Let
	\[ D:\cinf (V_{+})\rar\cinf (V_{-})\; ,\]
be an elliptic operator  with  symbol (\ref{m4})
	\[ \s_{D}(x,\xi ):V_{+x}\rar V_{-x} \; .\]
The projection map $\pi$ of the cotengent bundle
	\[ \pi :TM^{*}\rar M \]
can be used to pullback the bundles $V_{\pm}$ on $TM^{*}$ and the
symbol map (\ref{m6}) gives rise to the following triplet
	\be
	\pi^{*}(V_{+})\stackrel{\s_{D}}{\rar}\pi^{*}(V_{-})\; .
	\label{triplet}
	\ee
By definition of ellipticity $\s_{D}(x,\xi )$ is an isomorphism for
all $\xi\neq 0$. Thus $\s_{D}$ of (\ref{triplet}) fails to be an
isomorphism only on the zero section of the cotangent bundle. The
latter is canoniacally diffeomorphic to $M$.  Since $M$ is compact
and nonempty, $\s_{D}\in C(TM^{*})-C_{0}(TM^{*})=K(TM^{*})$.

As is argued in the preceding sections the index depends on the
symbol of the operator. In fact, it is a map of the form
	\[ index: K(TM^{*})\rar \Z \; .\]
This means that  it depends on the equivalence (homotopy) class of the
symbol. In other words, one may calculate the index of a given elliptic
operator by using any of the elements within the same equivalence class
in $K(TM^{*})$. Incidentally, one can show that $K(TM^{*})$ and $K(M)$
are indeed isomorphic, \cite{booss}.

On the other hand, one can show, \cite{gilkey,booss,atiyah-singer}, that within
every equivalence class of $K(M)$ there is a twisted spin complex (triplet).
This is to say that for any elliptic operator one can choose the auxiliary
bundle $V$ of the twisted spin complex such that its index equals to
that of the original operator. In this sense, a proof of the twisted
spin index theorem leads to a proof of the general index theorem. This
is the strategy used in the supersymmetric proofs of the index theorem
as well as the original {\em cobordism proof}, \cite{atiyah-singer},
and the heat  kernel proof, \cite{gilkey}. The latter is quite relevant to the
supersymmetric proofs of the index theorem. Hence, the next section
is devoted to a brief review of the heat kernel proof.
\section{~A Sketch of the Heat Kernel Proof}
\label{z2-A12}
There are four different proofs of the index theorem. These are
\begin{enumerate}
\item
Cobordism proof, \cite{palias};
\item
Embedding proof, \cite{atiyah-singer,shanahan};
\item
Heat kernel proof, \cite{abp,gilkey,booss,berline};
\item
Supersymmetric proofs \footnote{See chapter 2 for details and references.}.
\end{enumerate}
All of these proofs have a common feature, namely they use K-theory to
arrive at the index formula. In this section, I shall briefly introduce
the main idea behind the heat kernel proof.

Let $D:\cinf (V_{+})\to \cinf (V_{-})$ be an elliptic operator and
$\Delta_{\pm}$ be the corresponding Laplacians as in eq.\ (\ref{m21.1}).
Furthermore, let us define the eigenspaces of $\Delta_{\pm}$ by
	\be
	\Gamma^{\pm}_{\lambda} := \{ s\in\cinf (V_{\pm})\: :\; \Delta_{\pm}s
	=\lambda s\} \; .
	\ee
Then one has the following theorem.

{\bf Theorem}(Hodge): For all $\lambda \in Spec(\Delta_{+})\cap\R$,
$\Gamma_{\lambda}^{+}$ is finite dimensional and the space of square
integrable sections of $V_{+}$, $L_{2}(V_{+})$, has the following
decomposition
	\[ L_{2}(V_{+})\equiv \oplus_{\lambda} \Gamma_{\lambda}^{+}\; .\]
Furthermore, for all positive $\lambda\in \R$, $\lambda>0$
	\be
	\Gamma_{\lambda}^{+}\cong\Gamma_{\lambda}^{-}\; ,
	\label{hodge}
	\ee
where $\cong$ denotes the isomorphism of vector spaces.

According to eq.\ (\ref{m23}), the zero eigenspaces are
	\bea
	\Gamma_{0}^{+}&=:& Ker(\Delta_{+})\: =\: Ker(D) \label{kernel} \\
	\Gamma_{0}^{-}&=:& Ker(\Delta_{-})\: =\: Ker(D^{\dagger}) \; .\nn
	\eea
Thus using eqs.\ (\ref{hodge}) and (\ref{kernel}), one has the identities:
	\bea
	index(D)&=&dim(\Gamma_{0}^{+})-dim(\Gamma_{0}^{-}) \nn \\
	&=&\sum_{\lambda}e^{-t\lambda}\ll dim(\Gamma_{\lambda}^{+})
	-dim(\Gamma_{\lambda}^{-})\rr\nn \\
	&=&h_{t}(\Delta_{+})-h_{t}(\Delta_{-}) \; ,
	\label{heat1}
	\eea
where $t$ is some complex parameter with positive real part and
	\be
	h_{t}(\Delta_{\pm}):= \sum_{\lambda}e^{-t\lambda}
	dim(\Gamma_{\lambda}^{\pm}) \; .
	\label{heat2}
	\ee
One can show, \cite{booss}, that there are differential forms $\m_{0}^{\pm}$
over $M$ such that
	\[ h_{t}(\Delta_{\pm})=\int_{M}\m_{t}^{\pm} \; .\]
Thus the index is given by
	\be
	index(D)=\int_{M}\lll \m_{0}^{+}-\m_{0}^{-}\rrr \; .
	\label{heat3}
	\ee
In the heat kernel proof of the index theorem one tries to obtain
the integrand of eq.\ (\ref{heat3}) in terms of the coefficients of
the operator $D$, the curvature 2-forms of $V_{\pm}$, and the
Riemann curvature tensor of $M$. This can be achieved for classical
elliptic complexes and in particular for the {\em twisted signature}
complex, \cite{atiyah-singer,abp,shanahan,gilkey,egh,booss,nash}.
This is a short complex with the same K-theoretic properties as
the twisted spin complex. In fact, for each twisted spin complex there
is an associated twisted signature complex and vice versa, such that
their indices are equal, \cite{gl}. The computation of this integrand
for arbitrary elliptic complexes turned out to be extremely difficult and
has been an open problem since late sixties, \cite{abp}.

The name {\em heat kernel} is attached to this method because one can
write $\m_{t}^{\pm}$ of eq.\ (\ref{heat2}) in terms of the kernel
of the heat equation. Let us consider the operator
	\[ H_{t}^{+}:=e^{-t\Delta_{+}} \; .\]
$H_{t}^{+}$ satisfies the heat equation
	\[ \frac{d}{dt}H_{t}^{+}+\Delta_{+}H_{t}^{+}=0 \]
	\[ H_{0}^{+}=1\; .\]
Evidently,
	\[ h_{t}(\Delta_{+})= tr\lll e^{-t\Delta_{+}}\rrr
	=tr\lll H_{t}^{+}\rrr \; .\]
Moreover, one can represent $H_{t}^{+}$ by its kernel $H_{t}^{+}(x,y)$:
	\[ \lll H_{t}^{+}(s)\rrr (x) =: \int_{M} H_{t}(x,y)\, s(y)\vol \; .\]
Then one has
	\[ \m_{t}^{+}(x)=tr\ll H_{t}(x,x)\vol\rr \; .\]
The same applies for $\Delta_{-}$. Hence, the knowledge of the heat equation
can be used to yield a derivation of the index formula for classical
elliptic complexes directly and the general elliptic complexes indirectly,
\cite{abp,gilkey,berline}.

The supersymmetric proofs of the index theorem benefit from the
arguments presented above. The advantage of these proofs is that
one can use the analogy between the heat equation and the Schr\"{o}dinger
equation to employ the path integral technique in the  evaluation of the
index density, $\m_{t}^{+}-\m_{t}^{-}$. There are two main reasons
that make the supersymmetric proofs so effective. Firstly, in these
proofs one avoides the computation of $\m_{t}^{+}$ and $\m_{t}^{-}$
seperately and calculates their difference directly. Secondly, since
the index is a continuous function onto integers, it is independent
of the parameter $t$. Thus, one can simply perform the computations
in the limit $t\to 0$. Thus the WKB approximation for the
path integral is indeed exact. This simplfies the calculations a
great deal.

\chapter{A Brief Review of Supersymmetric Quantum Mechanics}
\label{z2-B}
\vspace{1cm}
\bcc
{\bf Abstract}
\ecc
\bcc
\parbox[b]{4.5in}{\small
A review of Witten's supersymmetric quantum mechanics is
presented.  The Witten index in defined and its topological invariance
is shown. A general supersymmetric proof of the index theorem is
outlined. The holomorphic or coherent state representation
is briefly discussed and the supernumbers and Berezin integration
are introduced.}
\ecc
\section{Witten's Definition of Supersymmetric Quantum Mechanics}
\label{Z2-B1}
Consider a quantum mechanical system consisting of a Hilbert (Fock)
space ${\cal F}$ and the Hamiltonian $H$. This system is said to
be a {\em supersymmetric quantum mechanical} (SUSYQM) system if
\begin{enumerate}
\item $\cuf$ consists of two subspaces $\cuf_+$ and $\cuf_-$ and
there is an operator $(-1)^f$ with $\ll (-1)^f\rr^2=1$, such that
	\[ (-1)^f\ll \Psi \rr=\pm\Psi \;\;\;\;\mbox{if}\;\; \;
\Psi\in\cuf_\pm .\]
$f$ and $(-1)^f$ are called the {\em fermion number operator} and the
{\em chirality operator}. The quantum states represented by the elements
of  $\cuf_+$ and $\cuf_-$ are called the {\em bosonic} and the
{\em fermionic} states, respectively.
\item There are $N$ operators ${\cal Q}^I$, $I=1,\cdots ,N$, such that
	\bea
	\cuq^I \lll\cuf_\pm\rrr&=&\cuf_\mp\nn\\
	\cuq^{I\dagger}\lll\cuf_\pm\rrr&=&\cuf_\mp\label{qb1}\\
	\llc (-1)^f, \cuq^I\rrc&=&\llc (-1)^f,\cuq^{I\dagger}\rrc\: =\: 0
	\; .\nn
	\eea
The operators $\cuq$ are called the (nonhermitian) supersymmetry (SUSY)
{\em charges} or {\em generators}.
\item The SUSY  generators satisfy the following general superalgebra
conditions:
	\bea
	\llc \cuq^I,\cuq^{J\dagger}\rrc&=&2\d^{IJ}H
	\label{qb2} \\
	\llc \cuq^I,\cuq^J\rrc&=&\llc\cuq^{I\dagger},\cuq^{J\dagger}\rrc
	\: =\: 0 \; ,
	\label{qb3}
	\eea
where $I,J=1,\cdots,N$. These equations can also be written in the form
	\bea
	H&=&\oot\lll\cuq^I\cuq^{I\dagger}+\cuq^{I\dagger}\cuq^I\rrr
	\label{qb4} \\
	\lll\cuq^I\rrr^2\!\!&=&\lll\cuq^{I\dagger}\rrr^2\: =\: 0 \; ,
	\label{qb5}
	\eea
for all $I=1,\cdots,N$.
\end{enumerate}
A quantum system that satisfies these conditions is said
to have a {\em type N supersymmetry}. We can also define (2N)
{\em hermitian SUSY charges} simply by
	\be
	\begin{array}{c}
	Q_{1}^{I}:=\frac{1}{\sqrt{2}}\lll \cuq^I+\cuq^{I\dagger}\rrr\\
	Q_{2}^{I}:=\frac{i}{\sqrt{2}}\lll \cuq^I-\cuq^{I\dagger}\rrr\; .
	\end{array}
	\label{qb6}
	\ee
Equations (\ref{qb2}) and (\ref{qb3}) can be written in terms of $Q_\a^I$,
with $\a=1,2$. The result is:
	\be
	\llc Q_{\a}^{I},Q_{\b}^{J}\rrc=2\d_{\a\b}\d^{IJ}H\; ,
	\label{qb7}
	\ee
where $I,J=1,\cdots N$ and $\a,\b=1,2$. Alternatively, one has
	\be
	\begin{array}{ccc}
	\hspace{8mm} H=\lll Q_1^I\rrr^2= \lll Q_2^I\rrr^2 &\hspace{.5cm}&
	\forall I=1,\cdots ,N \\
	\llc Q_1^I,Q_2^J\rrc =0 &\hspace{.5cm}&
	\forall I,J=1,\cdots ,N\; .
	\end{array}
	\label{qb8}
	\ee

In a SUSYQM system the SUSY charges (and their adjoints)
map the bosonic state vectors to the fermionic state vectors and vice versa,
hence the name {\em Supersymmetry}. It turns out that there are many
quantum systems that display this type of supersymmetry
\cite{witten,susy}. For such systems the knowledge of
supersymmetry allows for a great deal of simplifications. In fact, there are
purely algebraic methods to obtain the energy spectrum of these systems,
e.g., \cite{susy-berry}. For
more details see \cite{susy} and references therein. A classification
of some SUSYQM systems is provided in \cite{susy-classification}.
The extension to field theory is discussed in \cite{sohnius,wess,west}.

\section{Degeneracy of Energy Levels}
\label{z2-B2}
According to (\ref{qb8}), for all $I=1,\cdots,N$, one has
	\[ \ll Q_{1}^{I},H\rr =0 \; .\]
Hence $H$ and $Q_{1}^{I}$ have simtultaneous eigenstates. If we
denote the eigenvectors by $|E,q_1\kt$, we have
	\bea
	H|q_1,E\kt &=& E|q_1,E\kt \label{qb9} \\
	Q_{1}^{I}|q_1,E\kt &=& q_{1}^{I}|q_1,E\kt \; .
	\label{qb10}
	\eea
Moreover, since $H=(Q_{1}^{I})^2$
	\[ E=(q_{1}^{I})^2 \; .\]
Thus we can simplify the notation by setting $q_1:=q_{1}^{I}$,
$|q_1\kt :=|q_1,E\kt $,
and write (\ref{qb9}) and (\ref{qb10}) in the following form
	\bea
	H|q_1\kt&=&q_{1}^{2}|q_1\kt \label{qb11}\\
	Q_{1}^{I}|q_1\kt &=& q_{1}|q_1\kt \; .\label{qb12}
	\eea
Let us choose $J\in\{ 1,\dots,N\}$ arbitrarily and consider
	\be
	|q_{2}\kt :=Q_{2}^{J}|q_1\kt \; .
	\label{qb13}
	\ee
Then, it is easy to show that
	\bea
	Q_{1}^{I}|q_2\kt&=&Q_{1}^{I}Q_{2}^{J}|q_1\kt \nn \\
	&=& -Q_{2}^{J}Q_{1}^{I}|q_1\kt \label{qb14} \\
	&=&-q_{1}Q_{2}^{J}|q_1\kt \nn \\
	&=&-q_{1}|q_2\kt \; ,\nn
	\eea
where in the second, third, and fourth equalities use has been made of
(\ref{qb8}), (\ref{qb12}), and (\ref{qb13}).
Equation~(\ref{qb14}) indicates that $|q_2\kt$ is also an eigenstate
vector of $Q_{1}^{I}$. We have
	\bea
	Q_{1}^{I}|q_2\kt&=&-q_1|q_2\kt \label{qb15} \\
	H|q_2\kt&=&(Q_{1}^{I})^2 |q_2\kt\: =\:q_{1}^{2}|q_2\kt \; .\label{qb16}
	\eea
Equations (\ref{qb11}), (\ref{qb12}), (\ref{qb15}), and (\ref{qb16})
indicate that both $|q_1\kt$ and $|q_2\kt$ are energy eigenstate vectors.
Since $[Q_{1}^{I},Q_{2}^{J}]$ do not commute for all possible $I$ and
$J$, then
	\[ |q_1\kt\neq|q_2\kt \;\;\;\;\; \mbox{for}\;\;\; q_1\neq 0
	\neq q_2 .\]
Thus, the energy eigenstates are doubly degenerate unless the energy
eigenvalue equals zero. In fact, since $H$ is the square of some
hermitian operators its spectrum is nonnegative. Therefore, if the
ground state has positive energy, then all the energy eigenstates
are doubly degenerate. Furthermore, according to equations (\ref{qb1})
and (\ref{qb13}) the two state vectors $|q_1\kt$ and $|q_2\kt$ have
opposite chiralities, i.e., if $|q_1\kt$ is a bosonic (positive chirality)
state vector, $|q_2\kt$ will be a fermionic (negative chirality)
state vector and vice versa. $|q_1\kt$ and $|q_2\kt$ are called {\em
superpartners} associated with the energy eigenvalue $E=q_{1}^2$.

\section{Spontaneous Supersymmetry Breaking and the Witten Index}
\label{z2-B3}
If the ground state of a SUSYQM system has positive energy, then
the supersymmetry is said to be {\em spontaneously broken}.\footnote{This
is applied for other types of symmetry as well.} If, on the other
hand, there exist some state vector(s) $|\emptyset\kt$ such that
	\[ H|\emptyset\kt=(Q_{\a}^{I})^2|\emptyset\kt=0 \]
then the supersymmetry is said to be {\em unbroken} or {\em exact},
\cite{susy}.

A simple way of determinig whether a SUSYQM system is spontaneously
broken is to compare the number of bosonic ($n_{b}$) and fermionic states
($n_f$). If the SUSY is broken then all the energy states have positive
energy and thus there must be equal number of bosonic and fermionic
states (superpartners), i.e., the difference
	\be
	index_{\rm W}:=n_{b}-n_{f}
	\label{witten-index}
	\ee
vanishes.
The number $index_{\rm W}$ is called the {\em Witten index} of the
SUSYQM system. If $index_{\rm W}\neq 0$ then the supersymmetry
is definitely exact. Obviously, the numbers $n_b$ and $n_f$ are
in general infinite whereas their difference remains finite.\footnote{
I assume that the the ground state is not infinitely degerenerate.}
In fact, due to supersymmetry,
	\be
	index_{\rm W}=n_{b,0}-n_{f,0}\; ,
	\label{witten-index0}
	\ee
where $n_{b,0}$ and $n_{f,0}$ are the numbers of (linearly independent)
bosonic and fermionic zero energy eigenstate vectors, respectively.

A practicle way of computing $index_{\rm W}$ is through the following
set of identities:
	\bea
	index_{\rm W}&=& tr\ll (-1)^f\rr \nn \\
	&=& tr\ll (-1)^f e^{-i\b H}\rr \nn \\
	&=:&str\ll e^{-i\b H}\rr \; ,\label{qb17}
	\eea
where $tr$ and $str$ stand for {\em trace} and {\em supertrace} \cite{bd1},
$\b$ is a complex parameter with negative imaginary part.
The operator $\exp (-i\b H)$ plays the role of a ``regulator'' for the
trace. Note that, its insertion in the second equality does not
influence the result. It can contribute when the
trace is taken over the nonzero energy eigenstates, but these are
doubly degenerate and the contribution of the fermionic states cancels
the contribution of the bosonic states (due to the presence of $(-1)^f$.)
Incidentally, the right hand side of (\ref{qb17}) has a well-known
path integral representation. See \cite{bd1} and Sections~\ref{I-2-2}
and~\ref{I-2-3} for further details.

Probably, the most important hint that suggests
the relevance of $index_{\rm W}$
to the index theorem of mathematics is its ``topological invariance.''
Let us suppose that the Hamiltonian of a
SUSYQM system undergoes a continuous deformation such that the
SUSY remains intact. \footnote{This can be achieved
by changing the coefficients of different terms in the Hamiltonian
in an appropriate fashion.} In this case, some of the ground states may
be lifted to higher energies and vice versa. However, due to
supersymmetry the excited states can only exist in pairs. This
requires $index_{\rm W}$ to remain fixed while the total degree
of degeneracy of the ground state changes. To make contact with
topology, one can imagine a SUSYQM Hamiltonian that depends on the
geometric quantities of a differentiable manifold or a fibre bundle.
Under a smooth deformation of the geometric structure, i.e., under
a diffeomorphism, the Witten index remains unchanged, so in this sense
it is a topological invariant of the manifold or the bundle.

\section{Witten Index as the Index of an Elliptic Operator}
\label{z2-B4}
Consider an $N=1$ SUSYQM system and represent the elements of the
Fock space by
	\[ \Psi =\lll \begin{array}{c} \Psi_+ \\ \Psi_-
	\end{array} \rrr \; , \]
where $\Psi_{\pm}\in\cuf_{\pm}$. Then the operator $Q_1$ can be
represented by the matrix
	\be
	Q_1=\lll \begin{array}{cc}
	0&D^\dagger \\
	D&0 \end{array} \rrr\; ,
	\label{qb41}
	\ee
for some operator $D:\cuf_+\to\cuf_-$.\footnote{The adjoint operator
$D^\dagger$ maps $\cuf_-$ into $\cuf_+$.}
The Hamiltonian is then given by
	\be
	H=Q_1^2=\lll\begin{array}{cc}
	D^{\dagger}D&0\\
	0&DD^\dagger \end{array} \rrr\; .
	\label{qb42}
	\ee
Let us also  introduce
	\bea
	H_+&:=&D^\dagger D:\cuf_+\rar\cuf_+ \nn \\
	H_-&:=&DD^\dagger :\cuf_-\rar\cuf_- \; .\nn
	\eea
Evidently the kernels of $H_\pm$ are the subspaces of the zero energy
eigenspace. We have
	\bea
	dim\ll ker(H_+)\rr &=&n_{b,0}\label{qb43} \\
	dim\ll ker(H_-)\rr &=&n_{f,0}\; .\label{qb44}
	\eea
On the other hand, one can easily show that
	\bea
	ker(H_+)&=&ker(D^\dagger D)\: =\: ker(D) \label{qb45}\\
	ker(H_-)&=&ker(DD^\dagger )\: =\: ker(D^\dagger )\; .\label{qb46}
	\eea
Clearly, $ker(H_+)\subset ker(D)$, so to prove (\ref{qb45}), one must
show $ker(D)\subset ker(H_+)$. To see the latter, let
$|\Psi_+\kt\in ker(H_+)$ be an arbitrary element. Then,
$H_+|\Psi_+\kt =0$, so
	\[ 0=\br\Psi_+|H_+|\Psi_+\kt =\br\Psi_+|D^\dagger D|\Psi_+\kt
	=\parallel D|\Psi_+\kt\parallel \; .\]
This is sufficient to show that $D|\Psi_+\kt=0$, i.e., $\Psi_+\in ker(D)$,
and  so $ker(H_+)\subset ker(D)$. The proof of (\ref{qb46}) is similar.

Equations (\ref{m11}), (\ref{witten-index0}), and (\ref{qb43})--(\ref{qb46})
imply that
	\be
	index_{\rm W}=ker(D)-ker(D^\dagger)=index(D)\; .
	\label{witten-index3}
	\ee

The last equation describes the relation between the SUSYQM and the
Atiyah-Singer index theorem.
For general index problem, however, one first starts with
a given elliptic operator $D:\cinf (V_+)\to\cinf (V_-)$.
\footnote{See Appendix~\ref{z2-A} for a discussion of the index theorem.}
Thus, in priciple, a direct supersymmetric
proof of the index theorem involves the following two steps:
\begin{enumerate}
\item Construct a SUSYQM system  whose bosonic and fermionic Fock
subspaces $\cuf_\pm$ are identified with (the completions of) the spaces
of sections $\cinf (V_\pm)$ of $V_\pm$ and whose supersymmetric charge $Q_1$
is related to the elliptic operator as described above.
\item Use the path integral representation of the supertrace of the
evolution operator (\ref{qb17}) to compute the index.
\end{enumerate}

There are further similarities between the constructions of
Appendix~\ref{z2-A} and those encountered here. Following
Appendix~\ref{z2-A5}, let us extend the operator $D$ of (\ref{qb41}) to
	\[ d:\cuf\rar\cuf\; ,\]
by defining
	\bea
	\lpt d\right|_{\cuf_+}&:=& D \nn \\
	\lpt d\right|_{\cuf_-}&:=& 0 \nn \\
	\lpt d^\dagger\right|_{\cuf_+}&:=&0 \nn\\
	\lpt d^\dagger\right|_{\cuf_-}&:=& D^\dagger \; .\nn
	\eea
Then, we have the trivial identity
	\[ d^2 =(d^\dagger )^2=0\; .\]
In terms of $d$ and $d^\dagger$, the Hamiltonian is written in the form
	\[ H=d^\dagger d+dd^\dagger \; .\]
This suggests that the operators $d$ is related to the
non-hermitian supersymmetric charge $\cuq$. In fact, we have
	\bea
	d&=&\lpt\frac{1}{\sqrt{2}}(\cuq +\cuq^\dagger )\right|_{\cuf_+}\nn \\
	d^\dagger&=&\lpt\frac{1}{\sqrt{2}}(\cuq +\cuq^\dagger )
	\right|_{\cuf_-}\; .\nn
	\eea
Then,
	\bea
	H_+&=&D^\dagger D=\lpt\oot (\cuq +\cuq^\dagger )^2\right|_{\cuf_+}
	\: =\: \lpt\oot (\cuq\cuq^\dagger +\cuq^\dagger\cuq )\right|_{\cuf_+}
	\nn \\
	H_-&=&DD^\dagger=\lpt\oot (\cuq +\cuq^\dagger )^2\right|_{\cuf_-}
	\: =\: \lpt\oot (\cuq\cuq^\dagger +\cuq^\dagger\cuq )\right|_{
	\cuf_-}\; .\nn
	\eea
Consequently, we recover the familiar equation
	\bea
	H&=&H_+ +H_-\: =\: \oot (\cuq^\dagger\cuq +\cuq\cuq^\dagger )\nn \\
	Q_1&=&d+d^\dagger\: =\: \frac{1}{\sqrt{2}}(\cuq +\cuq^\dagger )
	\; ,\nn
	\eea
that were labelled (\ref{qb4}) and (\ref{qb6}), respectively.

A simple comparision of the quantities defined here with those of
Appendix~\ref{z2-A5}, suggests the following additional identifications:
	\[ \Omega\equiv \cuf\; ,\;\;\;\;\;\Omega_{\pm}\equiv\cuf_{\pm}\; ,\]
	\[ \Delta\equiv H\; ,\;\;\;\;\;\Delta_\pm\equiv H_\pm\; .\]

In spite of the existence of all these identities, the construction of a
SUSYQM system for arbitrary elliptic operators is a difficult
and sometimes an impossible task. An obvious problem is related to the
fibre dimension of the corresponding vector bundles.
For, the Fock space of SUSYQM systems cannot be made to take
arbitrary dimensions whereas there are no restrictions on
the fibre dimension of $V$. A more important problem with a direct
supersymmetric proof of the index theorem involves the second step
in the above list. Namely, the path integral techniques presently known
do not allow for a reasonable supersummetric derivation
of the index of second or higher order elliptic operators. These
difficulties have led the physicsts to restrict to the supersymmetric
proofs of the twisted Dirac operator which implies a proof of the
general index theorem via K-theory.

\section{Coherent State (Holomorphic) Representation of Quantum
Mechanics, Supernumbers, and Berezin Integration}
\label{z2-B5}
In the coherent state representation \cite{susy,bd1,coherent-state}
of a quantum system, the Hilbert (Fock) space is generated from
a ground state vector $|\emptyset\kt$ by  repeated action of the
{\em creation operators}. Let us first consider a bosonic system
with one degree of freedom. The creation operator is
denoted by $\hat{a}^\dagger$. It is related to the coordinate operator
$\hat{x}$
and momentum operator $\hat{p}$ of the one-dimensional quantum mechanics via
the familiar folmula:
	\be
	\hat{a}^\dagger :=\frac{1}{\sqrt{2}}(\hat{x}-i\hat{p})\; .
	\label{creation}
	\ee
The operator $\hat{a}$ is called the {\em annihilation operator}
	\be
	\hat{a}=\frac{1}{\sqrt{2}}(\hat{x}+i\hat{p})\; .
	\label{annihilation}
	\ee
Creation and annihilation operators satisfy a set of  commutation
relations  given by
	\bea
	\ll \hat{a},\hat{a}^{\dagger}\rr &=&1  \nn\\
	\ll \hat{a},\hat{a}\rr&=& 0 \label{qb51} \\
	\ll \hat{a}^\dagger ,\hat{a}^\dagger \rr &=& 0 \; .\nn
	\eea
The collection of the following state vectors
	\be
	(\hat{a}^\dagger)^r|\emptyset\kt\; ,\;\;\;\;\;r=0,1,2,\cdots
	\; .
	\label{qb52}
	\ee
form a basis. Thus, any state vector $|\Psi_+\kt$
can be wirtten as a linear combination of them, i.e.,
	\be
	|\Psi_+\kt =\sum_{r=0}^{\infty}c_r\, (\hat{a}^\dagger )^r|\emptyset\kt
	\; ,
	\label{qb53}
	\ee
where $c_r$ are complex coefficients. In fact, we can view (\ref{qb53})
as the Taylor series expnasion of a function $\hat{F}=F(\hat{a}^\dagger )$.
Alternatively, we define a {\em holomorphic} (analytic) function
of a single complex variable $z^*$ that is associated
with the state vector $\Psi_+$, namely
	\[ F(z^*):=\sum_{r=0}^{\infty}c_r\, (z^*)^{r}\; ,\;\;\;\; z^*\in\C
	\; .\]
Clearly, the correspondence is one-to-one. The choice of parameter $z^*$
rather than $z$ has conventional reasons.

The same  approach can be directly generalized to arbitrary $n$-dimensional
quantum systems. In that case, one introduces the creation operators
$\hat{a}^{i\dagger}$ and the annihilation operators $\hat{a}^{i}$
associated with pairs of coordinate and momentum operators.
They satisfy
	\bea
	\ll \hat{a}^{i},\hat{a}^{j\dagger}\rr&=&\d_i^j \nn\\
	\ll \hat{a}^i,\hat{a}^j\rr&=&0 \label{qb54} \\
	\ll \hat{a}^{i\dagger},\hat{a}^{j\dagger}\rr&=& 0\; ,\nn
	\eea
where $i,j=1,\cdots n$.

Again the Hilbert (Fock) space is spanned by monomials in
$\hat{a}^{i\dagger}$, i.e., every state vector $|\Psi_+\kt$ can be expressed
as
	\[ |\Psi_+\kt = \sum_{r_1,\cdots,r_n=0}^{\infty}c_{r_1\cdots r_n}\,
	(\hat{a}^{n\dagger})^{r_n}\cdots (\hat{a}^{1\dagger})^{r_1}
	|\emptyset\kt\; .\]
Thus, every state vector corresponds to a holomorphic function $F$ of
$n$ complex variable $z^{i*}$,
	\be
	|\Psi_+\kt\equiv F(z^{1*},\cdots,z^{n*}):=
	\sum_{r_1,\cdots,r_n=0}^{\infty}c_{r_1\cdots r_n}\, (z^{n*})^{r_n}
	\cdots (z^{1*})^{r_1}\; .
	\label{qb55}
	\ee
The validity of this identification relies heavily on the fact that
the creation operators commute among themselves, (\ref{qb54}).
Otherwise, one could not have replaced the operators by numbers. In fact,
for the fermionic systems the creation opertaors do not commute
among themselves, so one cannot use the holomorphic representation
based on complex numbers. Instead, one uses the anticommutativity of
the fermionic creation operators and replaces the role of complex variables
$z^*$ in the above construction by {\em anticommuting Grassman numbers}. These
are the ``odd'' elements of a {\em Grassman algebra} with an infinite
number of generators. They are also called {\em supernumbers}, \cite{bd1}.

The coherent states are indeed the eigenstates of the annihilation operators.
Let us consider the state vectors
	\bea
	|z\kt &:=& e^{-\frac{1}{2}z^{i*}z^i+\hat{a}^{i\dagger}z^i}|\emptyset\kt
	\; ,
	\label{qb56}\\
	\br z^*|&:=&|z\kt^\dagger \; , \nn
	\eea
where the sum over repeated indices is understood. Then, it is a matter
of simple algebra to show that
	\be
	\hat{a}^{i}|z\kt =z^i|z\kt \; ,\;\;\;\;\;\;\forall i=1,\cdots ,n\; .
	\label{qb57}
	\ee
In view of (\ref{qb55}) and (\ref{qb57}), one has
	\be
	\br z^*|\Psi_+\kt = F(z^{1*},\cdots ,z^{n*})\; ,
	\label{qb58}
	\ee
up to a common factor of $\br z^*|\emptyset\kt$.
Equation (\ref{qb58}) is called the {\em coherent state or holomorphic
representation} of the state vector $|\Psi_+\kt$.

Let us examine the expression for the inner product of two state vectors
$|\Psi_{+1}\kt$ and $|\Psi_{+2}\kt$. Let us introduce the basic
states
	\bea
	\br z^*|i,r\kt&:=&\frac{(z^{i*})^r}{\sqrt{r!}} \nn \\
	\br i,r|z\kt&:=&\br z^*|i,r\kt^\dagger\: =\: \frac{(z^i)^r}{
	\sqrt{r!}} \; ,\nn
	\eea
where $i=1,\cdots,n$ and $r=0,1,2,\cdots$. Then one can formally write
	\be
	\br\Psi_{+1}|\Psi_{+2}\kt = \int \br\Psi_{+1}|z\kt
	\br z|z^{'*}\kt\br z^{'*}|\Psi_{+2}\kt \prod_{i=1}^{n}dz^{i}
	dz^{i*}dz^{'i}dz^{'i*}\; .
	\label{qb59}
	\ee
Note that the quantities $\br z|$, $|z^{'*}\kt$, and $\br z|z^{'*}\kt$
have not been defined yet. In fact, we need to define the latter to have a
coherent state representation of the inner product. This quantity
behaves like a ``measure'' for the  integral depicted in (\ref{qb59}).
It is naturally chosen by requiring that the basic state vectors are
orthonormal, namely
	\[ \br i,r|j,s\kt=\d_{ij}\d_{rs} \; .\]
This suggests immediately
	\be
	 \br z|z^{'*}\kt :=\frac{e^{-z^{k*}z^{k}}}{(2\pi i)^n}\d (z-z')
	\; .
	\label{bosonic}
	\ee
Finally, one finds
	\be
	\br\Psi_{+1}|\Psi_{+2}\kt :=\int \br\Psi_{+1}|z\kt
	\br z^*|\Psi_{+2}\kt e^{-z^{k*}z^k}\prod_{l=1}^{n}
	\frac{dz^ldz^{l*}}{2\pi i} \; .
	\label{qb60}
	\ee
Here the multiple integral is over whole $\C^n=\R^{2n}$. It is performed
by changing to real variables.

The situation is quite similar for fermionic systems. The only difference
is in the use of complex anticommuting  supernumbers $\zeta^{i*}$ in the
place of the ordinary complex numbers $z^{i*}$. The creation operators
$\hat{\a}^{i\dagger}$ and the annihilation operators $\hat{\a}^i$ satisfy the
following anticommutation relations
	\bea
	\llc \hat{\a}^{i},\hat{\a}^{i\dagger}\rrc &=&\d^{ij}\nn\\
	\llc \hat{\a}^i,\hat{\a}^j\rrc&=&0\label{qb61}\\
	\llc\hat{\a}^{i\dagger},\hat{\a}^{j\dagger}\rrc&=&0\; .\nn
	\eea
The state vectors $|\Psi_-\kt$ are written as
	\[ |\Psi_-\kt =\sum_{r_1,\cdots,r_m=0}^{1}\g_{r_1,\cdots,r_m}\,
	(\hat{\a}^{m\dagger})^{r_m}\cdots (\hat{\a}^{1\dagger})^{r_1}
	|\emptyset\kt \; ,\]
where $m$ is the number of degrees of freedom, $r_i=0,1$, and $\g_{r_1\cdots
r_m}$ are complex coefficients. Follwing the
same line of reasoning as for the bosonic case, one represents $|\Psi_-\kt$
by a polynomial in the odd supernumbers $\zeta^{i*}$,
	\be
	|\Psi_-\kt\equiv\sum_{r_1,\cdots,r_m=0}^{1}\g_{r_1\cdots r_m}
	(\zeta^{m*})^{r_m}\cdots (\zeta^{1*})^{r_1}\; .
	\label{qb62}
	\ee
$\zeta^{i*}$ and their complex conjugate $\zeta^i$ anticommute
among themselves and commute with the ordinarly numbers and other
even supernumbers, \cite{bd1,berezin}. In particular,
	\bea
	\llc \zeta^{i*},\zeta^{j*}\rrc &=&0\nn\\
	\llc \zeta^i,\zeta^j\rrc&=&0\; .\nn
	\eea
Thus, $(\zeta^{i*})^2=(\zeta^i)^2=0$. It turns out that most of the
mathematical constructions which are based on ordinary real or complex
numbers can be generalized to supernumbers. A large number of these
are described in \cite{bd1}.

The holomorphic representation for fermionic systems is further
clarified by introducing the state vector
	\[ |\zeta\kt :=e^{-\frac{1}{2}\zeta^{i*}\zeta^{i}+
	\hat{\a}^{i\dagger}\zeta^i}|\emptyset\kt\;\]
Again, one can easily show that
	\[ \hat{\a}^i|\zeta\kt =\zeta^i|\zeta\kt\; . \]
Then, the right hand side of (\ref{qb62}) is conveniently denoted
by $\br\zeta^*|\Psi_-\kt$, where $\br\zeta^*|:=|\zeta\kt^\dagger$.

Next, one needs to devise an expression for the inner product of the
state vectors in the coherent state representation.  This requires the
notion of integration over the odd supernumbers, \cite{bd1}. The latter was
originally discovered by Berezin \cite{berezin}. I shall
follow the natural correspondence with the bosonic case. The inner
product of two state vectors $|\Psi_{-1}\kt$ and $|\Psi_{-2}\kt$
is formally given by:
	\be
	\br\Psi_{-1}|\Psi_{-2}\kt=\int \br\Psi_{-1}|\zeta\kt\br\zeta|
	\zeta^{'*}\kt\br\zeta^{'*}|\Psi_{-2}\kt\prod_{i=1}^{m}d\zeta^{'i*}
	d\zeta^{'i}d\zeta^{i*}d\zeta^i \; .
	\label{qb63}
	\ee
Once more, one needs to define $\br\zeta|\zeta^{'*}\kt$. The natural
choice is~\footnote{Here I assume that the delta function for odd
supernumbers $\zeta$ satisfies the familiar property:
	\[ \int f(\zeta,\zeta')\d (\zeta -\zeta')d\zeta' d\zeta^{'*} =
	f(\zeta,\zeta )\; .\]}
	\be
	\br\zeta|\zeta^{'*}\kt:=\frac{e^{-\zeta^{k*}\zeta^k}}{(2\pi i)^m}
	\d (\zeta -\zeta' )\; .
	\label{qb64}
	\ee
This choice leads to
	\be
	\br\Psi_{-1}|\Psi_{-2}\kt :=\int \br\Psi_{-1}|\zeta\kt\br\zeta^*|
	\Psi_{-2}\kt e^{-\zeta^{k*}\zeta^k}\prod_{l=1}^{m}
	\frac{d\zeta^{l*}d\zeta^l}{2\pi i}\; .
	\label{qb65}
	\ee
Here the integrals over $\zeta$ and $\zeta^*$'s are Berezin integrals which
is simply defined by requiring the basic state vectors to be orthonormal.
For simplicity, let us concentrate on the case, $m=1$. Then there
are obviously two basic state vectors;
	\[
	|\emptyset\kt\; ,\;\;\;\;\; |1\kt :=\hat{\a}^\dagger
	|\emptyset\kt\; .\]
In the coherent state representation, they are expressed by:
	\[ \br\zeta^*|\emptyset\kt =1\; ,\;\;\;\;\;
	\br\zeta^*|1\kt=\zeta^*\; .\]
Then, the orthonormality condition
	\bea
	\br\emptyset |\emptyset\kt&=&\br 1|1\kt\: =\: 1\nn\\
	\br\emptyset |1\kt&=&\br 1|\emptyset\kt\: =\: 0\;  .\nn
	\eea
together with (\ref{qb65}) lead to the following rules:
	\bea
	\int \zeta\, d\zeta &=&\int \zeta^*d\zeta^* \: =\:\sqrt{2\pi i}
	\label{qb66}\\
	\int \zeta\, d\zeta &=&\int \zeta^*d\zeta^* \: =\: 0\; .
	\label{qb67}
	\eea
This confirms DeWitt's convention of the choice of the factor $\sqrt{2\pi i}$
in (\ref{qb66}), \cite{bd1}. In Berezin's original work this factor
was chosen to be $1$, \cite{berezin}.
The latter corresponds to the following alternative
choice for (\ref{qb64}):
	\[ \br\zeta |\zeta^{'*}\kt =e^{-\zeta^{*k}\zeta^k}\d (\zeta'-\zeta )
	\; ,\]
which has less resemblance to its bosonic counterpart, i.e.,
(\ref{bosonic}). The Berezin integration satisfies many familiar
properties such as linearity. For further details, see \cite{bd1}
and the references therein.

\chapter{Derivation of the Scalar Curvature Factor in the Operator
Formalism}
\label{z2-C}
\vspace{1cm}
\bcc
{\bf Abstract}
\ecc
\bcc
\parbox[b]{4.5in}{\small
The supersymmetry algebra is used to obtain the quantum mechanical
Hamiltonian operator of the system studied in Chapters~\ref{I-2}
and~\ref{I-3}. It is shown that this Hamiltonian includes a factor of
$\frac{\hbar^2}{8}R$. This factor originates from the supercommutation
relation involving the fermionic variables.}
\ecc
\vspace{.5in}
\newcommand{\gf}{g^{\frac{1}{4}}}
\newcommand{\gmf}{g^{-\frac{1}{4}}}
The computation of the Hamiltonian $H$ involves some simple but rather
tedious algebra. One starts from the formula for the supersymmetric charge
(\ref{eq34}):
	\be
	Q=\mbox{\small$\frac{1}{\sqrt{\hbar}}$}
	\p^\n g^{\frac{1}{4}}p_\n g^{-\frac{1}{4}}
	\; ,
	\label{qc1}
	\ee
and the superalgebra condition (\ref{eq10.1}):
	\[ H=Q^2\; .\]
The  latter can be written in the form
	\bea
	H&=&\oot\{ Q,Q\}\nn\\
	&=&\mbox{\small$\frac{1}{2\hbar}$}
	\gf\llc \p^\n p_\n ,\p^\m p_\m \rrc \gmf
	\label{qc2} \\
	&=&\mbox{\small$\frac{1}{2\hbar}$}
	\gf\lll\p^\n\p^\m\ll p_\n ,p_\m\rr +\p^\n \ll p_\n ,\p^\m\rr p_\m \rpt
	\nn \\
	&&\hspace{.8cm}\lpt + \p^\m \ll p_\m ,\p^\n\rr p_\n +
	\llc \p^\m ,\p^\n\rrc p_\m p_\n\rrr\gmf \; ,\nn
	\eea
where in the second and third equalities use has been made of
the supercommutation relations (\ref{eq33}) and the following
identities
	\bea
	\llc AB,C\rrc&=&A\ll B,C\rr +\llc A,C\rrc B
	\label{qc3} \\
	\ll AB,C\rr&=&A\ll B,C\rr +\ll A,C\rr B
	\label{qc4}\\
	\ll AB,C\rr &=&A\llc B,C\rrc -\llc A,C\rrc B \; .
	\eea
Implementing (\ref{eq33}) in (\ref{qc2}), one obtains
	\bea
	H&=&\oot\gf\ll -\oot R_{\m\n\s\t}\p^\m\p^\n\p^\s\p^\t
	    +2i\G^{\m}_{\n\l}\p^\n\p^\l p_\m +g^{\m\n}p_\m p_\n\rr\gmf
	\nn \\
	&&-\mbox{\small$\frac{\kappa}{2}$}F^{ab}_{\n\m}\p^\n\p^\m\e^{a*}\e^b
	\; .\label{qc5}
	\eea
Next step is to recognize the following identities
	\bea
	R_{\m\n\s\t}\p^\m\p^\n\p^\s\p^\t&=&-
	\mbox{\small$\frac{\hbar^2}{2}$}R \label{qc6}\\
	\gf g^{\m\n}p_\m p_\n&=& \gmf p_\m g^{\frac{1}{2}}g^{\m\n}p_\n
	-i\hbar\gf g^{\t\m}\G^{\n}_{\t\m}\, p_\n \label{qc7} \\
	2i\G^{\m}_{\n\l}\p^\n\p^\l&=&i\hbar g^{\n\l}\G^{\m}_{\n\l}\; .
	\label{qc8}
	\eea
Substituting (\ref{qc6}), (\ref{qc7}) and (\ref{qc8}) in (\ref{qc5})
and simplifying the result, one arrives at
	\[ H=\oot\gmf p_\m g^{\frac{1}{2}}g^{\m\n}p_\n\gmf +
	\mbox{\small$\frac{\hbar^2}{8}$}R+
	\kappa\ll -\oot F_{\l\g}^{ab}\p^\l\p^\g\e^{a*}
	\e^b\rr\; . \]

In the rest of this appendix, I shall present the derivation of the
equations (\ref{qc6}) and (\ref{qc7}). The derivation of (\ref{qc8})
is straightforward.

Consider the following set of identities:
	\bea
	R_{\m\n\s\t}\p^\m\p^\n\p^\s\p^\t&=&\mbox{\small$\frac{1}{3}$}\p^\m\lll
	R_{\m\n\s\t}\p^\n\p^\s\p^\t +R_{\m\t\n\s}\p^\t\p^\n\p^\s\rpt \nn \\
	&&\lpt \hspace{.8cm} +R_{\m\s\t\n}\p^\s\p^\t\p^\n\rrr \label{qc9} \\
	\p^\t\p^\n\p^\s &=&\p^\n\p^\s\p^\t -\hbar g^{\t\s}p^\n +
	\hbar g^{\t\n}\p^\s\; ,
	\label{qc10}
	\eea
and similarly
	\be
	\p^\s\p^\t\p^\n =\p^\n\p^\s\p^\t -\hbar g^{\n\s}\p^\t +
	\hbar g^{\t\n}\p^\s\; . \label{qc11}
	\ee
In (\ref{qc10}) and (\ref{qc11}), use has been made of (\ref{eq33}),
namely
	\[ \llc \p^\l ,\p^\g\rrc = \hbar g^{\l\g}\; .\]
Substituting (\ref{qc10}) and (\ref{qc11}) in the right and side of
(\ref{qc9}), one has
	\bea
	R_{\m\n\s\t}\p^\m\p^\n\p^\s\p^\t &=&\mbox{\small$\frac{1}{3}$}\p^\m
	\lll \ll R_{\m\n\s\t}+R_{\m\t\n\s}+R_{\m\s\t\n}\rr\p^\n\p^\s\p^\t
	\rpt  \label{qc12} \\
	&& \hspace{-2cm} + \lpt \hbar R_{\m\t\n\s}\ll
	-g^{\t\s}\p^\n +g^{\t\n}\p^\s\rr +
	\hbar R_{\m\s\t\n}\ll -g^{\n\s}\p^{\t}+g^{\t\n}\p^{\s}\rr\rrr \; .\nn
	\eea
The first bracket on the right hand side of (\ref{qc12}) vanishes due to
the cyclic identity of the Riemann curvature tensor, \cite[p.~57]{stephani}.
The remaining term are further simplified to yeild
	\bea
	R_{\m\n\s\t}\p^\m\p^\n\p^\s\p^\t&=&-\hbar R_{\m\n}\p^\m\p^\n \nn \\
	&=& -\frac{\hbar}{2} R_{\m\n}\llc \p^\m ,\p^\n\rrc \nn \\
	&=& -\frac{\hbar^2}{2}R_{\m\n}g^{\m\n}\nn \\
	&=&-\frac{\hbar^2}{2}R\; ,\nn
	\eea
where I have used (\ref{eq33}). $R_{\m\n}$ and $R$ are the Ricci curvature
tensor and scalar, respectively. This completes the derivation of (\ref{qc6}).

Dervation of (\ref{qc7}) uses the following set of identities
	\bea
	\gmf p_\m g^{\frac{1}{2}} g^{\m\n}p_{\n}&=&\gmf\lll g^{\frac{1}{2}}
	g^{\m\n}p_{\m}+\ll p_\m ,g^{\frac{1}{2}}g^{\m\n}\rr \rrr p_\n \nn \\
	&=&\gf g^{\m\n}p_{\m}p_\n +\gmf\ll -i\hbar (g^{\frac{1}{2}}g^{\m\n})_{
	,\m}\rr p_\n \label{qc13} \\
	&=&\gf g^{\m\n}p_\m p_\n -i\hbar\gf\lll \oot g^{\m\n}g^{\s\t}-
	g^{\m\t}g^{\n\s}\rrr g_{\t\s ,\m}p_\n \; , \nn
	\eea
where use has been made of (\ref{eq33}) and
	\bea
	(g^{\frac{1}{2}})_{,\m}&=&\oot g^{\frac{1}{2}}g^{\s\t}g_{\s\t ,\m}
	\nn \\
	(g^{\m\n})_{,\m}&=&-g^{\m\t}g^{\n\s}g_{\t\s ,\m} \; . \nn
	\eea
Moreover, one has
	\be
	g_{\t\s,\m}=\G_{\t\s\m}+\G_{\s\t\m}\; .
	\label{qc14}
	\ee
Substituting (\ref{qc14}) in (\ref{qc13}), contracting the Christoffel
symbols with the metric tensors and simplifying the result, one obtains
equation (\ref{qc7}).

\chapter{Coherent State Path Integral and Calculation of $\so$}
\label{z2-D}
\vspace{1cm}
\bcc
{\bf Abstract}
\ecc
\bcc
\parbox[b]{4.5in}{\small
The end point contributions to the coherent state path integral is
discussed. A method is introduced that includes these contributions
authomatically in the action functional. This is done by considering
paths with step function discontinuities at the end points. The
method developed here allows for a direct application of the ordinary
path integral formula in the coherent state represenation. This method
is used to compute the factor $S_{0}$ in the WKB expression for
the path integral analyzed in Chapter~\ref{I-2}.}
\ecc
\vspace{.5in}

In the coherent state path integral formula for the kernel, one
must also include the end point contributions to the action,
\cite{it-zu,faddeev}\footnote{In fact, the action function can be
shown to include the end point contributions automatically. This
was pointed out to the author by Cecile DeWitt-Morette.}.
This can be done implicitly by including
step functions at the end points of the paths. An example of this
is provided in \cite{bd1}.

\renewcommand{\z}{\zeta}
Consider a fermionic quantum system and let $\hat{\z}^{i\dagger}$ and
$\hat{\z}^{i}$ be the creation and annihilation operators:
\bea
\{\hat{\z}^{i\dagger},\hat{\z}^{j}\}&=&\d^{ij} \nn \\
\{\hat{\z}^{i},\hat{\z}^{j}\} & = & \{\hat{\z}^{i\dagger},\hat{\z}^{j\dagger}
\} = 0 \; . \nn
\eea
The coherent states are defined by:
\bea
\mid\!\z\kt &:=& e^{\frac{1}{2}\z^{i*}\z^{i}+\hat{\z}^{i\dagger}\z^{i}}
\mid\emptyset\kt \nn \\
\br \z^{*}\!\mid &:=& \mid\! \z \kt^{\dagger}, \nn
\eea
where $|\emptyset\kt$ is the vacuum and $\z \in \C$. \footnote{
See Appendix~\ref{z2-B} for further detail.} Then one has:
\[ \hat{\z}^{i}\mid\! \z\kt =\z^{i}\mid\!\z\kt \hspace{5mm}\mbox{and}
\hspace{5mm}\br\z^{*}\!\mid \hat{\z}^{i\dagger} = \z^{i*}\br\z^{*}\!\mid \; .\]
In the path integral formula for the kernel:
\be
K(\z^{*''},t''\mid\z ',t'):=\br\z^{*''};t''\mid\z ';t'\kt =
\int e^{iS}(sdet\, G^{+})^{-\frac{1}{2}}{\cal D}\z^{*}\, {\cal D}\z\:,
\label{eqa1}
\ee
one must note that $\z^{*''}\neq (\z '')^{*}$. In fact a priori
only
\be
\z (t')=\z ' \hspace{1cm}\mbox{and}\hspace{1cm}\z^{*}(t'')=\z^{*''}
\label{eqa2}
\ee
are fixed. The boundary conditions on
\[ \z (t'')=\z ''\hspace{1cm}\mbox{and}\hspace{1cm}\z^{*}(t')=\z^{*'}\]
depend on the specifics of the problem.
In (\ref{eqa1}), the action is given by:
\be
S=\int_{\z ',t'}^{\z '',t''}dt\ll \frac{1}{2i}(\z^{i*}\dot{\z}^{i}
-\dot{\z}^{i*}\z^{i})-h(\z^{*},\z ,t)\rr
\label{eqa3}
\ee
where $h(\z^{*},\z ,t)$ is the normal symbol of the Hamiltonian,
\cite{it-zu}. The paths are given by:
\be
\begin{array}{ccc}
\z^{*}(t) & = & \lim_{\epsilon\rightarrow 0}\ll \th (t-t'-\epsilon )
\z_{c}^{*}(t) + \th (t'-t+\epsilon )\z^{*'}\rr \\
\z (t) & = & \lim_{\epsilon\rightarrow 0}\ll \th (t''-t-\epsilon )
\z_{c}(t) + \th (t-t''+\epsilon )\z ''\rr \; .
\end{array}
\label{eqa4}
\ee
In (\ref{eqa4}), $\z_{c}^{*}(t)$ and $\z_{c}(t)$ are paths connected
to the end points, i.e. :
\be
\z_{c}^{*}(t'')=\z^{*''}\hspace{1cm}\mbox{and}\hspace{1cm}
\z_{c}(t')=\z '
\label{eqa5}
\ee
In general,
\[ \z_{c}^{*}(t')\neq\z^{* '}\hspace{1cm}\mbox{and}\hspace{1cm}
\z_{c}(t'')\neq\z '' \: .\]
To compute $\so$, which appears in (\ref{eq112}), one needs to
specialize to the classical paths. In the limit $\b\rightarrow 0$,
these are given by (\ref{eq115}) and (\ref{eq116}).
Equations~(\ref{eq116}) are easily solved to give:
\bea
\tilde{\e}_{0c}^{a}(s) & = & \ll e^{i(\tcuf +\a\In )s} \rr^{ab}\e^{b'}
\label{eqa6} \\
\tilde{\e}^{a*}_{0c}(s) & = & \e^{b*''} \ll e^{i(\tcuf +\a\In )(1-s)}\rr^{ba}
\label{eqa7} \\
\tilde{\p}_{1c}^{\g}(s) & = & ig_{0}^{\g \l}F_{\l \d}^{ab}(\xo )
\tilde{\p}_{0}^{\d} E^{ab}(s) + C^{\g} \; , \label{eqa8}
\eea
where
\[ E^{ab}(s):=\int_{0}^{1} ds \, \tilde{\e}_{0c}^{a*}(s)\tilde{\e}_{0c}^{b}(s)
\, .\]
In particular, one has:
\be
\mbox{\fs$\frac{1}{2}$}g_{0\l\g}\tilde{\p}_{0}^{\l}\tilde{\p}_{1c}^{\g}(s)
=i\e^{a*''}\ll e^{i(\tcuf +\a\In )}\tcuf\rr^{ab}\e^{b'}s+C\; .
\label{eqa9}
\ee
In (\ref{eqa8}) and (\ref{eqa9}) $C^{\l}$ and $C$ are constants to be
determined by the boundary conditions on $\p$'s.

In view of equations (\ref{eq115}) and (\ref{eq17}), one has
\bea
\so &=& \int_{0}^{1}ds\ll \mbox{\fs$\frac{i}{2}$}g_{0\l\g}\,\tilde{\p}_{0}^{\l}
\,\dot{\tilde{\p}}_{1}^{\g}(s)+\mbox{\fs$\frac{i}{2}$} \left(
\tilde{\e}_{0}^{a*}(s)\,\dot{\tilde{\e}}_{0}^{a}(s)
-\dot{\tilde{\e}}_{0}^{a*}(s)\,\tilde{\e}_{0}^{a}(s) \right) + \right. \nn \\
& & \hspace{1.5cm} \left. + \tcuf^{ab}\,\tilde{\e}_{0}^{a*}(s)\,
\tilde{\e}_{0}^{b}(s) \rr + O(\b )\; .
\label{eqa10}
\eea
Defining
\[ \tilde{\xi}^{i}:=\sqrt{\b}\xi^{i}\; ,\]
and using (\ref{eq36}) and (\ref{eq38}), one may express
the first term on the right hand side of (\ref{eqa10}) in the form:
% ------------------------------
\newcommand{\tx}{\tilde{\xi}}
\newcommand{\dtx}{\dot{\tx}}
\renewcommand{\iot}{\mbox{\fs$\frac{i}{2}$}}
\newcommand{\tp}{\tilde{\p}}
% ------------------------------
\bea
\int_{0}^{1} ds \ll \iot g_{0\l\g}\, \tp_{0}^{\l}\, \dot{\tp}_{1}^{\g}
(s) \rr &=& \int_{0}^{\b} dt \ll \iot
g_{0\l\g}\, \p_{0}^{\l}(t)\, \dot{\p}_{0}^{\g}(t) \rr + O(\b ) \label{eqa11} \\
&=& \int_{0}^{\b} dt \ll \iot \left( \xi_{0}^{i*}(t)\, \dot{\xi}_{0}^{i}(t)
-\dot{\xi}_{0}^{i*}(t)\, \xi_{0}^{i}(t) \right) \rr + O(\b )\nn \\
&=&
\int_{0}^{1} ds \ll \iot \left( \tx_{0}^{i*}(s)\, \dtx_{0}^{i}(s)-
\dtx_{0}^{i*}(s) \, \tx_{0}^{i}(s) \right) \rr + O(\b ) \; . \nn
\eea
In view of (\ref{eqa11}), it is clear that (\ref{eqa10}) is already in
the form demanded by (\ref{eqa3}).
Next step is to evaluate (\ref{eqa10}). Let us define:
\bea
I_{1}&:=&\int_{0}^{1}ds \ll \iot g_{0\l\g}\tp_{0}^{\l}\dot{\tp}_{1}^{\g}(s)
\rr \label{eqa12} \\
&\stackrel{\b\rightarrow 0}{=}&\int_{0}^{1}ds\ll \iot \left(
\tx_{0}^{i*}(s)\, \dtx_{0}^{i}(s) - \dtx_{0}^{i*}(s)\, \tx_{0}^{i}(s)
\right) \rr \nn \\
I_{2}&:=&\int_{0}^{1}ds\ll \iot\left( \tilde{\e}_{0}^{a*}(s)\,
\dot{\tilde{\e}}_{0}^{a}(s)-\dot{\tilde{\e}}_{0}^{a*}(s)\,
\tilde{\e}_{0}^{a}(s)\right)+\tcuf^{ab}\tilde{\e}_{0}^{a*}(s)\,
\tilde{\e}_{0}^{b}(s) \rr \; ,\label{eqa13}
\eea
so that
\be
\so = I_{1}+I_{2} \; .
\label{eqa14}
\ee
Replacing $\z$'s by $\tx_{0}$'s and setting $t'=0$ and $t''=1$ in
(\ref{eqa4}), it is a matter of simple algebra to show that:
\bea
I_{1}&=& I_{1c}+\partial \nn \\
I_{1c}&:=& \int_{0}^{1}ds\ll \iot \left( \tx_{0c}^{i*}(s)\,\dtx_{0c}^{i}(s)
-\dtx_{0c}^{i*}(s)\,\tx_{0c}^{i}(s)\right)\rr \nn \\
\partial &:=& -\iot \ll \tx_{0}^{i*''}\left( \tx_{0c}^{i}(1)-\tx_{0}^{i''}
\right) + \left( \tx_{0c}^{i*}(0)-\tx_{0}^{i*'}\right)\tx_{0}^{i'} \rr \; .\nn
\eea
Imposing periodic boundary conditions, (\ref{eq81}),(\ref{eq114}):
\be
\begin{array}{ccc}
\tx_{0}'=\tx_{0}''=:\tx_{0} &\hspace{1cm}& \e '=\e ''=:\e \\
\tx_{0}^{*'}=\tx_{0}^{*''}=:\tx_{0}^{*} &\hspace{1cm}& \e^{*'}=\e^{*''}
=:\e^{*}\; ,
\end{array}
\label{eqa15}
\ee
and using (\ref{eqa9}) and (\ref{eqa11}), one has:
\bea
I_{1c}&=&\int_{0}^{1}ds \ll \iot g_{0\l\g}\tp_{0}^{\l}\dot{\tp}_{c1}^{\g}(s)\rr
=-\e^{a*}\ll e^{i(\tcuf +\a\In)} \tcuf \rr^{ab} \e^{b} \nn \\
\partial &=& -\iot\ll \tx_{0}^{i*}\left( \tx_{0c}^{i}(1)-\tx_{0}^{i} \right)
+\left( \tx_{0}^{i*}(0)-\tx_{0}^{i*}\right) \tx_{0}^{i}\rr \; . \nn
\eea
Next one needs to use equations (\ref{eq36}),(\ref{eq38}), and (\ref{eqa8})
to compute $\tx_{0c}^{i}(1)$ and $\tx_{0c}^{i*}(0)$. In (\ref{eqa8}), the
boundary conditions must be chosen appropriately. The correct choice is
the following:
\[ \begin{array}{ccc}
\mbox{choose:}&\p (0)=\p_{0}=\frac{\tp_{0}}{\sqrt{\b}}\,\Rightarrow\,
\tp_{c1}(0)=0 & \mbox{ to compute:  $\tx_{0c}(1)$} \\
\mbox{choose:}&\p (\b )=\p_{0}=\frac{\tp_{0}}{\sqrt{\b}}\, \Rightarrow\,
\tp_{c1}(1)=0 & \mbox{ to compute:  $\tx_{0c}^{i*}(0)$}\; .
\end{array} \]
This leads to :
\[ \partial = \e^{a*}\ll e^{i(\tcuf +\a\In )}\tcuf \rr^{ab}\e^{b} \; ,\]
and hence
\be
I_{1}=0
\label{eqa16.9}
\ee

The computation of $I_{2}$ is straightforward. One replaces $\z$'s by
$\e_{0}$'s in (\ref{eqa4}), and sets $t'=0$ and $t''=1$. The final
result is obtained using (\ref{eqa6}),(\ref{eqa7}), and (\ref{eqa15}):
\be
I_{2}=-i\e^{a*}\ll e^{i(\tcuf +\a\In )} -\In \rr^{ab}\e^{b}\; .
\label{eqa17}
\ee
Combinning (\ref{eqa14}),(\ref{eqa16.9}), and (\ref{eqa17}), one has:
\[\so = -i\e^{a*}\ll e^{i(\tcuf +\a\In )}-\In \rr^{ab}\e^{b} \; .\]
This is used in Section~\ref{I-2-5}, (\ref{eq119}).
Taking $\tcuf = \e =\e^{*} =0$ ,
the situation reduces to the case of $\k =0$. In this case, one has
$\so = 0$.

\end{document}